\documentclass{article}

\usepackage{microtype}
\usepackage{graphicx}
\usepackage{subcaption}
\usepackage{booktabs} 

\usepackage{multirow}
\usepackage{array}
\usepackage{tikz}
\usetikzlibrary{arrows.meta, patterns, decorations.pathreplacing}
\usepackage{pgfplots}
\pgfplotsset{compat=1.17}

\usepackage{hyperref}

\usepackage[preprint]{icml2027}

\usepackage{amsmath}
\usepackage{amssymb}
\usepackage{mathtools}
\usepackage{amsthm}

\usepackage[capitalize,noabbrev]{cleveref}
\RemoveFromHook{label}[firstaid/cleveref]
\theoremstyle{plain}
\newtheorem{theorem}{Theorem}[section]
\newtheorem{proposition}[theorem]{Proposition}
\newtheorem{lemma}[theorem]{Lemma}
\newtheorem{corollary}[theorem]{Corollary}
\theoremstyle{definition}
\newtheorem{definition}[theorem]{Definition}

\theoremstyle{remark}

\usepackage[textsize=tiny]{todonotes}

\newcommand{\TBD}{\textcolor{red}{\textbf{[?]}}}
\makeatletter
\newcommand{\setval}[2]{\expandafter\gdef\csname val@#1\endcsname{#2}}
\newcommand{\val}[1]{\@ifundefined{val@#1}{\TBD}{\csname val@#1\endcsname}}
\makeatother
\IfFileExists{numbers.tex}{
\setval{abs.sharehigh}{103}
\setval{abs.sharelow}{73}
\setval{dep.narratives.0.25.forward}{0.13}
\setval{dep.narratives.0.25.indep}{0.24}
\setval{dep.narratives.0.25.reverse}{0.01}
\setval{dep.narratives.0.25.shared}{0.01}
\setval{dep.narratives.0.5.forward}{0.17}
\setval{dep.narratives.0.5.indep}{0.26}
\setval{dep.narratives.0.5.reverse}{0.01}
\setval{dep.narratives.0.5.shared}{0.00}
\setval{dep.narratives.0.75.forward}{0.16}
\setval{dep.narratives.0.75.indep}{0.29}
\setval{dep.narratives.0.75.reverse}{0.01}
\setval{dep.narratives.0.75.shared}{0.02}
\setval{dep.podcast.0.25.forward}{0.36}
\setval{dep.podcast.0.25.indep}{0.10}
\setval{dep.podcast.0.25.reverse}{0.02}
\setval{dep.podcast.0.25.shared}{0.03}
\setval{dep.podcast.0.5.forward}{0.42}
\setval{dep.podcast.0.5.indep}{0.09}
\setval{dep.podcast.0.5.reverse}{0.02}
\setval{dep.podcast.0.5.shared}{0.03}
\setval{dep.podcast.0.75.forward}{0.38}
\setval{dep.podcast.0.75.indep}{0.16}
\setval{dep.podcast.0.75.reverse}{0.02}
\setval{dep.podcast.0.75.shared}{0.06}
\setval{ds.narratives.scans}{661}
\setval{ds.narratives.stories}{17}
\setval{ds.podcast.stories}{1}
\setval{e.160.0.2.1320.corrected}{0.47}
\setval{e.160.0.2.1320.noisy}{0.43}
\setval{e.160.0.2.1320.normalised}{0.55}
\setval{e.160.0.2.1320.raw}{0.38}
\setval{e.160.0.2.1320.signal}{0.73}
\setval{e.160.0.2.20000.corrected}{0.63}
\setval{e.160.0.2.20000.noisy}{0.43}
\setval{e.160.0.2.20000.normalised}{0.65}
\setval{e.160.0.2.20000.raw}{0.43}
\setval{e.160.0.2.20000.signal}{0.74}
\setval{e.160.0.2.500.corrected}{0.40}
\setval{e.160.0.2.500.noisy}{0.43}
\setval{e.160.0.2.500.normalised}{0.49}
\setval{e.160.0.2.500.raw}{0.35}
\setval{e.160.0.2.500.signal}{0.74}
\setval{e.160.0.2.5000.corrected}{0.56}
\setval{e.160.0.2.5000.noisy}{0.43}
\setval{e.160.0.2.5000.normalised}{0.61}
\setval{e.160.0.2.5000.raw}{0.41}
\setval{e.160.0.2.5000.signal}{0.73}
\setval{e.160.0.5.1320.corrected}{0.58}
\setval{e.160.0.5.1320.noisy}{0.57}
\setval{e.160.0.5.1320.normalised}{0.61}
\setval{e.160.0.5.1320.raw}{0.52}
\setval{e.160.0.5.1320.signal}{0.74}
\setval{e.160.0.5.20000.corrected}{0.71}
\setval{e.160.0.5.20000.noisy}{0.57}
\setval{e.160.0.5.20000.normalised}{0.69}
\setval{e.160.0.5.20000.raw}{0.56}
\setval{e.160.0.5.20000.signal}{0.74}
\setval{e.160.0.5.500.corrected}{0.52}
\setval{e.160.0.5.500.noisy}{0.57}
\setval{e.160.0.5.500.normalised}{0.55}
\setval{e.160.0.5.500.raw}{0.48}
\setval{e.160.0.5.500.signal}{0.74}
\setval{e.160.0.5.5000.corrected}{0.66}
\setval{e.160.0.5.5000.noisy}{0.57}
\setval{e.160.0.5.5000.normalised}{0.67}
\setval{e.160.0.5.5000.raw}{0.55}
\setval{e.160.0.5.5000.signal}{0.74}
\setval{e.160.0.9.1320.corrected}{0.68}
\setval{e.160.0.9.1320.noisy}{0.70}
\setval{e.160.0.9.1320.normalised}{0.68}
\setval{e.160.0.9.1320.raw}{0.65}
\setval{e.160.0.9.1320.signal}{0.74}
\setval{e.160.0.9.20000.corrected}{0.73}
\setval{e.160.0.9.20000.noisy}{0.69}
\setval{e.160.0.9.20000.normalised}{0.73}
\setval{e.160.0.9.20000.raw}{0.69}
\setval{e.160.0.9.20000.signal}{0.73}
\setval{e.160.0.9.500.corrected}{0.60}
\setval{e.160.0.9.500.noisy}{0.69}
\setval{e.160.0.9.500.normalised}{0.62}
\setval{e.160.0.9.500.raw}{0.60}
\setval{e.160.0.9.500.signal}{0.74}
\setval{e.160.0.9.5000.corrected}{0.72}
\setval{e.160.0.9.5000.noisy}{0.70}
\setval{e.160.0.9.5000.normalised}{0.72}
\setval{e.160.0.9.5000.raw}{0.69}
\setval{e.160.0.9.5000.signal}{0.74}
\setval{e.40.0.2.1320.corrected}{0.35}
\setval{e.40.0.2.1320.noisy}{0.24}
\setval{e.40.0.2.1320.normalised}{0.38}
\setval{e.40.0.2.1320.raw}{0.22}
\setval{e.40.0.2.1320.signal}{0.45}
\setval{e.40.0.2.20000.corrected}{0.44}
\setval{e.40.0.2.20000.noisy}{0.24}
\setval{e.40.0.2.20000.normalised}{0.42}
\setval{e.40.0.2.20000.raw}{0.24}
\setval{e.40.0.2.20000.signal}{0.46}
\setval{e.40.0.2.500.corrected}{0.29}
\setval{e.40.0.2.500.noisy}{0.23}
\setval{e.40.0.2.500.normalised}{0.36}
\setval{e.40.0.2.500.raw}{0.20}
\setval{e.40.0.2.500.signal}{0.45}
\setval{e.40.0.2.5000.corrected}{0.38}
\setval{e.40.0.2.5000.noisy}{0.23}
\setval{e.40.0.2.5000.normalised}{0.40}
\setval{e.40.0.2.5000.raw}{0.23}
\setval{e.40.0.2.5000.signal}{0.45}
\setval{e.40.0.5.1320.corrected}{0.39}
\setval{e.40.0.5.1320.noisy}{0.35}
\setval{e.40.0.5.1320.normalised}{0.41}
\setval{e.40.0.5.1320.raw}{0.33}
\setval{e.40.0.5.1320.signal}{0.45}
\setval{e.40.0.5.20000.corrected}{0.45}
\setval{e.40.0.5.20000.noisy}{0.35}
\setval{e.40.0.5.20000.normalised}{0.44}
\setval{e.40.0.5.20000.raw}{0.35}
\setval{e.40.0.5.20000.signal}{0.45}
\setval{e.40.0.5.500.corrected}{0.35}
\setval{e.40.0.5.500.noisy}{0.35}
\setval{e.40.0.5.500.normalised}{0.38}
\setval{e.40.0.5.500.raw}{0.31}
\setval{e.40.0.5.500.signal}{0.45}
\setval{e.40.0.5.5000.corrected}{0.43}
\setval{e.40.0.5.5000.noisy}{0.34}
\setval{e.40.0.5.5000.normalised}{0.42}
\setval{e.40.0.5.5000.raw}{0.33}
\setval{e.40.0.5.5000.signal}{0.44}
\setval{e.40.0.9.1320.corrected}{0.43}
\setval{e.40.0.9.1320.noisy}{0.44}
\setval{e.40.0.9.1320.normalised}{0.44}
\setval{e.40.0.9.1320.raw}{0.42}
\setval{e.40.0.9.1320.signal}{0.45}
\setval{e.40.0.9.20000.corrected}{0.43}
\setval{e.40.0.9.20000.noisy}{0.42}
\setval{e.40.0.9.20000.normalised}{0.43}
\setval{e.40.0.9.20000.raw}{0.42}
\setval{e.40.0.9.20000.signal}{0.43}
\setval{e.40.0.9.500.corrected}{0.39}
\setval{e.40.0.9.500.noisy}{0.43}
\setval{e.40.0.9.500.normalised}{0.39}
\setval{e.40.0.9.500.raw}{0.38}
\setval{e.40.0.9.500.signal}{0.45}
\setval{e.40.0.9.5000.corrected}{0.44}
\setval{e.40.0.9.5000.noisy}{0.43}
\setval{e.40.0.9.5000.normalised}{0.44}
\setval{e.40.0.9.5000.raw}{0.42}
\setval{e.40.0.9.5000.signal}{0.44}
\setval{f.0.6.0.5.max}{0.016}
\setval{f.0.6.0.5.mean}{0.003}
\setval{f.0.6.0.7.max}{0.017}
\setval{f.0.6.0.7.mean}{0.003}
\setval{f.0.6.0.9.max}{0.038}
\setval{f.0.6.0.9.mean}{0.004}
\setval{f.0.6.1.max}{0.018}
\setval{f.0.6.1.mean}{0.003}
\setval{f.1.0.5.max}{0.028}
\setval{f.1.0.5.mean}{0.004}
\setval{f.1.0.7.max}{0.036}
\setval{f.1.0.7.mean}{0.004}
\setval{f.1.0.9.max}{0.029}
\setval{f.1.0.9.mean}{0.004}
\setval{f.1.1.max}{0.036}
\setval{f.1.1.mean}{0.005}
\setval{f.1.4.0.5.max}{0.061}
\setval{f.1.4.0.5.mean}{0.004}
\setval{f.1.4.0.7.max}{0.044}
\setval{f.1.4.0.7.mean}{0.005}
\setval{f.1.4.0.9.max}{0.037}
\setval{f.1.4.0.9.mean}{0.004}
\setval{f.1.4.1.max}{0.054}
\setval{f.1.4.1.mean}{0.004}
\setval{f.all.above}{0.000}
\setval{f.all.max}{0.061}
\setval{f.all.mean}{0.004}
\setval{f.all.n}{672}
\setval{fam.monkey.pc.bottomi}{0.35}
\setval{fam.monkey.pc.bottomindep}{0.00}
\setval{fam.monkey.pc.bottomrev}{$-$0.01}
\setval{fam.monkey.pc.bottoms}{0.31}
\setval{fam.monkey.pc.bottomshared}{$-$0.01}
\setval{fam.monkey.pc.dimension}{166.4}
\setval{fam.monkey.pc.directions}{35}
\setval{fam.monkey.pc.forward}{0.62}
\setval{fam.monkey.pc.forwardi}{0.40}
\setval{fam.monkey.pc.forwardraw}{0.44}
\setval{fam.monkey.pc.forwards}{0.40}
\setval{fam.monkey.pc.indep}{0.01}
\setval{fam.monkey.pc.ranki}{0.31}
\setval{fam.monkey.pc.ranks}{0.26}
\setval{fam.monkey.pc.reverse}{0.00}
\setval{fam.monkey.pc.shared}{0.00}
\setval{fam.monkey.pc.sharei}{99}
\setval{fam.monkey.pc.shares}{100}
\setval{fam.monkey.pc.windep}{0.15}
\setval{fam.monkey.pc.wreverse}{0.11}
\setval{fam.monkey.pc.wshared}{0.08}
\setval{fam.narratives.pc.bottomi}{0.20}
\setval{fam.narratives.pc.bottomindep}{0.06}
\setval{fam.narratives.pc.bottomrev}{0.00}
\setval{fam.narratives.pc.bottoms}{0.24}
\setval{fam.narratives.pc.bottomshared}{0.00}
\setval{fam.narratives.pc.dimension}{194.2}
\setval{fam.narratives.pc.directions}{10}
\setval{fam.narratives.pc.forward}{0.14}
\setval{fam.narratives.pc.forwardi}{0.04}
\setval{fam.narratives.pc.forwardraw}{0.04}
\setval{fam.narratives.pc.forwards}{0.04}
\setval{fam.narratives.pc.indep}{0.05}
\setval{fam.narratives.pc.ranki}{0.04}
\setval{fam.narratives.pc.ranks}{0.08}
\setval{fam.narratives.pc.reverse}{0.00}
\setval{fam.narratives.pc.shared}{0.00}
\setval{fam.narratives.pc.sharei}{95}
\setval{fam.narratives.pc.shares}{100}
\setval{fam.narratives.pc.windep}{0.29}
\setval{fam.narratives.pc.wreverse}{0.01}
\setval{fam.narratives.pc.wshared}{0.02}
\setval{fam.narratives.sae.bottomi}{0.28}
\setval{fam.narratives.sae.bottomindep}{0.03}
\setval{fam.narratives.sae.bottomrev}{$-$0.06}
\setval{fam.narratives.sae.bottoms}{0.27}
\setval{fam.narratives.sae.bottomshared}{$-$0.04}
\setval{fam.narratives.sae.dimension}{347.5}
\setval{fam.narratives.sae.directions}{10}
\setval{fam.narratives.sae.forward}{0.19}
\setval{fam.narratives.sae.forwardi}{0.05}
\setval{fam.narratives.sae.forwardraw}{0.05}
\setval{fam.narratives.sae.forwards}{0.05}
\setval{fam.narratives.sae.indep}{0.05}
\setval{fam.narratives.sae.ranki}{0.41}
\setval{fam.narratives.sae.ranks}{0.24}
\setval{fam.narratives.sae.reverse}{$-$0.01}
\setval{fam.narratives.sae.shared}{$-$0.01}
\setval{fam.narratives.sae.sharei}{94}
\setval{fam.narratives.sae.shares}{100}
\setval{fam.podcast.pc.bottomi}{0.21}
\setval{fam.podcast.pc.bottomindep}{0.00}
\setval{fam.podcast.pc.bottomrev}{0.00}
\setval{fam.podcast.pc.bottoms}{0.24}
\setval{fam.podcast.pc.bottomshared}{0.00}
\setval{fam.podcast.pc.dimension}{257.3}
\setval{fam.podcast.pc.directions}{13}
\setval{fam.podcast.pc.forward}{0.34}
\setval{fam.podcast.pc.forwardi}{0.03}
\setval{fam.podcast.pc.forwardraw}{0.03}
\setval{fam.podcast.pc.forwards}{0.04}
\setval{fam.podcast.pc.indep}{0.01}
\setval{fam.podcast.pc.ranki}{0.06}
\setval{fam.podcast.pc.ranks}{0.08}
\setval{fam.podcast.pc.reverse}{0.00}
\setval{fam.podcast.pc.shared}{0.00}
\setval{fam.podcast.pc.sharei}{99}
\setval{fam.podcast.pc.shares}{100}
\setval{fam.podcast.pc.windep}{0.10}
\setval{fam.podcast.pc.wreverse}{0.03}
\setval{fam.podcast.pc.wshared}{0.02}
\setval{fam.podcast.sae.bottomi}{0.74}
\setval{fam.podcast.sae.bottomindep}{$-$0.06}
\setval{fam.podcast.sae.bottomrev}{$-$0.07}
\setval{fam.podcast.sae.bottoms}{0.73}
\setval{fam.podcast.sae.bottomshared}{$-$0.06}
\setval{fam.podcast.sae.dimension}{361.3}
\setval{fam.podcast.sae.directions}{15}
\setval{fam.podcast.sae.forward}{0.42}
\setval{fam.podcast.sae.forwardi}{0.04}
\setval{fam.podcast.sae.forwardraw}{0.04}
\setval{fam.podcast.sae.forwards}{0.04}
\setval{fam.podcast.sae.indep}{$-$0.01}
\setval{fam.podcast.sae.ranki}{0.71}
\setval{fam.podcast.sae.ranks}{0.62}
\setval{fam.podcast.sae.reverse}{$-$0.01}
\setval{fam.podcast.sae.shared}{$-$0.01}
\setval{fam.podcast.sae.sharei}{100}
\setval{fam.podcast.sae.shares}{100}
\setval{k.complete.default.indep}{0.15}
\setval{k.complete.default.raw}{0.15}
\setval{k.complete.default.shared}{0.12}
\setval{k.complete.default.true}{1.00}
\setval{k.complete.depth-0.25.indep}{0.14}
\setval{k.complete.depth-0.25.raw}{0.14}
\setval{k.complete.depth-0.25.shared}{0.11}
\setval{k.complete.depth-0.25.true}{1.00}
\setval{k.complete.depth-0.75.indep}{0.25}
\setval{k.complete.depth-0.75.raw}{0.25}
\setval{k.complete.depth-0.75.shared}{0.22}
\setval{k.complete.depth-0.75.true}{1.00}
\setval{k.complete.items-1000.indep}{0.12}
\setval{k.complete.items-1000.raw}{0.12}
\setval{k.complete.items-1000.shared}{0.08}
\setval{k.complete.items-1000.true}{1.00}
\setval{k.complete.items-4000.indep}{0.17}
\setval{k.complete.items-4000.raw}{0.17}
\setval{k.complete.items-4000.shared}{0.16}
\setval{k.complete.items-4000.true}{1.00}
\setval{k.complete.noise-0.2.indep}{0.08}
\setval{k.complete.noise-0.2.raw}{0.08}
\setval{k.complete.noise-0.2.shared}{0.06}
\setval{k.complete.noise-0.2.true}{1.00}
\setval{k.complete.noise-0.8.indep}{0.20}
\setval{k.complete.noise-0.8.raw}{0.19}
\setval{k.complete.noise-0.8.shared}{0.17}
\setval{k.complete.noise-0.8.true}{1.00}
\setval{k.complete.sampling-redundant.indep}{0.15}
\setval{k.complete.sampling-redundant.raw}{0.06}
\setval{k.complete.sampling-redundant.shared}{0.06}
\setval{k.complete.sampling-redundant.true}{1.00}
\setval{k.complete.sampling-regional.indep}{0.15}
\setval{k.complete.sampling-regional.raw}{0.15}
\setval{k.complete.sampling-regional.shared}{0.11}
\setval{k.complete.sampling-regional.true}{1.00}
\setval{k.complete.sampling-sparse.indep}{0.15}
\setval{k.complete.sampling-sparse.raw}{0.16}
\setval{k.complete.sampling-sparse.shared}{0.12}
\setval{k.complete.sampling-sparse.true}{1.00}
\setval{k.complete.sites-1024.indep}{0.29}
\setval{k.complete.sites-1024.raw}{0.29}
\setval{k.complete.sites-1024.shared}{0.18}
\setval{k.complete.sites-1024.true}{1.00}
\setval{k.complete.sites-64.indep}{0.07}
\setval{k.complete.sites-64.raw}{0.07}
\setval{k.complete.sites-64.shared}{0.07}
\setval{k.complete.sites-64.true}{1.00}
\setval{k.notop.default.indep}{0.13}
\setval{k.notop.default.raw}{0.03}
\setval{k.notop.default.shared}{0.11}
\setval{k.notop.default.true}{0.72}
\setval{k.notop.depth-0.25.indep}{0.12}
\setval{k.notop.depth-0.25.raw}{0.03}
\setval{k.notop.depth-0.25.shared}{0.10}
\setval{k.notop.depth-0.25.true}{0.73}
\setval{k.notop.depth-0.75.indep}{0.23}
\setval{k.notop.depth-0.75.raw}{0.02}
\setval{k.notop.depth-0.75.shared}{0.21}
\setval{k.notop.depth-0.75.true}{0.61}
\setval{k.notop.items-1000.indep}{0.09}
\setval{k.notop.items-1000.raw}{0.01}
\setval{k.notop.items-1000.shared}{0.06}
\setval{k.notop.items-1000.true}{0.72}
\setval{k.notop.items-4000.indep}{0.17}
\setval{k.notop.items-4000.raw}{0.05}
\setval{k.notop.items-4000.shared}{0.16}
\setval{k.notop.items-4000.true}{0.72}
\setval{k.notop.noise-0.2.indep}{0.06}
\setval{k.notop.noise-0.2.raw}{0.01}
\setval{k.notop.noise-0.2.shared}{0.05}
\setval{k.notop.noise-0.2.true}{0.72}
\setval{k.notop.noise-0.8.indep}{0.18}
\setval{k.notop.noise-0.8.raw}{0.05}
\setval{k.notop.noise-0.8.shared}{0.16}
\setval{k.notop.noise-0.8.true}{0.72}
\setval{k.notop.sampling-redundant.indep}{0.14}
\setval{k.notop.sampling-redundant.raw}{0.00}
\setval{k.notop.sampling-redundant.shared}{0.05}
\setval{k.notop.sampling-redundant.true}{0.72}
\setval{k.notop.sampling-regional.indep}{0.13}
\setval{k.notop.sampling-regional.raw}{0.03}
\setval{k.notop.sampling-regional.shared}{0.11}
\setval{k.notop.sampling-regional.true}{0.72}
\setval{k.notop.sampling-sparse.indep}{0.13}
\setval{k.notop.sampling-sparse.raw}{0.04}
\setval{k.notop.sampling-sparse.shared}{0.11}
\setval{k.notop.sampling-sparse.true}{0.72}
\setval{k.notop.sites-1024.indep}{0.27}
\setval{k.notop.sites-1024.raw}{0.10}
\setval{k.notop.sites-1024.shared}{0.17}
\setval{k.notop.sites-1024.true}{0.72}
\setval{k.notop.sites-64.indep}{0.06}
\setval{k.notop.sites-64.raw}{0.01}
\setval{k.notop.sites-64.shared}{0.06}
\setval{k.notop.sites-64.true}{0.72}
\setval{m.gemma2-2b.blocks}{26}
\setval{m.gemma2-2b.layer.0.25}{7}
\setval{m.gemma2-2b.layer.0.5}{13}
\setval{m.gemma2-2b.layer.0.75}{20}
\setval{m.gemma2-2b.params}{2{,}614}
\setval{m.gemma2-2b.width}{2{,}304}
\setval{m.gemma2-9b.blocks}{42}
\setval{m.gemma2-9b.layer.0.25}{11}
\setval{m.gemma2-9b.layer.0.5}{21}
\setval{m.gemma2-9b.layer.0.75}{32}
\setval{m.gemma2-9b.params}{9{,}242}
\setval{m.gemma2-9b.width}{3{,}584}
\setval{m.gpt2-large.blocks}{36}
\setval{m.gpt2-large.layer.0.25}{9}
\setval{m.gpt2-large.layer.0.5}{18}
\setval{m.gpt2-large.layer.0.75}{27}
\setval{m.gpt2-large.params}{774}
\setval{m.gpt2-large.width}{1{,}280}
\setval{m.gpt2-medium.blocks}{24}
\setval{m.gpt2-medium.layer.0.25}{6}
\setval{m.gpt2-medium.layer.0.5}{12}
\setval{m.gpt2-medium.layer.0.75}{18}
\setval{m.gpt2-medium.params}{355}
\setval{m.gpt2-medium.width}{1{,}024}
\setval{m.gpt2-small-random.blocks}{12}
\setval{m.gpt2-small-random.layer.0.25}{3}
\setval{m.gpt2-small-random.layer.0.5}{6}
\setval{m.gpt2-small-random.layer.0.75}{9}
\setval{m.gpt2-small-random.params}{124}
\setval{m.gpt2-small-random.width}{768}
\setval{m.gpt2-small.blocks}{12}
\setval{m.gpt2-small.layer.0.25}{3}
\setval{m.gpt2-small.layer.0.5}{6}
\setval{m.gpt2-small.layer.0.75}{9}
\setval{m.gpt2-small.params}{124}
\setval{m.gpt2-small.width}{768}
\setval{m.gpt2-xl.blocks}{48}
\setval{m.gpt2-xl.layer.0.25}{12}
\setval{m.gpt2-xl.layer.0.5}{24}
\setval{m.gpt2-xl.layer.0.75}{36}
\setval{m.gpt2-xl.params}{1{,}558}
\setval{m.gpt2-xl.width}{1{,}600}
\setval{m.llama3.1-8b.blocks}{32}
\setval{m.llama3.1-8b.layer.0.25}{8}
\setval{m.llama3.1-8b.layer.0.5}{16}
\setval{m.llama3.1-8b.layer.0.75}{24}
\setval{m.llama3.1-8b.params}{7{,}505}
\setval{m.llama3.1-8b.width}{4{,}096}
\setval{m.llama3.2-1b-random.blocks}{16}
\setval{m.llama3.2-1b-random.layer.0.25}{4}
\setval{m.llama3.2-1b-random.layer.0.5}{8}
\setval{m.llama3.2-1b-random.layer.0.75}{12}
\setval{m.llama3.2-1b-random.params}{1{,}236}
\setval{m.llama3.2-1b-random.width}{2{,}048}
\setval{m.llama3.2-1b.blocks}{16}
\setval{m.llama3.2-1b.layer.0.25}{4}
\setval{m.llama3.2-1b.layer.0.5}{8}
\setval{m.llama3.2-1b.layer.0.75}{12}
\setval{m.llama3.2-1b.params}{1{,}236}
\setval{m.llama3.2-1b.width}{2{,}048}
\setval{m.llama3.2-3b.blocks}{28}
\setval{m.llama3.2-3b.layer.0.25}{7}
\setval{m.llama3.2-3b.layer.0.5}{14}
\setval{m.llama3.2-3b.layer.0.75}{21}
\setval{m.llama3.2-3b.params}{3{,}213}
\setval{m.llama3.2-3b.width}{3{,}072}
\setval{m.olmo2-1b.blocks}{16}
\setval{m.olmo2-1b.layer.0.25}{4}
\setval{m.olmo2-1b.layer.0.5}{8}
\setval{m.olmo2-1b.layer.0.75}{12}
\setval{m.olmo2-1b.params}{1{,}279}
\setval{m.olmo2-1b.width}{2{,}048}
\setval{m.olmo2-7b.blocks}{32}
\setval{m.olmo2-7b.layer.0.25}{8}
\setval{m.olmo2-7b.layer.0.5}{16}
\setval{m.olmo2-7b.layer.0.75}{24}
\setval{m.olmo2-7b.params}{6{,}888}
\setval{m.olmo2-7b.width}{4{,}096}
\setval{m.pythia-1.4b.blocks}{24}
\setval{m.pythia-1.4b.layer.0.25}{6}
\setval{m.pythia-1.4b.layer.0.5}{12}
\setval{m.pythia-1.4b.layer.0.75}{18}
\setval{m.pythia-1.4b.params}{1{,}312}
\setval{m.pythia-1.4b.width}{2{,}048}
\setval{m.pythia-160m.blocks}{12}
\setval{m.pythia-160m.layer.0.25}{3}
\setval{m.pythia-160m.layer.0.5}{6}
\setval{m.pythia-160m.layer.0.75}{9}
\setval{m.pythia-160m.params}{124}
\setval{m.pythia-160m.width}{768}
\setval{m.pythia-1b.blocks}{16}
\setval{m.pythia-1b.layer.0.25}{4}
\setval{m.pythia-1b.layer.0.5}{8}
\setval{m.pythia-1b.layer.0.75}{12}
\setval{m.pythia-1b.params}{909}
\setval{m.pythia-1b.width}{2{,}048}
\setval{m.pythia-2.8b.blocks}{32}
\setval{m.pythia-2.8b.layer.0.25}{8}
\setval{m.pythia-2.8b.layer.0.5}{16}
\setval{m.pythia-2.8b.layer.0.75}{24}
\setval{m.pythia-2.8b.params}{2{,}646}
\setval{m.pythia-2.8b.width}{2{,}560}
\setval{m.pythia-410m-random.blocks}{24}
\setval{m.pythia-410m-random.layer.0.25}{6}
\setval{m.pythia-410m-random.layer.0.5}{12}
\setval{m.pythia-410m-random.layer.0.75}{18}
\setval{m.pythia-410m-random.params}{354}
\setval{m.pythia-410m-random.width}{1{,}024}
\setval{m.pythia-410m.blocks}{24}
\setval{m.pythia-410m.layer.0.25}{6}
\setval{m.pythia-410m.layer.0.5}{12}
\setval{m.pythia-410m.layer.0.75}{18}
\setval{m.pythia-410m.params}{354}
\setval{m.pythia-410m.width}{1{,}024}
\setval{m.pythia-6.9b.blocks}{32}
\setval{m.pythia-6.9b.layer.0.25}{8}
\setval{m.pythia-6.9b.layer.0.5}{16}
\setval{m.pythia-6.9b.layer.0.75}{24}
\setval{m.pythia-6.9b.params}{6{,}651}
\setval{m.pythia-6.9b.width}{4{,}096}
\setval{m.pythia-70m.blocks}{6}
\setval{m.pythia-70m.layer.0.25}{2}
\setval{m.pythia-70m.layer.0.5}{3}
\setval{m.pythia-70m.layer.0.75}{5}
\setval{m.pythia-70m.params}{45}
\setval{m.pythia-70m.width}{512}
\setval{mk.count}{41}
\setval{mk.m2.kept}{101}
\setval{mk.m2.repeats}{83}
\setval{mk.m2.stored}{107}
\setval{mk.models}{CORnet-RT\_t4, CORnet-S\_t0, alexnet, convnext, convnext\_ssl, cornetS\_regularized, densenet101, densenet121, densenet161, densenet161\_robust, densenet169, inceptionv1, inceptionv3, mobilenet, mobilenet\_robust, mobilenet\_v2, nasnet, pnasnet, resnet101, resnet101\_ssl, resnet152, resnet18, resnet18\_robust, resnet18\_ssl, resnet50, resnet50\_eps1\_robust, resnet50\_eps3\_robust, resnet50\_robust, resnet50\_ssl, resnet50\_swsl, resnetSIN, shufflenet, shufflenet\_robust, squeezenet, swin, swin\_ssl, vgg16, vgg16\_robust, vgg16\_robust\_eps3, vit, vit\_ssl}
\setval{ob.default.all.base.all.indep}{0.04}
\setval{ob.default.all.base.all.pubrev}{17.7}
\setval{ob.default.all.base.all.raw}{0.13}
\setval{ob.default.all.base.all.shared}{0.03}
\setval{ob.default.all.base.all.units}{1{,}024}
\setval{ob.default.all.base.top.indep}{0.09}
\setval{ob.default.all.base.top.pubrev}{28.4}
\setval{ob.default.all.base.top.raw}{0.23}
\setval{ob.default.all.base.top.shared}{0.07}
\setval{ob.default.all.base.top.units}{275}
\setval{ob.default.all.diffi}{0.01}
\setval{ob.default.all.diffs}{0.03}
\setval{ob.default.all.directions}{122.3}
\setval{ob.default.all.ed}{630.5}
\setval{ob.default.all.forward}{0.42}
\setval{ob.default.all.fwdi}{0.42}
\setval{ob.default.all.fwds}{0.43}
\setval{ob.default.all.gain0.25.all.indep}{0.04}
\setval{ob.default.all.gain0.25.all.pubrev}{0.5}
\setval{ob.default.all.gain0.25.all.raw}{0.00}
\setval{ob.default.all.gain0.25.all.shared}{0.03}
\setval{ob.default.all.gain0.25.all.units}{1{,}024}
\setval{ob.default.all.gain0.25.top.indep}{0.09}
\setval{ob.default.all.gain0.25.top.pubrev}{1.0}
\setval{ob.default.all.gain0.25.top.raw}{0.00}
\setval{ob.default.all.gain0.25.top.shared}{0.07}
\setval{ob.default.all.gain0.25.top.units}{283}
\setval{ob.default.all.gain0.5.all.indep}{0.04}
\setval{ob.default.all.gain0.5.all.pubrev}{3.5}
\setval{ob.default.all.gain0.5.all.raw}{0.02}
\setval{ob.default.all.gain0.5.all.shared}{0.03}
\setval{ob.default.all.gain0.5.all.units}{1{,}024}
\setval{ob.default.all.gain0.5.top.indep}{0.08}
\setval{ob.default.all.gain0.5.top.pubrev}{6.7}
\setval{ob.default.all.gain0.5.top.raw}{0.04}
\setval{ob.default.all.gain0.5.top.shared}{0.06}
\setval{ob.default.all.gain0.5.top.units}{253}
\setval{ob.default.all.gain0.all.indep}{0.04}
\setval{ob.default.all.gain0.all.pubrev}{0.2}
\setval{ob.default.all.gain0.all.raw}{0.00}
\setval{ob.default.all.gain0.all.shared}{0.03}
\setval{ob.default.all.gain0.all.units}{1{,}024}
\setval{ob.default.all.gain0.top.indep}{0.09}
\setval{ob.default.all.gain0.top.pubrev}{0.2}
\setval{ob.default.all.gain0.top.raw}{0.00}
\setval{ob.default.all.gain0.top.shared}{0.06}
\setval{ob.default.all.gain0.top.units}{266}
\setval{ob.default.all.gain1.all.indep}{0.05}
\setval{ob.default.all.gain1.all.pubrev}{18.0}
\setval{ob.default.all.gain1.all.raw}{0.13}
\setval{ob.default.all.gain1.all.shared}{0.03}
\setval{ob.default.all.gain1.all.units}{1{,}024}
\setval{ob.default.all.gain1.top.indep}{0.10}
\setval{ob.default.all.gain1.top.pubrev}{29.0}
\setval{ob.default.all.gain1.top.raw}{0.23}
\setval{ob.default.all.gain1.top.shared}{0.07}
\setval{ob.default.all.gain1.top.units}{278}
\setval{ob.default.all.gapbottomadded}{1.00}
\setval{ob.default.all.indep}{0.04}
\setval{ob.default.all.pubbottom}{0.42}
\setval{ob.default.all.pubbottomadded}{1.00}
\setval{ob.default.all.pubfwd}{67.9}
\setval{ob.default.all.pubrev}{2.7}
\setval{ob.default.all.pubrevmean}{8.0}
\setval{ob.default.all.ranki}{0.23}
\setval{ob.default.all.ranks}{0.22}
\setval{ob.default.all.raw}{0.06}
\setval{ob.default.all.shared}{0.03}
\setval{ob.default.all.sharei}{101}
\setval{ob.default.all.shares}{103}
\setval{ob.default.all.true}{0.80}
\setval{ob.default.all.units}{5{,}120}
\setval{ob.default.base.base.all.indep}{0.13}
\setval{ob.default.base.base.all.pubrev}{17.7}
\setval{ob.default.base.base.all.raw}{0.13}
\setval{ob.default.base.base.all.shared}{0.08}
\setval{ob.default.base.base.all.units}{1{,}024}
\setval{ob.default.base.base.top.indep}{0.22}
\setval{ob.default.base.base.top.pubrev}{28.4}
\setval{ob.default.base.base.top.raw}{0.23}
\setval{ob.default.base.base.top.shared}{0.14}
\setval{ob.default.base.base.top.units}{275}
\setval{ob.default.base.diffi}{0.00}
\setval{ob.default.base.diffs}{0.05}
\setval{ob.default.base.directions}{62.7}
\setval{ob.default.base.ed}{205.1}
\setval{ob.default.base.forward}{0.21}
\setval{ob.default.base.fwdi}{0.21}
\setval{ob.default.base.fwds}{0.21}
\setval{ob.default.base.indep}{0.13}
\setval{ob.default.base.pubbottom}{1.00}
\setval{ob.default.base.pubfwd}{21.9}
\setval{ob.default.base.pubrev}{15.2}
\setval{ob.default.base.pubrevmean}{17.7}
\setval{ob.default.base.ranki}{0.87}
\setval{ob.default.base.ranks}{0.84}
\setval{ob.default.base.raw}{0.13}
\setval{ob.default.base.shared}{0.08}
\setval{ob.default.base.sharei}{100}
\setval{ob.default.base.shares}{106}
\setval{ob.default.base.true}{1.00}
\setval{ob.default.base.units}{1{,}024}
\setval{ob.default.full.edfwd}{1.00}
\setval{ob.default.full.edpubfwd}{0.94}
\setval{ob.default.full.edpubrev}{$-$0.94}
\setval{ob.default.full.edrev}{$-$0.94}
\setval{ob.default.full.models}{6}
\setval{ob.default.full.revindep}{0.96}
\setval{ob.default.full.revshared}{0.96}
\setval{ob.default.gain0-256.base.all.indep}{0.10}
\setval{ob.default.gain0-256.base.all.pubrev}{17.7}
\setval{ob.default.gain0-256.base.all.raw}{0.13}
\setval{ob.default.gain0-256.base.all.shared}{0.06}
\setval{ob.default.gain0-256.base.all.units}{1{,}024}
\setval{ob.default.gain0-256.base.top.indep}{0.18}
\setval{ob.default.gain0-256.base.top.pubrev}{28.4}
\setval{ob.default.gain0-256.base.top.raw}{0.23}
\setval{ob.default.gain0-256.base.top.shared}{0.11}
\setval{ob.default.gain0-256.base.top.units}{275}
\setval{ob.default.gain0-256.diffi}{0.02}
\setval{ob.default.gain0-256.diffs}{0.06}
\setval{ob.default.gain0-256.directions}{57.7}
\setval{ob.default.gain0-256.ed}{272.3}
\setval{ob.default.gain0-256.forward}{0.19}
\setval{ob.default.gain0-256.fwdi}{0.19}
\setval{ob.default.gain0-256.fwds}{0.19}
\setval{ob.default.gain0-256.gain0.all.indep}{0.03}
\setval{ob.default.gain0-256.gain0.all.pubrev}{0.2}
\setval{ob.default.gain0-256.gain0.all.raw}{0.00}
\setval{ob.default.gain0-256.gain0.all.shared}{0.02}
\setval{ob.default.gain0-256.gain0.all.units}{256}
\setval{ob.default.gain0-256.gain0.top.indep}{0.05}
\setval{ob.default.gain0-256.gain0.top.pubrev}{0.2}
\setval{ob.default.gain0-256.gain0.top.raw}{0.00}
\setval{ob.default.gain0-256.gain0.top.shared}{0.03}
\setval{ob.default.gain0-256.gain0.top.units}{66}
\setval{ob.default.gain0-256.gapbottomadded}{0.74}
\setval{ob.default.gain0-256.indep}{0.09}
\setval{ob.default.gain0-256.pubbottom}{0.00}
\setval{ob.default.gain0-256.pubbottomadded}{1.00}
\setval{ob.default.gain0-256.pubfwd}{12.9}
\setval{ob.default.gain0-256.pubrev}{13.1}
\setval{ob.default.gain0-256.pubrevmean}{14.2}
\setval{ob.default.gain0-256.ranki}{0.88}
\setval{ob.default.gain0-256.ranks}{0.85}
\setval{ob.default.gain0-256.raw}{0.11}
\setval{ob.default.gain0-256.shared}{0.05}
\setval{ob.default.gain0-256.sharei}{102}
\setval{ob.default.gain0-256.shares}{106}
\setval{ob.default.gain0-256.true}{0.80}
\setval{ob.default.gain0-256.units}{1{,}280}
\setval{ob.default.gain0-512.base.all.indep}{0.08}
\setval{ob.default.gain0-512.base.all.pubrev}{17.7}
\setval{ob.default.gain0-512.base.all.raw}{0.13}
\setval{ob.default.gain0-512.base.all.shared}{0.05}
\setval{ob.default.gain0-512.base.all.units}{1{,}024}
\setval{ob.default.gain0-512.base.top.indep}{0.15}
\setval{ob.default.gain0-512.base.top.pubrev}{28.4}
\setval{ob.default.gain0-512.base.top.raw}{0.23}
\setval{ob.default.gain0-512.base.top.shared}{0.09}
\setval{ob.default.gain0-512.base.top.units}{275}
\setval{ob.default.gain0-512.diffi}{0.02}
\setval{ob.default.gain0-512.diffs}{0.05}
\setval{ob.default.gain0-512.directions}{56.0}
\setval{ob.default.gain0-512.ed}{318.0}
\setval{ob.default.gain0-512.forward}{0.18}
\setval{ob.default.gain0-512.fwdi}{0.18}
\setval{ob.default.gain0-512.fwds}{0.17}
\setval{ob.default.gain0-512.gain0.all.indep}{0.04}
\setval{ob.default.gain0-512.gain0.all.pubrev}{0.2}
\setval{ob.default.gain0-512.gain0.all.raw}{0.00}
\setval{ob.default.gain0-512.gain0.all.shared}{0.02}
\setval{ob.default.gain0-512.gain0.all.units}{512}
\setval{ob.default.gain0-512.gain0.top.indep}{0.08}
\setval{ob.default.gain0-512.gain0.top.pubrev}{0.2}
\setval{ob.default.gain0-512.gain0.top.raw}{0.00}
\setval{ob.default.gain0-512.gain0.top.shared}{0.04}
\setval{ob.default.gain0-512.gain0.top.units}{128}
\setval{ob.default.gain0-512.gapbottomadded}{0.98}
\setval{ob.default.gain0-512.indep}{0.07}
\setval{ob.default.gain0-512.pubbottom}{0.00}
\setval{ob.default.gain0-512.pubbottomadded}{1.00}
\setval{ob.default.gain0-512.pubfwd}{5.7}
\setval{ob.default.gain0-512.pubrev}{10.9}
\setval{ob.default.gain0-512.pubrevmean}{11.9}
\setval{ob.default.gain0-512.ranki}{0.71}
\setval{ob.default.gain0-512.ranks}{0.67}
\setval{ob.default.gain0-512.raw}{0.09}
\setval{ob.default.gain0-512.shared}{0.04}
\setval{ob.default.gain0-512.sharei}{102}
\setval{ob.default.gain0-512.shares}{106}
\setval{ob.default.gain0-512.true}{0.67}
\setval{ob.default.gain0-512.units}{1{,}536}
\setval{ob.default.gain0.25.base.all.indep}{0.06}
\setval{ob.default.gain0.25.base.all.pubrev}{17.7}
\setval{ob.default.gain0.25.base.all.raw}{0.13}
\setval{ob.default.gain0.25.base.all.shared}{0.03}
\setval{ob.default.gain0.25.base.all.units}{1{,}024}
\setval{ob.default.gain0.25.base.top.indep}{0.11}
\setval{ob.default.gain0.25.base.top.pubrev}{28.4}
\setval{ob.default.gain0.25.base.top.raw}{0.23}
\setval{ob.default.gain0.25.base.top.shared}{0.06}
\setval{ob.default.gain0.25.base.top.units}{275}
\setval{ob.default.gain0.25.diffi}{0.01}
\setval{ob.default.gain0.25.diffs}{0.04}
\setval{ob.default.gain0.25.directions}{56.0}
\setval{ob.default.gain0.25.ed}{353.3}
\setval{ob.default.gain0.25.forward}{0.17}
\setval{ob.default.gain0.25.fwdi}{0.17}
\setval{ob.default.gain0.25.fwds}{0.17}
\setval{ob.default.gain0.25.gain0.25.all.indep}{0.06}
\setval{ob.default.gain0.25.gain0.25.all.pubrev}{0.5}
\setval{ob.default.gain0.25.gain0.25.all.raw}{0.00}
\setval{ob.default.gain0.25.gain0.25.all.shared}{0.03}
\setval{ob.default.gain0.25.gain0.25.all.units}{1{,}024}
\setval{ob.default.gain0.25.gain0.25.top.indep}{0.12}
\setval{ob.default.gain0.25.gain0.25.top.pubrev}{1.0}
\setval{ob.default.gain0.25.gain0.25.top.raw}{0.00}
\setval{ob.default.gain0.25.gain0.25.top.shared}{0.06}
\setval{ob.default.gain0.25.gain0.25.top.units}{283}
\setval{ob.default.gain0.25.gapbottomadded}{1.00}
\setval{ob.default.gain0.25.indep}{0.06}
\setval{ob.default.gain0.25.pubbottom}{1.00}
\setval{ob.default.gain0.25.pubbottomadded}{1.00}
\setval{ob.default.gain0.25.pubfwd}{5.7}
\setval{ob.default.gain0.25.pubrev}{3.5}
\setval{ob.default.gain0.25.pubrevmean}{9.1}
\setval{ob.default.gain0.25.ranki}{0.25}
\setval{ob.default.gain0.25.ranks}{0.24}
\setval{ob.default.gain0.25.raw}{0.07}
\setval{ob.default.gain0.25.shared}{0.03}
\setval{ob.default.gain0.25.sharei}{101}
\setval{ob.default.gain0.25.shares}{104}
\setval{ob.default.gain0.25.true}{1.00}
\setval{ob.default.gain0.25.units}{2{,}048}
\setval{ob.default.gain0.5.base.all.indep}{0.07}
\setval{ob.default.gain0.5.base.all.pubrev}{17.7}
\setval{ob.default.gain0.5.base.all.raw}{0.13}
\setval{ob.default.gain0.5.base.all.shared}{0.04}
\setval{ob.default.gain0.5.base.all.units}{1{,}024}
\setval{ob.default.gain0.5.base.top.indep}{0.14}
\setval{ob.default.gain0.5.base.top.pubrev}{28.4}
\setval{ob.default.gain0.5.base.top.raw}{0.23}
\setval{ob.default.gain0.5.base.top.shared}{0.08}
\setval{ob.default.gain0.5.base.top.units}{275}
\setval{ob.default.gain0.5.diffi}{0.00}
\setval{ob.default.gain0.5.diffs}{0.03}
\setval{ob.default.gain0.5.directions}{69.3}
\setval{ob.default.gain0.5.ed}{356.4}
\setval{ob.default.gain0.5.forward}{0.23}
\setval{ob.default.gain0.5.fwdi}{0.23}
\setval{ob.default.gain0.5.fwds}{0.23}
\setval{ob.default.gain0.5.gain0.5.all.indep}{0.07}
\setval{ob.default.gain0.5.gain0.5.all.pubrev}{3.5}
\setval{ob.default.gain0.5.gain0.5.all.raw}{0.02}
\setval{ob.default.gain0.5.gain0.5.all.shared}{0.04}
\setval{ob.default.gain0.5.gain0.5.all.units}{1{,}024}
\setval{ob.default.gain0.5.gain0.5.top.indep}{0.13}
\setval{ob.default.gain0.5.gain0.5.top.pubrev}{6.7}
\setval{ob.default.gain0.5.gain0.5.top.raw}{0.04}
\setval{ob.default.gain0.5.gain0.5.top.shared}{0.08}
\setval{ob.default.gain0.5.gain0.5.top.units}{253}
\setval{ob.default.gain0.5.gapbottomadded}{1.00}
\setval{ob.default.gain0.5.indep}{0.07}
\setval{ob.default.gain0.5.pubbottom}{1.00}
\setval{ob.default.gain0.5.pubbottomadded}{1.00}
\setval{ob.default.gain0.5.pubfwd}{8.8}
\setval{ob.default.gain0.5.pubrev}{8.0}
\setval{ob.default.gain0.5.pubrevmean}{10.6}
\setval{ob.default.gain0.5.ranki}{0.46}
\setval{ob.default.gain0.5.ranks}{0.44}
\setval{ob.default.gain0.5.raw}{0.07}
\setval{ob.default.gain0.5.shared}{0.04}
\setval{ob.default.gain0.5.sharei}{100}
\setval{ob.default.gain0.5.shares}{104}
\setval{ob.default.gain0.5.true}{1.00}
\setval{ob.default.gain0.5.units}{2{,}048}
\setval{ob.default.gain0.base.all.indep}{0.05}
\setval{ob.default.gain0.base.all.pubrev}{17.7}
\setval{ob.default.gain0.base.all.raw}{0.13}
\setval{ob.default.gain0.base.all.shared}{0.03}
\setval{ob.default.gain0.base.all.units}{1{,}024}
\setval{ob.default.gain0.base.top.indep}{0.10}
\setval{ob.default.gain0.base.top.pubrev}{28.4}
\setval{ob.default.gain0.base.top.raw}{0.23}
\setval{ob.default.gain0.base.top.shared}{0.05}
\setval{ob.default.gain0.base.top.units}{275}
\setval{ob.default.gain0.diffi}{0.01}
\setval{ob.default.gain0.diffs}{0.04}
\setval{ob.default.gain0.directions}{52.7}
\setval{ob.default.gain0.ed}{354.7}
\setval{ob.default.gain0.forward}{0.15}
\setval{ob.default.gain0.fwdi}{0.15}
\setval{ob.default.gain0.fwds}{0.15}
\setval{ob.default.gain0.gain0.all.indep}{0.05}
\setval{ob.default.gain0.gain0.all.pubrev}{0.2}
\setval{ob.default.gain0.gain0.all.raw}{0.00}
\setval{ob.default.gain0.gain0.all.shared}{0.03}
\setval{ob.default.gain0.gain0.all.units}{1{,}024}
\setval{ob.default.gain0.gain0.top.indep}{0.10}
\setval{ob.default.gain0.gain0.top.pubrev}{0.2}
\setval{ob.default.gain0.gain0.top.raw}{0.00}
\setval{ob.default.gain0.gain0.top.shared}{0.05}
\setval{ob.default.gain0.gain0.top.units}{266}
\setval{ob.default.gain0.gapbottomadded}{1.00}
\setval{ob.default.gain0.indep}{0.05}
\setval{ob.default.gain0.pubbottom}{0.00}
\setval{ob.default.gain0.pubbottomadded}{1.00}
\setval{ob.default.gain0.pubfwd}{4.8}
\setval{ob.default.gain0.pubrev}{2.5}
\setval{ob.default.gain0.pubrevmean}{9.0}
\setval{ob.default.gain0.ranki}{0.22}
\setval{ob.default.gain0.ranks}{0.19}
\setval{ob.default.gain0.raw}{0.07}
\setval{ob.default.gain0.shared}{0.03}
\setval{ob.default.gain0.sharei}{102}
\setval{ob.default.gain0.shares}{104}
\setval{ob.default.gain0.true}{0.50}
\setval{ob.default.gain0.units}{2{,}048}
\setval{ob.default.gain1.base.all.indep}{0.13}
\setval{ob.default.gain1.base.all.pubrev}{17.7}
\setval{ob.default.gain1.base.all.raw}{0.13}
\setval{ob.default.gain1.base.all.shared}{0.10}
\setval{ob.default.gain1.base.all.units}{1{,}024}
\setval{ob.default.gain1.base.top.indep}{0.23}
\setval{ob.default.gain1.base.top.pubrev}{28.4}
\setval{ob.default.gain1.base.top.raw}{0.23}
\setval{ob.default.gain1.base.top.shared}{0.18}
\setval{ob.default.gain1.base.top.units}{275}
\setval{ob.default.gain1.diffi}{0.00}
\setval{ob.default.gain1.diffs}{0.03}
\setval{ob.default.gain1.directions}{132.7}
\setval{ob.default.gain1.ed}{352.9}
\setval{ob.default.gain1.forward}{0.51}
\setval{ob.default.gain1.fwdi}{0.51}
\setval{ob.default.gain1.fwds}{0.51}
\setval{ob.default.gain1.gain1.all.indep}{0.13}
\setval{ob.default.gain1.gain1.all.pubrev}{18.0}
\setval{ob.default.gain1.gain1.all.raw}{0.13}
\setval{ob.default.gain1.gain1.all.shared}{0.11}
\setval{ob.default.gain1.gain1.all.units}{1{,}024}
\setval{ob.default.gain1.gain1.top.indep}{0.23}
\setval{ob.default.gain1.gain1.top.pubrev}{29.0}
\setval{ob.default.gain1.gain1.top.raw}{0.23}
\setval{ob.default.gain1.gain1.top.shared}{0.19}
\setval{ob.default.gain1.gain1.top.units}{278}
\setval{ob.default.gain1.gapbottomadded}{0.49}
\setval{ob.default.gain1.indep}{0.13}
\setval{ob.default.gain1.pubbottom}{1.00}
\setval{ob.default.gain1.pubbottomadded}{0.50}
\setval{ob.default.gain1.pubfwd}{44.2}
\setval{ob.default.gain1.pubrev}{15.3}
\setval{ob.default.gain1.pubrevmean}{17.9}
\setval{ob.default.gain1.ranki}{0.87}
\setval{ob.default.gain1.ranks}{0.86}
\setval{ob.default.gain1.raw}{0.13}
\setval{ob.default.gain1.shared}{0.10}
\setval{ob.default.gain1.sharei}{100}
\setval{ob.default.gain1.shares}{103}
\setval{ob.default.gain1.true}{1.00}
\setval{ob.default.gain1.units}{2{,}048}
\setval{ob.default.top128.base.all.indep}{0.39}
\setval{ob.default.top128.base.all.pubrev}{33.9}
\setval{ob.default.top128.base.all.raw}{0.28}
\setval{ob.default.top128.base.all.shared}{0.30}
\setval{ob.default.top128.base.all.units}{128}
\setval{ob.default.top128.base.top.indep}{0.39}
\setval{ob.default.top128.base.top.pubrev}{33.9}
\setval{ob.default.top128.base.top.raw}{0.28}
\setval{ob.default.top128.base.top.shared}{0.30}
\setval{ob.default.top128.base.top.units}{128}
\setval{ob.default.top128.diffi}{$-$0.11}
\setval{ob.default.top128.diffs}{$-$0.02}
\setval{ob.default.top128.directions}{26.0}
\setval{ob.default.top128.ed}{16.8}
\setval{ob.default.top128.forward}{0.06}
\setval{ob.default.top128.fwdi}{0.06}
\setval{ob.default.top128.fwds}{0.06}
\setval{ob.default.top128.indep}{0.39}
\setval{ob.default.top128.pubbottom}{1.00}
\setval{ob.default.top128.pubfwd}{9.6}
\setval{ob.default.top128.pubrev}{34.0}
\setval{ob.default.top128.pubrevmean}{33.9}
\setval{ob.default.top128.ranki}{0.82}
\setval{ob.default.top128.ranks}{0.73}
\setval{ob.default.top128.raw}{0.28}
\setval{ob.default.top128.shared}{0.30}
\setval{ob.default.top128.sharei}{85}
\setval{ob.default.top128.shares}{97}
\setval{ob.default.top128.true}{1.00}
\setval{ob.default.top128.units}{128}
\setval{ob.default.top256.base.all.indep}{0.31}
\setval{ob.default.top256.base.all.pubrev}{29.0}
\setval{ob.default.top256.base.all.raw}{0.24}
\setval{ob.default.top256.base.all.shared}{0.22}
\setval{ob.default.top256.base.all.units}{256}
\setval{ob.default.top256.base.top.indep}{0.31}
\setval{ob.default.top256.base.top.pubrev}{29.0}
\setval{ob.default.top256.base.top.raw}{0.24}
\setval{ob.default.top256.base.top.shared}{0.22}
\setval{ob.default.top256.base.top.units}{256}
\setval{ob.default.top256.diffi}{$-$0.07}
\setval{ob.default.top256.diffs}{0.01}
\setval{ob.default.top256.directions}{37.7}
\setval{ob.default.top256.ed}{40.9}
\setval{ob.default.top256.forward}{0.11}
\setval{ob.default.top256.fwdi}{0.11}
\setval{ob.default.top256.fwds}{0.11}
\setval{ob.default.top256.indep}{0.31}
\setval{ob.default.top256.pubbottom}{1.00}
\setval{ob.default.top256.pubfwd}{16.6}
\setval{ob.default.top256.pubrev}{28.2}
\setval{ob.default.top256.pubrevmean}{29.0}
\setval{ob.default.top256.ranki}{0.86}
\setval{ob.default.top256.ranks}{0.87}
\setval{ob.default.top256.raw}{0.24}
\setval{ob.default.top256.shared}{0.22}
\setval{ob.default.top256.sharei}{90}
\setval{ob.default.top256.shares}{101}
\setval{ob.default.top256.true}{1.00}
\setval{ob.default.top256.units}{256}
\setval{ob.default.top512.base.all.indep}{0.22}
\setval{ob.default.top512.base.all.pubrev}{23.4}
\setval{ob.default.top512.base.all.raw}{0.18}
\setval{ob.default.top512.base.all.shared}{0.15}
\setval{ob.default.top512.base.all.units}{512}
\setval{ob.default.top512.base.top.indep}{0.28}
\setval{ob.default.top512.base.top.pubrev}{28.4}
\setval{ob.default.top512.base.top.raw}{0.23}
\setval{ob.default.top512.base.top.shared}{0.19}
\setval{ob.default.top512.base.top.units}{275}
\setval{ob.default.top512.diffi}{$-$0.03}
\setval{ob.default.top512.diffs}{0.04}
\setval{ob.default.top512.directions}{53.7}
\setval{ob.default.top512.ed}{96.6}
\setval{ob.default.top512.forward}{0.17}
\setval{ob.default.top512.fwdi}{0.17}
\setval{ob.default.top512.fwds}{0.17}
\setval{ob.default.top512.indep}{0.22}
\setval{ob.default.top512.pubbottom}{1.00}
\setval{ob.default.top512.pubfwd}{22.2}
\setval{ob.default.top512.pubrev}{21.3}
\setval{ob.default.top512.pubrevmean}{23.4}
\setval{ob.default.top512.ranki}{0.87}
\setval{ob.default.top512.ranks}{0.86}
\setval{ob.default.top512.raw}{0.18}
\setval{ob.default.top512.shared}{0.15}
\setval{ob.default.top512.sharei}{96}
\setval{ob.default.top512.shares}{105}
\setval{ob.default.top512.true}{1.00}
\setval{ob.default.top512.units}{512}
\setval{ob.default.top64.base.all.indep}{0.44}
\setval{ob.default.top64.base.all.pubrev}{37.3}
\setval{ob.default.top64.base.all.raw}{0.31}
\setval{ob.default.top64.base.all.shared}{0.37}
\setval{ob.default.top64.base.all.units}{64}
\setval{ob.default.top64.base.top.indep}{0.44}
\setval{ob.default.top64.base.top.pubrev}{37.3}
\setval{ob.default.top64.base.top.raw}{0.31}
\setval{ob.default.top64.base.top.shared}{0.37}
\setval{ob.default.top64.base.top.units}{64}
\setval{ob.default.top64.diffi}{$-$0.13}
\setval{ob.default.top64.diffs}{$-$0.05}
\setval{ob.default.top64.directions}{18.3}
\setval{ob.default.top64.ed}{7.7}
\setval{ob.default.top64.forward}{0.03}
\setval{ob.default.top64.fwdi}{0.04}
\setval{ob.default.top64.fwds}{0.03}
\setval{ob.default.top64.indep}{0.44}
\setval{ob.default.top64.pubbottom}{1.00}
\setval{ob.default.top64.pubfwd}{5.4}
\setval{ob.default.top64.pubrev}{38.7}
\setval{ob.default.top64.pubrevmean}{37.3}
\setval{ob.default.top64.ranki}{0.80}
\setval{ob.default.top64.ranks}{0.57}
\setval{ob.default.top64.raw}{0.31}
\setval{ob.default.top64.shared}{0.37}
\setval{ob.default.top64.sharei}{82}
\setval{ob.default.top64.shares}{92}
\setval{ob.default.top64.true}{1.00}
\setval{ob.default.top64.units}{64}
\setval{ob.group.dims}{512}
\setval{ob.group.units}{1{,}024}
\setval{ob.noise-0.5.all.base.all.indep}{0.03}
\setval{ob.noise-0.5.all.base.all.pubrev}{17.7}
\setval{ob.noise-0.5.all.base.all.raw}{0.09}
\setval{ob.noise-0.5.all.base.all.shared}{0.02}
\setval{ob.noise-0.5.all.base.all.units}{1{,}024}
\setval{ob.noise-0.5.all.base.top.indep}{0.06}
\setval{ob.noise-0.5.all.base.top.pubrev}{28.0}
\setval{ob.noise-0.5.all.base.top.raw}{0.17}
\setval{ob.noise-0.5.all.base.top.shared}{0.04}
\setval{ob.noise-0.5.all.base.top.units}{275}
\setval{ob.noise-0.5.all.diffi}{0.01}
\setval{ob.noise-0.5.all.diffs}{0.02}
\setval{ob.noise-0.5.all.directions}{87.0}
\setval{ob.noise-0.5.all.ed}{630.5}
\setval{ob.noise-0.5.all.forward}{0.25}
\setval{ob.noise-0.5.all.fwdi}{0.25}
\setval{ob.noise-0.5.all.fwds}{0.27}
\setval{ob.noise-0.5.all.gain0.25.all.indep}{0.03}
\setval{ob.noise-0.5.all.gain0.25.all.pubrev}{0.7}
\setval{ob.noise-0.5.all.gain0.25.all.raw}{0.00}
\setval{ob.noise-0.5.all.gain0.25.all.shared}{0.02}
\setval{ob.noise-0.5.all.gain0.25.all.units}{1{,}024}
\setval{ob.noise-0.5.all.gain0.25.top.indep}{0.06}
\setval{ob.noise-0.5.all.gain0.25.top.pubrev}{1.0}
\setval{ob.noise-0.5.all.gain0.25.top.raw}{0.00}
\setval{ob.noise-0.5.all.gain0.25.top.shared}{0.04}
\setval{ob.noise-0.5.all.gain0.25.top.units}{283}
\setval{ob.noise-0.5.all.gain0.5.all.indep}{0.03}
\setval{ob.noise-0.5.all.gain0.5.all.pubrev}{3.7}
\setval{ob.noise-0.5.all.gain0.5.all.raw}{0.01}
\setval{ob.noise-0.5.all.gain0.5.all.shared}{0.02}
\setval{ob.noise-0.5.all.gain0.5.all.units}{1{,}024}
\setval{ob.noise-0.5.all.gain0.5.top.indep}{0.06}
\setval{ob.noise-0.5.all.gain0.5.top.pubrev}{6.8}
\setval{ob.noise-0.5.all.gain0.5.top.raw}{0.02}
\setval{ob.noise-0.5.all.gain0.5.top.shared}{0.04}
\setval{ob.noise-0.5.all.gain0.5.top.units}{253}
\setval{ob.noise-0.5.all.gain0.all.indep}{0.03}
\setval{ob.noise-0.5.all.gain0.all.pubrev}{0.5}
\setval{ob.noise-0.5.all.gain0.all.raw}{0.00}
\setval{ob.noise-0.5.all.gain0.all.shared}{0.02}
\setval{ob.noise-0.5.all.gain0.all.units}{1{,}024}
\setval{ob.noise-0.5.all.gain0.top.indep}{0.06}
\setval{ob.noise-0.5.all.gain0.top.pubrev}{0.5}
\setval{ob.noise-0.5.all.gain0.top.raw}{0.00}
\setval{ob.noise-0.5.all.gain0.top.shared}{0.04}
\setval{ob.noise-0.5.all.gain0.top.units}{266}
\setval{ob.noise-0.5.all.gain1.all.indep}{0.03}
\setval{ob.noise-0.5.all.gain1.all.pubrev}{18.3}
\setval{ob.noise-0.5.all.gain1.all.raw}{0.09}
\setval{ob.noise-0.5.all.gain1.all.shared}{0.02}
\setval{ob.noise-0.5.all.gain1.all.units}{1{,}024}
\setval{ob.noise-0.5.all.gain1.top.indep}{0.06}
\setval{ob.noise-0.5.all.gain1.top.pubrev}{28.9}
\setval{ob.noise-0.5.all.gain1.top.raw}{0.17}
\setval{ob.noise-0.5.all.gain1.top.shared}{0.04}
\setval{ob.noise-0.5.all.gain1.top.units}{278}
\setval{ob.noise-0.5.all.gapbottomadded}{1.00}
\setval{ob.noise-0.5.all.indep}{0.03}
\setval{ob.noise-0.5.all.pubbottom}{0.49}
\setval{ob.noise-0.5.all.pubbottomadded}{1.00}
\setval{ob.noise-0.5.all.pubfwd}{70.0}
\setval{ob.noise-0.5.all.pubrev}{2.9}
\setval{ob.noise-0.5.all.pubrevmean}{8.2}
\setval{ob.noise-0.5.all.ranki}{0.22}
\setval{ob.noise-0.5.all.ranks}{0.21}
\setval{ob.noise-0.5.all.raw}{0.04}
\setval{ob.noise-0.5.all.shared}{0.02}
\setval{ob.noise-0.5.all.sharei}{101}
\setval{ob.noise-0.5.all.shares}{102}
\setval{ob.noise-0.5.all.true}{0.80}
\setval{ob.noise-0.5.all.units}{5{,}120}
\setval{ob.noise-0.5.base.base.all.indep}{0.09}
\setval{ob.noise-0.5.base.base.all.pubrev}{17.7}
\setval{ob.noise-0.5.base.base.all.raw}{0.09}
\setval{ob.noise-0.5.base.base.all.shared}{0.06}
\setval{ob.noise-0.5.base.base.all.units}{1{,}024}
\setval{ob.noise-0.5.base.base.top.indep}{0.16}
\setval{ob.noise-0.5.base.base.top.pubrev}{28.0}
\setval{ob.noise-0.5.base.base.top.raw}{0.17}
\setval{ob.noise-0.5.base.base.top.shared}{0.11}
\setval{ob.noise-0.5.base.base.top.units}{275}
\setval{ob.noise-0.5.base.diffi}{0.00}
\setval{ob.noise-0.5.base.diffs}{0.03}
\setval{ob.noise-0.5.base.directions}{54.0}
\setval{ob.noise-0.5.base.ed}{205.1}
\setval{ob.noise-0.5.base.forward}{0.14}
\setval{ob.noise-0.5.base.fwdi}{0.14}
\setval{ob.noise-0.5.base.fwds}{0.14}
\setval{ob.noise-0.5.base.indep}{0.09}
\setval{ob.noise-0.5.base.pubbottom}{1.00}
\setval{ob.noise-0.5.base.pubfwd}{21.7}
\setval{ob.noise-0.5.base.pubrev}{15.7}
\setval{ob.noise-0.5.base.pubrevmean}{17.7}
\setval{ob.noise-0.5.base.ranki}{0.85}
\setval{ob.noise-0.5.base.ranks}{0.80}
\setval{ob.noise-0.5.base.raw}{0.09}
\setval{ob.noise-0.5.base.shared}{0.06}
\setval{ob.noise-0.5.base.sharei}{100}
\setval{ob.noise-0.5.base.shares}{104}
\setval{ob.noise-0.5.base.true}{1.00}
\setval{ob.noise-0.5.base.units}{1{,}024}
\setval{ob.noise-0.95.all.base.all.indep}{0.06}
\setval{ob.noise-0.95.all.base.all.pubrev}{17.2}
\setval{ob.noise-0.95.all.base.all.raw}{0.16}
\setval{ob.noise-0.95.all.base.all.shared}{0.04}
\setval{ob.noise-0.95.all.base.all.units}{1{,}024}
\setval{ob.noise-0.95.all.base.top.indep}{0.11}
\setval{ob.noise-0.95.all.base.top.pubrev}{27.1}
\setval{ob.noise-0.95.all.base.top.raw}{0.26}
\setval{ob.noise-0.95.all.base.top.shared}{0.08}
\setval{ob.noise-0.95.all.base.top.units}{275}
\setval{ob.noise-0.95.all.diffi}{0.01}
\setval{ob.noise-0.95.all.diffs}{0.03}
\setval{ob.noise-0.95.all.directions}{148.7}
\setval{ob.noise-0.95.all.ed}{630.5}
\setval{ob.noise-0.95.all.forward}{0.57}
\setval{ob.noise-0.95.all.fwdi}{0.61}
\setval{ob.noise-0.95.all.fwds}{0.58}
\setval{ob.noise-0.95.all.gain0.25.all.indep}{0.06}
\setval{ob.noise-0.95.all.gain0.25.all.pubrev}{0.5}
\setval{ob.noise-0.95.all.gain0.25.all.raw}{0.00}
\setval{ob.noise-0.95.all.gain0.25.all.shared}{0.04}
\setval{ob.noise-0.95.all.gain0.25.all.units}{1{,}024}
\setval{ob.noise-0.95.all.gain0.25.top.indep}{0.11}
\setval{ob.noise-0.95.all.gain0.25.top.pubrev}{0.9}
\setval{ob.noise-0.95.all.gain0.25.top.raw}{0.00}
\setval{ob.noise-0.95.all.gain0.25.top.shared}{0.09}
\setval{ob.noise-0.95.all.gain0.25.top.units}{283}
\setval{ob.noise-0.95.all.gain0.5.all.indep}{0.06}
\setval{ob.noise-0.95.all.gain0.5.all.pubrev}{3.3}
\setval{ob.noise-0.95.all.gain0.5.all.raw}{0.02}
\setval{ob.noise-0.95.all.gain0.5.all.shared}{0.04}
\setval{ob.noise-0.95.all.gain0.5.all.units}{1{,}024}
\setval{ob.noise-0.95.all.gain0.5.top.indep}{0.11}
\setval{ob.noise-0.95.all.gain0.5.top.pubrev}{6.1}
\setval{ob.noise-0.95.all.gain0.5.top.raw}{0.05}
\setval{ob.noise-0.95.all.gain0.5.top.shared}{0.08}
\setval{ob.noise-0.95.all.gain0.5.top.units}{253}
\setval{ob.noise-0.95.all.gain0.all.indep}{0.06}
\setval{ob.noise-0.95.all.gain0.all.pubrev}{0.2}
\setval{ob.noise-0.95.all.gain0.all.raw}{0.00}
\setval{ob.noise-0.95.all.gain0.all.shared}{0.04}
\setval{ob.noise-0.95.all.gain0.all.units}{1{,}024}
\setval{ob.noise-0.95.all.gain0.top.indep}{0.11}
\setval{ob.noise-0.95.all.gain0.top.pubrev}{0.2}
\setval{ob.noise-0.95.all.gain0.top.raw}{0.00}
\setval{ob.noise-0.95.all.gain0.top.shared}{0.09}
\setval{ob.noise-0.95.all.gain0.top.units}{266}
\setval{ob.noise-0.95.all.gain1.all.indep}{0.06}
\setval{ob.noise-0.95.all.gain1.all.pubrev}{17.9}
\setval{ob.noise-0.95.all.gain1.all.raw}{0.16}
\setval{ob.noise-0.95.all.gain1.all.shared}{0.04}
\setval{ob.noise-0.95.all.gain1.all.units}{1{,}024}
\setval{ob.noise-0.95.all.gain1.top.indep}{0.12}
\setval{ob.noise-0.95.all.gain1.top.pubrev}{28.7}
\setval{ob.noise-0.95.all.gain1.top.raw}{0.27}
\setval{ob.noise-0.95.all.gain1.top.shared}{0.09}
\setval{ob.noise-0.95.all.gain1.top.units}{278}
\setval{ob.noise-0.95.all.gapbottomadded}{1.00}
\setval{ob.noise-0.95.all.indep}{0.06}
\setval{ob.noise-0.95.all.pubbottom}{0.38}
\setval{ob.noise-0.95.all.pubbottomadded}{1.00}
\setval{ob.noise-0.95.all.pubfwd}{67.0}
\setval{ob.noise-0.95.all.pubrev}{2.6}
\setval{ob.noise-0.95.all.pubrevmean}{7.8}
\setval{ob.noise-0.95.all.ranki}{0.23}
\setval{ob.noise-0.95.all.ranks}{0.23}
\setval{ob.noise-0.95.all.raw}{0.07}
\setval{ob.noise-0.95.all.shared}{0.04}
\setval{ob.noise-0.95.all.sharei}{101}
\setval{ob.noise-0.95.all.shares}{103}
\setval{ob.noise-0.95.all.true}{0.80}
\setval{ob.noise-0.95.all.units}{5{,}120}
\setval{ob.noise-0.95.base.base.all.indep}{0.15}
\setval{ob.noise-0.95.base.base.all.pubrev}{17.2}
\setval{ob.noise-0.95.base.base.all.raw}{0.16}
\setval{ob.noise-0.95.base.base.all.shared}{0.10}
\setval{ob.noise-0.95.base.base.all.units}{1{,}024}
\setval{ob.noise-0.95.base.base.top.indep}{0.26}
\setval{ob.noise-0.95.base.base.top.pubrev}{27.1}
\setval{ob.noise-0.95.base.base.top.raw}{0.26}
\setval{ob.noise-0.95.base.base.top.shared}{0.17}
\setval{ob.noise-0.95.base.base.top.units}{275}
\setval{ob.noise-0.95.base.diffi}{0.00}
\setval{ob.noise-0.95.base.diffs}{0.06}
\setval{ob.noise-0.95.base.directions}{70.7}
\setval{ob.noise-0.95.base.ed}{205.1}
\setval{ob.noise-0.95.base.forward}{0.26}
\setval{ob.noise-0.95.base.fwdi}{0.25}
\setval{ob.noise-0.95.base.fwds}{0.25}
\setval{ob.noise-0.95.base.indep}{0.15}
\setval{ob.noise-0.95.base.pubbottom}{1.00}
\setval{ob.noise-0.95.base.pubfwd}{21.2}
\setval{ob.noise-0.95.base.pubrev}{14.9}
\setval{ob.noise-0.95.base.pubrevmean}{17.2}
\setval{ob.noise-0.95.base.ranki}{0.87}
\setval{ob.noise-0.95.base.ranks}{0.85}
\setval{ob.noise-0.95.base.raw}{0.16}
\setval{ob.noise-0.95.base.shared}{0.10}
\setval{ob.noise-0.95.base.sharei}{100}
\setval{ob.noise-0.95.base.shares}{107}
\setval{ob.noise-0.95.base.true}{1.00}
\setval{ob.noise-0.95.base.units}{1{,}024}
\setval{ob.repeats}{3}
\setval{ob.sites}{169}
\setval{ob.sites-42.all.base.all.indep}{0.01}
\setval{ob.sites-42.all.base.all.pubrev}{5.0}
\setval{ob.sites-42.all.base.all.raw}{0.04}
\setval{ob.sites-42.all.base.all.shared}{0.01}
\setval{ob.sites-42.all.base.all.units}{1{,}024}
\setval{ob.sites-42.all.base.top.indep}{0.02}
\setval{ob.sites-42.all.base.top.pubrev}{8.9}
\setval{ob.sites-42.all.base.top.raw}{0.07}
\setval{ob.sites-42.all.base.top.shared}{0.02}
\setval{ob.sites-42.all.base.top.units}{275}
\setval{ob.sites-42.all.diffi}{0.00}
\setval{ob.sites-42.all.diffs}{0.00}
\setval{ob.sites-42.all.directions}{42.0}
\setval{ob.sites-42.all.ed}{630.5}
\setval{ob.sites-42.all.forward}{0.42}
\setval{ob.sites-42.all.fwdi}{0.42}
\setval{ob.sites-42.all.fwds}{0.43}
\setval{ob.sites-42.all.gain0.25.all.indep}{0.01}
\setval{ob.sites-42.all.gain0.25.all.pubrev}{0.3}
\setval{ob.sites-42.all.gain0.25.all.raw}{0.00}
\setval{ob.sites-42.all.gain0.25.all.shared}{0.01}
\setval{ob.sites-42.all.gain0.25.all.units}{1{,}024}
\setval{ob.sites-42.all.gain0.25.top.indep}{0.02}
\setval{ob.sites-42.all.gain0.25.top.pubrev}{0.4}
\setval{ob.sites-42.all.gain0.25.top.raw}{0.00}
\setval{ob.sites-42.all.gain0.25.top.shared}{0.02}
\setval{ob.sites-42.all.gain0.25.top.units}{283}
\setval{ob.sites-42.all.gain0.5.all.indep}{0.01}
\setval{ob.sites-42.all.gain0.5.all.pubrev}{0.8}
\setval{ob.sites-42.all.gain0.5.all.raw}{0.00}
\setval{ob.sites-42.all.gain0.5.all.shared}{0.01}
\setval{ob.sites-42.all.gain0.5.all.units}{1{,}024}
\setval{ob.sites-42.all.gain0.5.top.indep}{0.02}
\setval{ob.sites-42.all.gain0.5.top.pubrev}{1.5}
\setval{ob.sites-42.all.gain0.5.top.raw}{0.01}
\setval{ob.sites-42.all.gain0.5.top.shared}{0.02}
\setval{ob.sites-42.all.gain0.5.top.units}{253}
\setval{ob.sites-42.all.gain0.all.indep}{0.01}
\setval{ob.sites-42.all.gain0.all.pubrev}{0.2}
\setval{ob.sites-42.all.gain0.all.raw}{0.00}
\setval{ob.sites-42.all.gain0.all.shared}{0.01}
\setval{ob.sites-42.all.gain0.all.units}{1{,}024}
\setval{ob.sites-42.all.gain0.top.indep}{0.02}
\setval{ob.sites-42.all.gain0.top.pubrev}{0.2}
\setval{ob.sites-42.all.gain0.top.raw}{0.00}
\setval{ob.sites-42.all.gain0.top.shared}{0.02}
\setval{ob.sites-42.all.gain0.top.units}{266}
\setval{ob.sites-42.all.gain1.all.indep}{0.01}
\setval{ob.sites-42.all.gain1.all.pubrev}{5.0}
\setval{ob.sites-42.all.gain1.all.raw}{0.04}
\setval{ob.sites-42.all.gain1.all.shared}{0.01}
\setval{ob.sites-42.all.gain1.all.units}{1{,}024}
\setval{ob.sites-42.all.gain1.top.indep}{0.02}
\setval{ob.sites-42.all.gain1.top.pubrev}{8.6}
\setval{ob.sites-42.all.gain1.top.raw}{0.07}
\setval{ob.sites-42.all.gain1.top.shared}{0.02}
\setval{ob.sites-42.all.gain1.top.units}{278}
\setval{ob.sites-42.all.gapbottomadded}{0.99}
\setval{ob.sites-42.all.indep}{0.01}
\setval{ob.sites-42.all.pubbottom}{0.56}
\setval{ob.sites-42.all.pubbottomadded}{1.00}
\setval{ob.sites-42.all.pubfwd}{68.0}
\setval{ob.sites-42.all.pubrev}{0.7}
\setval{ob.sites-42.all.pubrevmean}{2.3}
\setval{ob.sites-42.all.ranki}{0.19}
\setval{ob.sites-42.all.ranks}{0.18}
\setval{ob.sites-42.all.raw}{0.01}
\setval{ob.sites-42.all.shared}{0.01}
\setval{ob.sites-42.all.sharei}{100}
\setval{ob.sites-42.all.shares}{100}
\setval{ob.sites-42.all.true}{0.80}
\setval{ob.sites-42.all.units}{5{,}120}
\setval{ob.sites-42.base.base.all.indep}{0.03}
\setval{ob.sites-42.base.base.all.pubrev}{5.0}
\setval{ob.sites-42.base.base.all.raw}{0.04}
\setval{ob.sites-42.base.base.all.shared}{0.03}
\setval{ob.sites-42.base.base.all.units}{1{,}024}
\setval{ob.sites-42.base.base.top.indep}{0.07}
\setval{ob.sites-42.base.base.top.pubrev}{8.9}
\setval{ob.sites-42.base.base.top.raw}{0.07}
\setval{ob.sites-42.base.base.top.shared}{0.06}
\setval{ob.sites-42.base.base.top.units}{275}
\setval{ob.sites-42.base.diffi}{0.00}
\setval{ob.sites-42.base.diffs}{0.00}
\setval{ob.sites-42.base.directions}{40.7}
\setval{ob.sites-42.base.ed}{205.1}
\setval{ob.sites-42.base.forward}{0.22}
\setval{ob.sites-42.base.fwdi}{0.22}
\setval{ob.sites-42.base.fwds}{0.21}
\setval{ob.sites-42.base.indep}{0.03}
\setval{ob.sites-42.base.pubbottom}{1.00}
\setval{ob.sites-42.base.pubfwd}{21.9}
\setval{ob.sites-42.base.pubrev}{4.0}
\setval{ob.sites-42.base.pubrevmean}{5.0}
\setval{ob.sites-42.base.ranki}{0.66}
\setval{ob.sites-42.base.ranks}{0.62}
\setval{ob.sites-42.base.raw}{0.04}
\setval{ob.sites-42.base.shared}{0.03}
\setval{ob.sites-42.base.sharei}{100}
\setval{ob.sites-42.base.shares}{101}
\setval{ob.sites-42.base.true}{1.00}
\setval{ob.sites-42.base.units}{1{,}024}
\setval{ob.sites-676.all.base.all.indep}{0.15}
\setval{ob.sites-676.all.base.all.pubrev}{48.6}
\setval{ob.sites-676.all.base.all.raw}{0.37}
\setval{ob.sites-676.all.base.all.shared}{0.05}
\setval{ob.sites-676.all.base.all.units}{1{,}024}
\setval{ob.sites-676.all.base.top.indep}{0.28}
\setval{ob.sites-676.all.base.top.pubrev}{61.9}
\setval{ob.sites-676.all.base.top.raw}{0.53}
\setval{ob.sites-676.all.base.top.shared}{0.11}
\setval{ob.sites-676.all.base.top.units}{275}
\setval{ob.sites-676.all.diffi}{0.01}
\setval{ob.sites-676.all.diffs}{0.11}
\setval{ob.sites-676.all.directions}{198.3}
\setval{ob.sites-676.all.ed}{630.5}
\setval{ob.sites-676.all.forward}{0.42}
\setval{ob.sites-676.all.fwdi}{0.42}
\setval{ob.sites-676.all.fwds}{0.42}
\setval{ob.sites-676.all.gain0.25.all.indep}{0.16}
\setval{ob.sites-676.all.gain0.25.all.pubrev}{2.8}
\setval{ob.sites-676.all.gain0.25.all.raw}{0.00}
\setval{ob.sites-676.all.gain0.25.all.shared}{0.06}
\setval{ob.sites-676.all.gain0.25.all.units}{1{,}024}
\setval{ob.sites-676.all.gain0.25.top.indep}{0.28}
\setval{ob.sites-676.all.gain0.25.top.pubrev}{5.6}
\setval{ob.sites-676.all.gain0.25.top.raw}{0.01}
\setval{ob.sites-676.all.gain0.25.top.shared}{0.12}
\setval{ob.sites-676.all.gain0.25.top.units}{283}
\setval{ob.sites-676.all.gain0.5.all.indep}{0.15}
\setval{ob.sites-676.all.gain0.5.all.pubrev}{16.4}
\setval{ob.sites-676.all.gain0.5.all.raw}{0.07}
\setval{ob.sites-676.all.gain0.5.all.shared}{0.05}
\setval{ob.sites-676.all.gain0.5.all.units}{1{,}024}
\setval{ob.sites-676.all.gain0.5.top.indep}{0.27}
\setval{ob.sites-676.all.gain0.5.top.pubrev}{27.6}
\setval{ob.sites-676.all.gain0.5.top.raw}{0.15}
\setval{ob.sites-676.all.gain0.5.top.shared}{0.11}
\setval{ob.sites-676.all.gain0.5.top.units}{253}
\setval{ob.sites-676.all.gain0.all.indep}{0.16}
\setval{ob.sites-676.all.gain0.all.pubrev}{0.4}
\setval{ob.sites-676.all.gain0.all.raw}{0.00}
\setval{ob.sites-676.all.gain0.all.shared}{0.06}
\setval{ob.sites-676.all.gain0.all.units}{1{,}024}
\setval{ob.sites-676.all.gain0.top.indep}{0.28}
\setval{ob.sites-676.all.gain0.top.pubrev}{0.4}
\setval{ob.sites-676.all.gain0.top.raw}{0.00}
\setval{ob.sites-676.all.gain0.top.shared}{0.12}
\setval{ob.sites-676.all.gain0.top.units}{266}
\setval{ob.sites-676.all.gain1.all.indep}{0.16}
\setval{ob.sites-676.all.gain1.all.pubrev}{48.9}
\setval{ob.sites-676.all.gain1.all.raw}{0.38}
\setval{ob.sites-676.all.gain1.all.shared}{0.06}
\setval{ob.sites-676.all.gain1.all.units}{1{,}024}
\setval{ob.sites-676.all.gain1.top.indep}{0.28}
\setval{ob.sites-676.all.gain1.top.pubrev}{62.5}
\setval{ob.sites-676.all.gain1.top.raw}{0.53}
\setval{ob.sites-676.all.gain1.top.shared}{0.12}
\setval{ob.sites-676.all.gain1.top.units}{278}
\setval{ob.sites-676.all.gapbottomadded}{1.00}
\setval{ob.sites-676.all.indep}{0.15}
\setval{ob.sites-676.all.pubbottom}{0.23}
\setval{ob.sites-676.all.pubbottomadded}{1.00}
\setval{ob.sites-676.all.pubfwd}{67.9}
\setval{ob.sites-676.all.pubrev}{14.2}
\setval{ob.sites-676.all.pubrevmean}{23.4}
\setval{ob.sites-676.all.ranki}{0.25}
\setval{ob.sites-676.all.ranks}{0.24}
\setval{ob.sites-676.all.raw}{0.16}
\setval{ob.sites-676.all.shared}{0.06}
\setval{ob.sites-676.all.sharei}{101}
\setval{ob.sites-676.all.shares}{113}
\setval{ob.sites-676.all.true}{0.80}
\setval{ob.sites-676.all.units}{5{,}120}
\setval{ob.sites-676.base.base.all.indep}{0.37}
\setval{ob.sites-676.base.base.all.pubrev}{48.6}
\setval{ob.sites-676.base.base.all.raw}{0.37}
\setval{ob.sites-676.base.base.all.shared}{0.14}
\setval{ob.sites-676.base.base.all.units}{1{,}024}
\setval{ob.sites-676.base.base.top.indep}{0.52}
\setval{ob.sites-676.base.base.top.pubrev}{61.9}
\setval{ob.sites-676.base.base.top.raw}{0.53}
\setval{ob.sites-676.base.base.top.shared}{0.23}
\setval{ob.sites-676.base.base.top.units}{275}
\setval{ob.sites-676.base.diffi}{0.01}
\setval{ob.sites-676.base.diffs}{0.23}
\setval{ob.sites-676.base.directions}{81.3}
\setval{ob.sites-676.base.ed}{205.1}
\setval{ob.sites-676.base.forward}{0.21}
\setval{ob.sites-676.base.fwdi}{0.21}
\setval{ob.sites-676.base.fwds}{0.21}
\setval{ob.sites-676.base.indep}{0.37}
\setval{ob.sites-676.base.pubbottom}{1.00}
\setval{ob.sites-676.base.pubfwd}{21.9}
\setval{ob.sites-676.base.pubrev}{47.0}
\setval{ob.sites-676.base.pubrevmean}{48.6}
\setval{ob.sites-676.base.ranki}{0.96}
\setval{ob.sites-676.base.ranks}{0.91}
\setval{ob.sites-676.base.raw}{0.37}
\setval{ob.sites-676.base.shared}{0.14}
\setval{ob.sites-676.base.sharei}{101}
\setval{ob.sites-676.base.shares}{137}
\setval{ob.sites-676.base.true}{1.00}
\setval{ob.sites-676.base.units}{1{,}024}
\setval{ob.stimuli}{1{,}320}
\setval{rel.monkey.forward.dimension}{$-$0.01}
\setval{rel.monkey.indep.dimension}{$-$0.92}
\setval{rel.monkey.reverse.dimension}{$-$0.95}
\setval{rel.monkey.reverse.forward}{0.15}
\setval{rel.monkey.reverse.indep}{0.97}
\setval{rel.monkey.reverse.shared}{0.96}
\setval{rel.monkey.reverse.top}{0.83}
\setval{rel.monkey.shared.dimension}{$-$0.92}
\setval{rel.narratives.forward.dimension}{0.81}
\setval{rel.narratives.indep.dimension}{$-$0.86}
\setval{rel.narratives.reverse.dimension}{$-$0.93}
\setval{rel.narratives.reverse.forward}{$-$0.61}
\setval{rel.narratives.reverse.indep}{0.89}
\setval{rel.narratives.reverse.shared}{0.97}
\setval{rel.narratives.shared.dimension}{$-$0.95}
\setval{rel.podcast.forward.dimension}{0.72}
\setval{rel.podcast.indep.dimension}{$-$0.76}
\setval{rel.podcast.reverse.dimension}{$-$0.46}
\setval{rel.podcast.reverse.forward}{$-$0.12}
\setval{rel.podcast.reverse.indep}{0.78}
\setval{rel.podcast.reverse.shared}{0.64}
\setval{rel.podcast.reverse.top}{0.53}
\setval{rel.podcast.shared.dimension}{$-$0.81}
\setval{t1.a12.complete}{43.9}
\setval{t1.a12.extra}{37.1}
\setval{t1.a12.incomplete}{98.8}
\setval{t1.a12.partial}{52.1}
\setval{t1.a21.complete}{37.8}
\setval{t1.a21.extra}{32.7}
\setval{t1.a21.incomplete}{98.1}
\setval{t1.a21.partial}{45.2}
\setval{t1.agreement.complete}{0.86}
\setval{t1.agreement.extra}{0.84}
\setval{t1.agreement.incomplete}{0.40}
\setval{t1.agreement.partial}{0.91}
\setval{t1.cut.repeats}{5}
\setval{t1.cut.sitesfirst}{130}
\setval{t1.cut.sitessecond}{74}
\setval{t1.cut.stimuli}{300}
\setval{t1.cut.units}{819}
\setval{t1.cutfwdcommon.complete}{14.5}
\setval{t1.cutfwdcommon.extra}{10.2}
\setval{t1.cutfwdcommon.incomplete}{0.6}
\setval{t1.cutfwdcommon.partial}{8.5}
\setval{t1.cutfwdunique.complete}{77.8}
\setval{t1.cutfwdunique.extra}{49.5}
\setval{t1.cutfwdunique.incomplete}{99.8}
\setval{t1.cutfwdunique.partial}{58.8}
\setval{t1.cutrevcommon.complete}{2.5}
\setval{t1.cutrevcommon.extra}{1.8}
\setval{t1.cutrevcommon.incomplete}{0.2}
\setval{t1.cutrevcommon.partial}{1.9}
\setval{t1.cutrevunique.complete}{4.4}
\setval{t1.cutrevunique.extra}{3.1}
\setval{t1.cutrevunique.incomplete}{0.4}
\setval{t1.cutrevunique.partial}{4.7}
\setval{t1.cuttruecommon.complete}{1.00}
\setval{t1.cuttruecommon.extra}{1.00}
\setval{t1.cuttruecommon.incomplete}{0.47}
\setval{t1.cuttruecommon.partial}{0.84}
\setval{t1.cuttrueunique.complete}{1.00}
\setval{t1.cuttrueunique.extra}{1.00}
\setval{t1.cuttrueunique.incomplete}{0.00}
\setval{t1.cuttrueunique.partial}{0.37}
\setval{t1.edfwd.complete}{$-$0.69}
\setval{t1.edfwd.extra}{$-$0.74}
\setval{t1.edfwd.incomplete}{$-$0.62}
\setval{t1.edfwd.partial}{$-$0.59}
\setval{t1.edrev.complete}{$-$0.98}
\setval{t1.edrev.extra}{$-$0.96}
\setval{t1.edrev.incomplete}{$-$0.98}
\setval{t1.edrev.partial}{$-$0.93}
\setval{t1.family}{27}
\setval{t1.fix.models}{27}
\setval{t1.fix.repeats}{5}
\setval{t1.fixfwd.complete}{0.22}
\setval{t1.fixfwd.decay0.7}{0.36}
\setval{t1.fixfwd.decay1.3}{$-$0.19}
\setval{t1.fixfwd.dims8192}{$-$0.54}
\setval{t1.fixfwd.extra}{$-$0.54}
\setval{t1.fixfwd.incomplete}{$-$0.62}
\setval{t1.fixfwd.partial}{$-$0.55}
\setval{t1.fixrev.complete}{$-$1.00}
\setval{t1.fixrev.decay0.7}{$-$1.00}
\setval{t1.fixrev.decay1.3}{$-$1.00}
\setval{t1.fixrev.dims8192}{$-$1.00}
\setval{t1.fixrev.extra}{$-$1.00}
\setval{t1.fixrev.incomplete}{$-$0.96}
\setval{t1.fixrev.partial}{$-$1.00}
\setval{t1.fixtrue.complete}{1.00}
\setval{t1.fixtrue.decay0.7}{1.00}
\setval{t1.fixtrue.decay1.3}{1.00}
\setval{t1.fixtrue.dims8192}{1.00}
\setval{t1.fixtrue.extra}{1.00}
\setval{t1.fixtrue.incomplete}{0.16}
\setval{t1.fixtrue.partial}{0.74}
\setval{t1.forward.complete}{80.0}
\setval{t1.forward.extra}{51.9}
\setval{t1.forward.incomplete}{99.3}
\setval{t1.forward.partial}{59.9}
\setval{t1.fromunits.complete}{29.3}
\setval{t1.fromunits.extra}{29.6}
\setval{t1.fromunits.incomplete}{51.1}
\setval{t1.fromunits.partial}{31.4}
\setval{t1.inst.allfwd}{56.2}
\setval{t1.inst.allrev}{56.2}
\setval{t1.inst.seen}{169}
\setval{t1.inst.seenfwd}{56.1}
\setval{t1.inst.seenrev}{1.4}
\setval{t1.leastfirst.complete}{0.1}
\setval{t1.leastfirst.extra}{0.1}
\setval{t1.leastfirst.incomplete}{0.1}
\setval{t1.leastfirst.partial}{0.1}
\setval{t1.leastlast.complete}{1.2}
\setval{t1.leastlast.extra}{0.6}
\setval{t1.leastlast.incomplete}{0.2}
\setval{t1.leastlast.partial}{0.1}
\setval{t1.match.components}{288}
\setval{t1.matchfwd.complete}{60.0}
\setval{t1.matchfwd.extra}{49.6}
\setval{t1.matchfwd.incomplete}{98.2}
\setval{t1.matchfwd.partial}{64.1}
\setval{t1.matchrev.complete}{9.6}
\setval{t1.matchrev.extra}{7.3}
\setval{t1.matchrev.incomplete}{1.3}
\setval{t1.matchrev.partial}{11.1}
\setval{t1.paira12.complete}{10.5}
\setval{t1.paira12.extra}{8.1}
\setval{t1.paira12.incomplete}{34.3}
\setval{t1.paira12.partial}{11.4}
\setval{t1.paira21.complete}{9.3}
\setval{t1.paira21.extra}{7.7}
\setval{t1.paira21.incomplete}{31.5}
\setval{t1.paira21.partial}{10.7}
\setval{t1.pairfwd.complete}{14.0}
\setval{t1.pairfwd.extra}{12.7}
\setval{t1.pairfwd.incomplete}{27.7}
\setval{t1.pairfwd.partial}{14.7}
\setval{t1.pairrev.complete}{1.0}
\setval{t1.pairrev.extra}{1.0}
\setval{t1.pairrev.incomplete}{0.9}
\setval{t1.pairrev.partial}{1.0}
\setval{t1.pairrevmean.complete}{1.9}
\setval{t1.pairrevmean.extra}{1.8}
\setval{t1.pairrevmean.incomplete}{2.6}
\setval{t1.pairrevmean.partial}{1.9}
\setval{t1.randomfirst.complete}{0.4}
\setval{t1.randomfirst.extra}{0.4}
\setval{t1.randomfirst.incomplete}{0.2}
\setval{t1.randomfirst.partial}{0.4}
\setval{t1.randomlast.complete}{8.8}
\setval{t1.randomlast.extra}{6.6}
\setval{t1.randomlast.incomplete}{0.6}
\setval{t1.randomlast.partial}{9.1}
\setval{t1.repeats}{5}
\setval{t1.reverse.complete}{3.5}
\setval{t1.reverse.extra}{2.5}
\setval{t1.reverse.incomplete}{0.3}
\setval{t1.reverse.partial}{3.3}
\setval{t1.sites.first}{20}
\setval{t1.sites.last}{300}
\setval{t1.trueall.complete}{1.00}
\setval{t1.trueall.extra}{1.00}
\setval{t1.trueall.incomplete}{0.10}
\setval{t1.trueall.partial}{0.52}
\setval{t1.truebottom.complete}{1.00}
\setval{t1.truebottom.extra}{1.00}
\setval{t1.truebottom.incomplete}{0.00}
\setval{t1.truebottom.partial}{0.09}
\setval{t2.complete.bottom}{1.00}
\setval{t2.complete.corrected}{0.18}
\setval{t2.complete.forward}{0.36}
\setval{t2.complete.indep}{0.15}
\setval{t2.complete.noisy}{0.23}
\setval{t2.complete.normalised}{0.19}
\setval{t2.complete.ranki}{0.73}
\setval{t2.complete.ranks}{0.72}
\setval{t2.complete.raw}{0.15}
\setval{t2.complete.shared}{0.12}
\setval{t2.complete.sharei}{100}
\setval{t2.complete.shares}{104}
\setval{t2.complete.signal}{0.32}
\setval{t2.complete.true}{1.00}
\setval{t2.extra.bottom}{1.00}
\setval{t2.extra.corrected}{0.11}
\setval{t2.extra.forward}{0.11}
\setval{t2.extra.indep}{0.09}
\setval{t2.extra.noisy}{0.17}
\setval{t2.extra.normalised}{0.12}
\setval{t2.extra.ranki}{0.72}
\setval{t2.extra.ranks}{0.60}
\setval{t2.extra.raw}{0.09}
\setval{t2.extra.shared}{0.04}
\setval{t2.extra.sharei}{101}
\setval{t2.extra.shares}{105}
\setval{t2.extra.signal}{0.24}
\setval{t2.extra.true}{1.00}
\setval{t2.flatter.bottom}{1.00}
\setval{t2.flatter.corrected}{0.10}
\setval{t2.flatter.forward}{0.25}
\setval{t2.flatter.indep}{0.12}
\setval{t2.flatter.noisy}{0.16}
\setval{t2.flatter.normalised}{0.10}
\setval{t2.flatter.ranki}{0.49}
\setval{t2.flatter.ranks}{0.49}
\setval{t2.flatter.raw}{0.07}
\setval{t2.flatter.shared}{0.10}
\setval{t2.flatter.sharei}{95}
\setval{t2.flatter.shares}{97}
\setval{t2.flatter.signal}{0.22}
\setval{t2.flatter.true}{1.00}
\setval{t2.half.bottom}{0.48}
\setval{t2.half.corrected}{0.13}
\setval{t2.half.forward}{0.37}
\setval{t2.half.indep}{0.15}
\setval{t2.half.noisy}{0.18}
\setval{t2.half.normalised}{0.14}
\setval{t2.half.ranki}{0.52}
\setval{t2.half.ranks}{0.52}
\setval{t2.half.raw}{0.11}
\setval{t2.half.shared}{0.11}
\setval{t2.half.sharei}{96}
\setval{t2.half.shares}{100}
\setval{t2.half.signal}{0.25}
\setval{t2.half.true}{0.51}
\setval{t2.notop.bottom}{0.67}
\setval{t2.notop.corrected}{0.04}
\setval{t2.notop.forward}{0.30}
\setval{t2.notop.indep}{0.13}
\setval{t2.notop.noisy}{0.10}
\setval{t2.notop.normalised}{0.04}
\setval{t2.notop.ranki}{$-$0.29}
\setval{t2.notop.ranks}{$-$0.28}
\setval{t2.notop.raw}{0.03}
\setval{t2.notop.shared}{0.11}
\setval{t2.notop.sharei}{89}
\setval{t2.notop.shares}{92}
\setval{t2.notop.signal}{0.15}
\setval{t2.notop.true}{0.72}
\setval{t2.runs}{12}
\setval{t2.steeper.bottom}{1.00}
\setval{t2.steeper.corrected}{0.21}
\setval{t2.steeper.forward}{0.47}
\setval{t2.steeper.indep}{0.17}
\setval{t2.steeper.noisy}{0.26}
\setval{t2.steeper.normalised}{0.22}
\setval{t2.steeper.ranki}{0.80}
\setval{t2.steeper.ranks}{0.77}
\setval{t2.steeper.raw}{0.18}
\setval{t2.steeper.shared}{0.12}
\setval{t2.steeper.sharei}{102}
\setval{t2.steeper.shares}{108}
\setval{t2.steeper.signal}{0.38}
\setval{t2.steeper.true}{1.00}
\setval{t2.top.bottom}{0.22}
\setval{t2.top.corrected}{0.23}
\setval{t2.top.forward}{0.59}
\setval{t2.top.indep}{0.19}
\setval{t2.top.noisy}{0.24}
\setval{t2.top.normalised}{0.23}
\setval{t2.top.ranki}{0.83}
\setval{t2.top.ranks}{0.73}
\setval{t2.top.raw}{0.20}
\setval{t2.top.shared}{0.09}
\setval{t2.top.sharei}{101}
\setval{t2.top.shares}{113}
\setval{t2.top.signal}{0.28}
\setval{t2.top.true}{0.28}
\setval{t3.monkey.bottomi}{0.59}
\setval{t3.monkey.bottomindep}{0.04}
\setval{t3.monkey.bottomrev}{0.01}
\setval{t3.monkey.bottoms}{0.57}
\setval{t3.monkey.bottomshared}{0.01}
\setval{t3.monkey.ceiling}{0.70}
\setval{t3.monkey.columns}{101}
\setval{t3.monkey.corrected}{0.09}
\setval{t3.monkey.diffi}{$-$0.03}
\setval{t3.monkey.diffs}{0.03}
\setval{t3.monkey.dimension}{166.4}
\setval{t3.monkey.directions}{35}
\setval{t3.monkey.forward}{0.62}
\setval{t3.monkey.forwardi}{0.40}
\setval{t3.monkey.forwardraw}{0.44}
\setval{t3.monkey.forwards}{0.40}
\setval{t3.monkey.indep}{0.12}
\setval{t3.monkey.indep.hi}{0.14}
\setval{t3.monkey.indep.lo}{0.10}
\setval{t3.monkey.models}{41}
\setval{t3.monkey.normalised}{0.10}
\setval{t3.monkey.pi}{$<$0.001}
\setval{t3.monkey.ps}{$<$0.001}
\setval{t3.monkey.ranki}{0.78}
\setval{t3.monkey.ranks}{0.77}
\setval{t3.monkey.raw}{0.09}
\setval{t3.monkey.reverse}{0.09}
\setval{t3.monkey.reverse.hi}{0.10}
\setval{t3.monkey.reverse.lo}{0.08}
\setval{t3.monkey.rows}{1{,}320}
\setval{t3.monkey.shared}{0.06}
\setval{t3.monkey.shared.hi}{0.07}
\setval{t3.monkey.shared.lo}{0.05}
\setval{t3.monkey.sharei}{96}
\setval{t3.monkey.sharei.hi}{97}
\setval{t3.monkey.sharei.lo}{96}
\setval{t3.monkey.shares}{103}
\setval{t3.monkey.shares.hi}{103}
\setval{t3.monkey.shares.lo}{103}
\setval{t3.monkey.sites}{101}
\setval{t3.narratives.bottomi}{0.36}
\setval{t3.narratives.bottomindep}{0.25}
\setval{t3.narratives.bottomrev}{0.00}
\setval{t3.narratives.bottoms}{0.35}
\setval{t3.narratives.bottomshared}{0.00}
\setval{t3.narratives.ceiling}{0.27}
\setval{t3.narratives.columns}{14{,}990}
\setval{t3.narratives.corrected}{0.01}
\setval{t3.narratives.diffi}{$-$0.26}
\setval{t3.narratives.diffs}{$-$0.01}
\setval{t3.narratives.dimension}{194.2}
\setval{t3.narratives.directions}{10}
\setval{t3.narratives.forward}{0.14}
\setval{t3.narratives.forwardi}{0.04}
\setval{t3.narratives.forwardraw}{0.04}
\setval{t3.narratives.forwards}{0.04}
\setval{t3.narratives.indep}{0.27}
\setval{t3.narratives.indep.hi}{0.30}
\setval{t3.narratives.indep.lo}{0.24}
\setval{t3.narratives.models}{21}
\setval{t3.narratives.normalised}{0.01}
\setval{t3.narratives.pi}{$<$0.001}
\setval{t3.narratives.ps}{0.199}
\setval{t3.narratives.ranki}{0.50}
\setval{t3.narratives.ranks}{0.43}
\setval{t3.narratives.raw}{0.01}
\setval{t3.narratives.reverse}{0.01}
\setval{t3.narratives.reverse.hi}{0.01}
\setval{t3.narratives.reverse.lo}{0.01}
\setval{t3.narratives.rows}{9{,}329}
\setval{t3.narratives.shared}{0.01}
\setval{t3.narratives.shared.hi}{0.02}
\setval{t3.narratives.shared.lo}{0.01}
\setval{t3.narratives.sharei}{73}
\setval{t3.narratives.sharei.hi}{76}
\setval{t3.narratives.sharei.lo}{71}
\setval{t3.narratives.shares}{99}
\setval{t3.narratives.shares.hi}{100}
\setval{t3.narratives.shares.lo}{99}
\setval{t3.narratives.sites}{14{,}990}
\setval{t3.podcast.bottomi}{0.38}
\setval{t3.podcast.bottomindep}{0.06}
\setval{t3.podcast.bottomrev}{0.00}
\setval{t3.podcast.bottoms}{0.28}
\setval{t3.podcast.bottomshared}{0.01}
\setval{t3.podcast.ceiling}{0.09}
\setval{t3.podcast.columns}{1{,}018}
\setval{t3.podcast.diffi}{$-$0.06}
\setval{t3.podcast.diffs}{0.01}
\setval{t3.podcast.dimension}{257.3}
\setval{t3.podcast.directions}{13}
\setval{t3.podcast.forward}{0.34}
\setval{t3.podcast.forwardi}{0.03}
\setval{t3.podcast.forwardraw}{0.03}
\setval{t3.podcast.forwards}{0.04}
\setval{t3.podcast.indep}{0.08}
\setval{t3.podcast.indep.hi}{0.10}
\setval{t3.podcast.indep.lo}{0.06}
\setval{t3.podcast.models}{21}
\setval{t3.podcast.pi}{$<$0.001}
\setval{t3.podcast.ps}{0.315}
\setval{t3.podcast.ranki}{0.47}
\setval{t3.podcast.ranks}{0.28}
\setval{t3.podcast.reverse}{0.02}
\setval{t3.podcast.reverse.hi}{0.02}
\setval{t3.podcast.reverse.lo}{0.01}
\setval{t3.podcast.rows}{5{,}135}
\setval{t3.podcast.shared}{0.01}
\setval{t3.podcast.shared.hi}{0.02}
\setval{t3.podcast.shared.lo}{0.01}
\setval{t3.podcast.sharei}{94}
\setval{t3.podcast.sharei.hi}{96}
\setval{t3.podcast.sharei.lo}{92}
\setval{t3.podcast.shares}{101}
\setval{t3.podcast.shares.hi}{101}
\setval{t3.podcast.shares.lo}{99}
\setval{t3.podcast.sites}{358}
}{}

\pgfplotsset{
	paperaxis/.style={scale only axis, width=1.22in, height=0.95in,
		tick label style={font=\scriptsize}, label style={font=\footnotesize},
		axis line style={line width=0.5pt}, tick style={line width=0.4pt, black},
		xlabel style={yshift=2pt}, ylabel style={yshift=-3pt},
		legend style={font=\scriptsize, draw=none, fill=none, row sep=-2pt},
		legend cell align=left, scaled ticks=false,
		title style={font=\footnotesize, yshift=-5pt},
		yticklabel style={/pgf/number format/fixed, /pgf/number format/precision=2},
		xticklabel style={/pgf/number format/fixed, /pgf/number format/precision=2}},
}
\newcommand{\ifdata}[3]{\IfFileExists{figdata/#1}{#2}{#3}}
\newcommand{\nodata}{\node at (rel axis cs:0.5,0.5) {\TBD};}
\newcommand{\datarange}[3]{\IfFileExists{figdata/#1}{\pgfplotsset{#2/.style={}}}{\pgfplotsset{#2/.style={#3}}}}

\newcommand{\Cov}{\operatorname{Cov}}
\newcommand{\E}{\mathbb{E}}
\newcommand{\Hsp}{\mathcal{H}}
\newcommand{\Rsig}{R^{\circ}}
\newcommand{\Cind}{C^{\mathrm{ind}}}
\newcommand{\Csh}{C^{\mathrm{sh}}}

\icmltitlerunning{Absent from the Brain or Absent from the Recording?}

\begin{document}
	
	\twocolumn[
	\icmltitle{Absent from the Brain or Absent from the Recording? \\
		How Neural Sampling and Noise Shape Reverse Predictivity}
	
	\icmlsetsymbol{equal}{*}
	
		\begin{icmlauthorlist}
		\icmlauthor{Xin Zhang}{ttu}
		\icmlauthor{Hanyao Zeng}{ttu}
		\icmlauthor{Victor S. Sheng}{ttu}
	\end{icmlauthorlist}
	
	\icmlaffiliation{ttu}{Department of Computer Science, Texas Tech University, Lubbock, Texas, USA}
	
	\icmlcorrespondingauthor{Xin Zhang}{xin.zhang@ttu.edu}

	\icmlkeywords{reverse predictivity, model-brain comparison, neural recordings, language models, partial identification}
	
	\vskip 0.3in
	]

	\printAffiliationsAndNotice{}  

	\begin{abstract}
		Comparing neural networks and the brain focuses on how accurately the model can predict recorded brain activity. Recent research on reverse predictivity reverses this question and reads model units that brain activity fails to predict as a mismatch between the model and the brain. The control analyses were conducted to rule out the possibility of insufficient neuron sampling. We show that regardless of the recording length, the possibility that the brain has a unit that the recorded sites fail to predict cannot be ruled out. Reverse scores measure the proportion of a unit that the brain expresses at an intensity exceeding the recording resolution limit, which depends on the number of recorded sites and their noise level. Therefore, a brain possessing the entire model can be predicted well by the model, but it may predict the model poorly and give the same results in these control analyses. We suggest using two reference values for each reverse score, which are the scores the same recording would receive if the brain possessed the entire model. In brains built from four language models, where we know what each brain contains, the reverse score of the brain possessing the entire model is \val{t2.complete.raw}, with the two reference values being \val{t2.complete.indep} and \val{t2.complete.shared}. In recordings of the monkey visual cortex and human brains listening to stories, the reference values account for at least \val{abs.sharelow} percent of the reverse score's deviation from 1. Reference values also fail to identify missing units. In simulations with only one brain and multiple models, units missing from the brain lower both the reverse score and the reference values.
	\end{abstract}
		\newpage
	\section{Introduction}
	\label{sec:intro}
	
	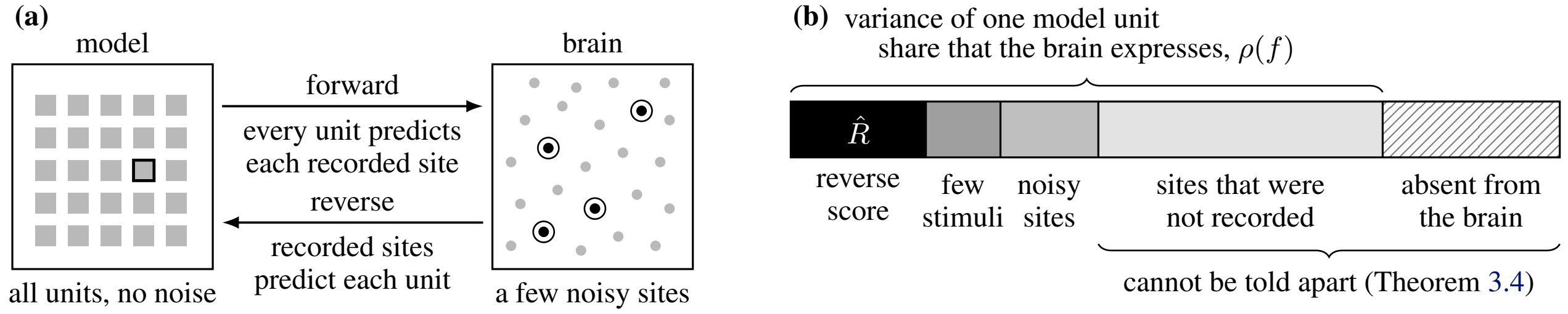
\begin{figure*}[t]
		\centering
		\begin{tikzpicture}[font=\footnotesize, >={Latex[length=1.8mm]}]
			\node[anchor=west, font=\footnotesize\bfseries] at (-0.1,3.05) {(a)};
			\draw[line width=0.6pt] (0,0.35) rectangle (2.15,2.55);
			\node[anchor=south] at (1.075,2.55) {model};
			\foreach \x in {0.25,0.6,0.95,1.3,1.65} {
				\foreach \y in {0.6,0.95,1.3,1.65,2.0} {
					\fill[gray!55] (\x,\y) rectangle ++(0.22,0.22);
				}
			}
			\draw[line width=0.9pt] (1.3,1.3) rectangle ++(0.22,0.22);
			\node[anchor=north, align=center] at (1.075,0.33) {all units, no noise};
			\draw[line width=0.6pt] (5.15,0.35) rectangle (7.3,2.55);
			\node[anchor=south] at (6.225,2.55) {brain};
			\foreach \x/\y in {5.35/0.6,5.7/0.75,6.1/0.55,6.5/0.7,6.9/0.6,5.45/1.05,5.85/1.2,6.25/1.0,6.7/1.15,7.05/1.0,
				5.35/1.5,5.75/1.65,6.15/1.45,6.55/1.6,6.95/1.5,5.5/1.95,5.9/2.1,6.3/1.9,6.75/2.05,7.05/1.95,
				5.6/2.35,6.45/2.35,7.0/2.35,6.05/2.3} {
				\fill[gray!55] (\x,\y) circle (1.6pt);
			}
			\foreach \x/\y in {5.7/0.75,6.25/1.0,5.75/1.65,6.75/2.05} {
				\fill[black] (\x,\y) circle (1.6pt);
				\draw[line width=0.5pt] (\x,\y) circle (3.2pt);
			}
			\node[anchor=north, align=center] at (6.225,0.33) {a few noisy sites};
			\draw[->, line width=0.7pt] (2.25,2.1) -- (5.05,2.1);
			\node[anchor=south] at (3.65,2.1) {forward};
			\node[anchor=north, align=center] at (3.65,2.08) {every unit predicts\\each recorded site};
			\draw[->, line width=0.7pt] (5.05,0.85) -- (2.25,0.85);
			\node[anchor=south] at (3.65,0.85) {reverse};
			\node[anchor=north, align=center] at (3.65,0.83) {recorded sites\\predict each unit};
			
			\begin{scope}[shift={(8.35,0)}]
				\node[anchor=west, font=\footnotesize\bfseries] at (-0.1,3.05) {(b)};
				\node[anchor=west] at (0.45,3.05) {variance of one model unit};
				\fill[black] (0,1.55) rectangle (1.45,2.15);
				\fill[gray!75] (1.45,1.55) rectangle (2.25,2.15);
				\fill[gray!50] (2.25,1.55) rectangle (3.3,2.15);
				\fill[gray!22] (3.3,1.55) rectangle (6.35,2.15);
				\draw[pattern=north east lines, pattern color=gray] (6.35,1.55) rectangle (8.25,2.15);
				\draw[line width=0.6pt] (0,1.55) rectangle (8.25,2.15);
				\foreach \x in {1.45,2.25,3.3,6.35} { \draw[line width=0.6pt] (\x,1.55) -- (\x,2.15); }
				\node[white] at (0.725,1.85) {$\hat R$};
				\draw[decorate, decoration={brace, amplitude=4pt}, line width=0.5pt] (0,2.25) -- (6.35,2.25);
				\node[anchor=south] at (3.175,2.42) {share that the brain expresses, $\rho(f)$};
				\node[anchor=north, align=center] at (0.725,1.5) {reverse\\score};
				\node[anchor=north, align=center] at (1.85,1.5) {few\\stimuli};
				\node[anchor=north, align=center] at (2.775,1.5) {noisy\\sites};
				\node[anchor=north, align=center] at (4.825,1.5) {sites that were\\not recorded};
				\node[anchor=north, align=center] at (7.3,1.5) {absent from\\the brain};
				\draw[decorate, decoration={brace, amplitude=4pt, mirror}, line width=0.5pt] (3.3,0.62) -- (8.25,0.62);
				\node[anchor=north] at (5.775,0.45) {cannot be told apart (\cref{thm:identified})};
			\end{scope}
		\end{tikzpicture}
		\caption{The two directions of a comparison between a model and a brain. (a) In the forward direction all units of the model predict each recorded site. In the reverse direction a few noisy sites predict each unit. (b) The variance of a unit that a reverse score leaves unpredicted has four sources. Three of them belong to the recording. No amount of data separates what the brain lacks from what the sites did not record.}
		\label{fig:overview}
	\end{figure*}

	A common approach to comparing neural networks and the brain is to use the network to predict brain activity. For example, both are shown the same stimuli, a linear mapping is fitted from the model's activation to the response at each recording site in the brain, and then the model is scored based on how much of the response it can predict on new stimuli. Models that perform well in this forward test are considered candidates to explain the brain, in vision \citep{yamins2014} and in language \citep{schrimpf2021,goldstein2022,caucheteux2022,antonello2023}.
	
	However, the forward test does not reveal what the model has beyond the brain. While a model can predict every recording site, it may also carry a large amount of information that no brain neuron expresses. \citet{muzellec2026} reversed this test. They used the recorded neuronal responses to predict each unit of the model, calling the result reverse predictivity. In the inferior temporal (IT) cortex of macaques, models that predicted about half of the neural variance contained many model units that were poorly predicted by neurons in the brain. These units are called unique. This measure aims to distinguish different models of the same brain. The authors note that a low reverse score does not necessarily mean a dimension is missing in the brain, but that the dimension cannot be linearly recovered from the sampled IT neuron population \citep[p.~475]{muzellec2026}. They also explored whether noise or undersampling of neurons could explain these unique units. The same unit scored consistently in both monkeys, the most difficult-to-predict units remained low even after increasing the number of recording sites, and both monkeys had roughly the same predictive ability for each other in bidirectional prediction. Based on these results and further control analyses, they concluded that these unique units reflect a mismatch between the model and the brain, and are not merely a product of noise or undersampling. They note that more neurons or stimuli may reveal further matching, but the gains are diminishing. Since then, reverse predictivity has been used as a training goal \citep{kostousov2026} and as a diagnostic tool outside of neuroscience \citep{kutluozen2026}. For language models, the underlying question is urgent. If we could list what language models have learned that the human brain does not express, it would tell us what humans can specifically learn from models.
	
	The question we need to ask is whether a single recording can provide such a list. The two directions use the recording differently (\cref{fig:overview}a). In the forward direction, every unit of the model is available, and none of them is noisy. Dividing by the reliability of the predicted brain site removes noise, and averaging across all recorded sites provides a reasonable estimate of the entire brain region. In the reverse direction, the prediction uses only a few hundred noisy sites out of millions of neurons, without considering unrecorded neurons. Therefore, if a unit cannot be predicted, there could be two different reasons. It might not exist in the brain, or it might have been missed in the recording.
	
	Our first result is that these two reasons cannot be told apart (\cref{fig:overview}b). Regardless of the specific recording or the amount of data used, the probability of a brain lacking a certain unit and a brain possessing that unit producing the same data is as close to 1 as we like (\cref{thm:identified}). The reverse score is a floor on how much of a unit the brain expresses, never a ceiling.
	
	Our second result explains what this score measures. If the sites are randomly selected from a brain region, the score roughly equals the proportion of that unit that the brain carries in dimensions with variance exceeding a certain threshold (\cref{prop:resolution}). We call this threshold the resolution limit of the recording. As the number of sites increases, the resolution limit decreases, just as the error of a learner decreases with increasing samples, but it increases with increasing noise. A low reverse score is exactly what a brain with the entire model would produce. Such a brain also yields the three observations described above and the results of the further control analyses, so these do not exclude limited sampling as the cause.
	
	The third result is a method for interpreting reverse scores. We compare each score to a reference value—the score the same recording would give if the brain possessed the entire model. The reference value is calculated from the model and the recording itself. Simulated sites mix model units at random, carry the noise of the real sites, and are predicted by the model as well as the real sites are. A single recording cannot explain whether its sites respond alike due to the brain or the measurement, so we use two reference values, representing two extremes. The reference values show how low a score can be when nothing is absent. They do not show that nothing is absent. For recordings with repeated stimuli, we also add a reverse score to remove site noise.
	
	We test these results in three scenarios. In simulated neuronal populations, the formula for the resolution limit differs from the exact value by an average of no more than \val{f.all.mean}, and a brain possessing the entire model reproduces the results of the control analyses. With one brain and several models, units that the brain expresses weakly score almost as low as units that it lacks (\cref{sec:simulation}). In brains constructed using four language models, the proportion of each unit possessed by the brain is known precisely. The average reverse score of the brain with the entire model is \val{t2.complete.raw}, with the two reference values being \val{t2.complete.indep} and \val{t2.complete.shared} (\cref{sec:known}). In recordings of the inferior temporal cortex of monkeys, human electrocorticography (ECoG), and human fMRI, at least \val{abs.sharelow} percent of the deviation of the reverse scores from 1 can be explained by the reference values (\cref{sec:recordings}).
	
	\section{Related Work}
	\label{sec:related}
	
	\textbf{Forward Comparison.} Using model activation to predict neural responses is a standard method for evaluating a brain model \citep{yamins2014,schrimpf2021,huth2016}. Forward scores are compared to a ceiling, which is the reliability of the site or the scores achievable by other animals \citep{nayebi2023}. Several studies have cautioned that forward scores are influenced by the model's spectrum as much as by its similarity to the brain \citep{elmoznino2024,schaeffer2024,canatar2023}. \citet{han2023} replaced the brain with a known network and found that forward scores often fail to select the correct model. We use the same approach in the reverse direction, using a brain with known content.
	
	\textbf{Reverse and Bidirectional Comparison.} \citet{muzellec2026} proposed reverse predictivity and reported that it is low where forward predictivity is high. They present it as a conservative criterion for a mismatch between models and brains, to be read alongside the forward score and the comparison between two animals. They expect a model that matches the cortex at the resolution of their recordings to be as symmetric as two animals are, and they suggest that future work normalize the two scores, for example by ceilings between brains or by dimensionality. \cref{prop:resolution} puts this resolution into a formula, and our reference values are a normalization of this kind. \citet{thobani2025} used the kind of transformation that connects two animals to map in both directions and pointed out that their method did not account for noise in the responses used for prediction, which they leave as an open problem. \citet{kutluozen2026} used the variance on the first few principal components of a fully observed system to define a bound for its reverse score. Our results discuss what remains unseen when only a subset of noisy sites are observed. Decoding studies also predict model variables from brain activity, but for a different purpose, the reconstruction of language \citep{tang2023,defossez2023}.
	
	\textbf{A small number of neurons.} \citet{kang2025} showed that when only a small number of neurons are recorded, the similarity measure between representations is underestimated, and they also corrected for this \citep[see also][]{chun2025}. \citet{canatar2023} linked the forward score to the model's spectrum and the number of stimuli. \citet{gao2015} explored when a small number of neuron samples are sufficient, and \citet{stringer2019} found that the variance of the population response decays very slowly across all dimensions, leaving a large portion of the population beyond the reach of small samples. We investigate the regression from a small number of noisy neurons to the model, and what this regression can tell us about absence.
	
	\textbf{Tools.} Our formula for the reverse score is a special case of deterministic equivalents. These results describe ridge regression with random features \citep{rahimi2007,bach2024,cheng2024,misiakiewicz2024,defilippis2024} and learning curves for kernel methods \citep{canatar2021,simon2023}. Correction for noise in predictor variables can be traced back to \citet{spearman1904} and errors-in-variables regression \citep{fuller1987}. The idea that data can only constrain a quantity from one side is the subject of partial identification \citep{manski2003}. Our impossibility result belongs to the category first proposed by \citet{bahadur1956}.
	
	\textbf{Language Models and the Brain.} Language models can predict brain activity during listening and reading \citep{schrimpf2021,goldstein2022,caucheteux2022,antonello2023}. Recent work uses the sparse features of the model to describe this alignment \citep{lepori2026} and finds that only a small portion of the model carries this alignment \citep{paris2026}. The reverse predictivity of language models has not yet been reported.
	
	\section{What a Recording Can Show}
	\label{sec:theory}
	
	\subsection{Setting}

	Stimuli $s$ are drawn from a fixed distribution. We consider functions of stimuli with zero mean and finite variance, under the inner product $\langle g,h\rangle=\E[g(s)\,h(s)]$ and the norm $\|g\|=\langle g,g\rangle^{1/2}$. A unit of the model is such a function $f$. It can be a neuron in a layer, a principal component of that layer, or a feature of a sparse autoencoder. A brain region is described by a population, which is a distribution $Q$ of site signals. The signal $r$ of a site is its average response to each stimulus, which is also a function of the stimulus. We assume $\E_Q\|r\|^2<\infty$. We use $\Hsp$ to denote the closed linear span of this population, which is the smallest closed subspace containing $r$ with probability 1, and $\Pi_{\Hsp}$ to denote the orthogonal projection onto it.
	
	\begin{definition}[Expressible share]
		\label{def:share}
		The share of a unit $f$ that the population expresses is $\rho(f)=\|\Pi_{\Hsp}f\|^2/\|f\|^2$. The population holds $f$ when $\rho(f)=1$ and lacks a part of it when $\rho(f)<1$.
	\end{definition}
	
	The definition is linear because the scores in use are linear.
	
	\begin{definition}[Recording and scores]
		\label{def:recording}
		A recording consists of $n$ sites whose signals $r_1,\dots,r_n$ are drawn independently from $Q$. At each presentation of a stimulus $s$, site $j$ returns $y_j=r_j(s)+\varepsilon_j$. The noise $\varepsilon_j$ has mean zero and variance $\sigma_j^2$ and is independent of the stimulus, of the other sites and of other presentations. The reverse score $R_n(f)$ is the largest share of the variance of $f$ that a linear readout of $y_1,\dots,y_n$ explains. The noise-free reverse score $\Rsig_n(f)$ is the same quantity for the signals $r_1,\dots,r_n$. The forward score $F(r)$ of a site is the share of its signal that lies in the span of the units of the model.
	\end{definition}
	
	Let $A$ be the matrix with entries $\langle r_j,r_l\rangle$, let $a$ be the vector with entries $\langle f,r_j\rangle$, and let $\Sigma$ be the diagonal matrix of the $\sigma_j^2$. Then
	\begin{equation}
		\label{eq:scores}
		R_n(f)=\frac{a^\top(A+\Sigma)^{-1}a}{\|f\|^2},\qquad \Rsig_n(f)=\frac{a^\top A^{+}a}{\|f\|^2},
	\end{equation}
	where $A^{+}$ is the pseudoinverse.
	
	\begin{proposition}[Four losses]
		\label{prop:losses}
		For every population, unit and recording,
		\begin{equation}
			\label{eq:chain}
			\E\hat R\;\le\;R_n(f)\;\le\;\Rsig_n(f)\;\le\;\rho(f)\;\le\;1,
		\end{equation}
		where $\hat R$ is the share of the variance of $f$ that a readout fitted on finitely many stimuli explains on new stimuli. Each inequality can be strict. In the forward direction, the largest share of the variance of $y_j$ that the units explain, divided by the reliability $\|r_j\|^2/(\|r_j\|^2+\sigma_j^2)$ of the site, equals $F(r_j)$, and the mean of $F(r_j)$ over the recorded sites has expectation $\E_Q F(r)$.
	\end{proposition}
	
	The four gaps in \cref{eq:chain} correspond to the four sources in \cref{fig:overview}b, originating from stimuli, noise, sites, and the brain, respectively. The forward direction loses nothing to noise or sites. Dividing by the reliability precisely removes the noise, and sampling sites does not alter the mean. See Appendix~\ref{app:proofs} for the proof.
	
	\subsection{Absence Cannot Be Shown}
	
	\begin{theorem}[Absence is not identified]
		\label{thm:identified}
		Fix a unit $f$ and a number of sites $n$.
		
		(i) Every population whose support contains the recorded signals satisfies $\rho(f)\ge\Rsig_n(f)$, and $\Rsig_n(f)$ is determined by the joint distribution of $f$ and two presentations of the recording.
		
		(ii) For every population $Q$ and every $\delta>0$ there is a population $Q'$ that holds $f$ and is such that the data of a recording of $n$ sites, with any number of stimuli and presentations, have distributions under $Q$ and under $Q'$ whose total variation distance is at most $\delta$.
	\end{theorem}
	
	\begin{corollary}[No test of absence]
		\label{cor:test}
		Consider a test that uses the data of a recording of $n$ sites and rejects the hypothesis $\rho(f)=1$ with probability at most $\alpha$ whenever the hypothesis is true. The test rejects with probability at most $\alpha$ under every population, including those that lack $f$.
	\end{corollary}
	
	Therefore, a test that rarely declares absence when the brain has $f$ rarely makes the same determination when the brain lacks $f$. The population $Q'$ incorporates a rare class of sites into $Q$, which carry the portion of $f$ missing in $Q$. A recording of $n$ sites misses this class of sites with a probability of at least $1-\delta$, and it is then a recording from $Q$. Therefore, the data can only confine $\rho(f)$ to a range of possible values, $\Rsig_n(f)\le\rho(f)\le1$, and Appendix~\ref{app:proofs} shows that every value within this range can be taken. Reverse scores can support claims of presence. A claim of absence relies on an assumption that the data cannot test, such as the assumption that everything the brain possesses is expressed with a variance higher than a certain level. The next result shows which level a single recording imposes.
	
	\subsection{What the Reverse Score Measures}
	
	Let $\lambda_1\ge\lambda_2\ge\dots$ be the eigenvalues of the covariance operator $\E_Q[r\otimes r]$ and $\varphi_1,\varphi_2,\dots$ its eigenfunctions. These are the principal components of the population, and $\lambda_i$ is the variance that a site carries on component $i$ on average. Let $\beta_i=\langle f,\varphi_i\rangle^2/\|\Pi_{\Hsp}f\|^2$ be the part of the expressed share of $f$ that lies on component $i$.
	
	\begin{proposition}[Resolution limit]
		\label{prop:resolution}
		Let all sites have noise variance $\sigma^2>0$.
		
		(a) For every $n$, $\;\E R_n(f)\le\rho(f)\sum_i\beta_i\,\lambda_i/(\lambda_i+\sigma^2/n)$.
		
		(b) Let the site signals be Gaussian, $r=\sum_i\sqrt{\lambda_i}\,\xi_i\varphi_i$ with independent standard normal $\xi_i$, and let $n$ and the number of dimensions of the population grow in proportion under the conditions of Appendix~\ref{app:proofs}. Then, almost surely,
		\begin{equation}
			\label{eq:resolved}
			R_n(f)=\big(1+o(1)\big)\,\rho(f)\sum_i\beta_i\,\frac{\lambda_i}{\lambda_i+\kappa_n},
		\end{equation}
		where the resolution limit $\kappa_n$ is the unique positive solution of
		\begin{equation}
			\label{eq:kappa}
			\sum_i\frac{\lambda_i}{\lambda_i+\kappa}+\frac{\sigma^2}{\kappa}=n .
		\end{equation}
	\end{proposition}
	
	Part (b) applies a known result about the sample covariance matrix to the sites of a single recording \citep{bach2024,ledoit2011}. The important point here is its meaning. Sites act as training samples, and the noise at those sites acts as a ridge penalty, so increasing the number of sites improves the reverse score, just as increasing the number of samples improves the learner \citep{canatar2021,simon2023}. A component enters \cref{eq:resolved} with a weight of $\lambda_i/(\lambda_i+\kappa_n)$. Components with variances much higher than the resolution limit are recovered, while those below the resolution limit are lost. The resolution limit is at least $\sigma^2/n$, decreasing as sites are added and increasing with increasing noise. \cref{eq:kappa} also shows that, whatever the population, there are at most $2n$ components higher than the resolution limit. Therefore, the reverse score is a joint property of the model, the brain, and the recording, and it is equal to the expressible share only when the resolution limit is zero.
	
	\begin{corollary}[A ceiling under an assumption]
		\label{cor:ceiling}
		In the setting of part (b), let every component that carries a part of $f$ have a variance of at least $v$, so that $\lambda_i\ge v$ whenever $\beta_i>0$. Then, almost surely,
		\begin{equation}
			\label{eq:ceiling}
			\rho(f)\le\big(1+o(1)\big)\Big(1+\frac{\kappa_n}{v}\Big)R_n(f),
		\end{equation}
		and the resolution limit satisfies $\kappa_n\le(\sum_i\lambda_i+\sigma^2)/n$.
	\end{corollary}
	
	The sum $\sum_i\lambda_i+\sigma^2$ represents the variance of a site response, a quantity visible in the recording. Once this assumption is made, low scores do indeed constrain the expressible share from above. In this sense, the statement that the brain lacks a certain unit is actually a statement about $v$, the weakest expression that still counts as holding the unit.
	
	Consider a brain that possesses the entire model with the model's own variances, so that the site signal is a random combination of model units. Each unit then has $\rho(f)=1$, each site has $F(r)=1$, and \cref{eq:resolved} becomes $R_n(f)=\sum_i\beta_i\mu_i/(\mu_i+\kappa_n)$, where $\mu_i$ is the variance of each principal component in this layer. For such a brain, six results can be derived. First, units that lie on components below the resolution limit have scores close to zero. Second, the scores depend only on the units and the resolution limit, so two recordings of equal size agree on which units score low. Third, as long as the number of recorded sites is less than $k/2$, the unit falling on the $k$-th component will always score below half. Fourth, low-scoring units can be predicted by other units in the model, and these units are numerous and noise-free. Fifth, one recording of this brain region predicts the sites of another at the level $\sum_i\mu_i^2/(\mu_i+\kappa_n)\big/\sum_i\mu_i$ times their reliability, which is less than one and is the same in both directions for two recordings of the same size. Sixth, the average score of a layer decreases as its spectrum flattens. These six phenomena have all been observed in actual recordings, and several of them have been taken to show that limited sampling does not explain the low-scoring units  \citep{muzellec2026}. Appendix~\ref{app:proofs} derives them, and \cref{sec:simulation} tests them in simulation.
	
	\subsection{Reading Reverse Scores}
	\label{sec:reading}
	
	\textbf{A score without noise.} Suppose every stimulus is presented twice, with responses $y^{(1)}$ and $y^{(2)}$. For a readout $w$, let $\hat f^{(1)}=w^\top y^{(1)}$ and $\hat f^{(2)}=w^\top y^{(2)}$ be its two predictions and $\hat f$ their mean.
	
	\begin{proposition}[A floor without noise]
		\label{prop:noise}
		Let $w$ be fitted on other stimuli, and let $\widehat{\Cov}$ denote sample covariances over new stimuli. Then
		\begin{equation}
			\label{eq:corrected}
			\E\big[2\,\widehat{\Cov}(\hat f,f)-\widehat{\Cov}(\hat f^{(1)},\hat f^{(2)})\big]=\|f\|^2-\|f-w^\top r\|^2 ,
		\end{equation}
		which is at most $\|f\|^2\,\Rsig_n(f)$, with equality for $w=A^{+}a$.
	\end{proposition}
	
	We divide the left side of \cref{eq:corrected} by the sample variance of $f$ and call the result the noise-corrected score. For any readout, it estimates the lower bound in \cref{thm:identified} from below. When fitting readouts, we replace the covariance among the predictors with the covariance between two presentations, so that as the number of stimuli increases, the readouts will approach $A^{+}a$ (Appendix~\ref{app:details}). Previous correction methods divided the square of the correlation coefficient of the standard readout by the reliability of its prediction \citep{nayebi2023,muzellec2026}. This approach removes noise when evaluating readouts, but the readouts themselves are still fitted to noisy sites.
	
	\textbf{Two reference values.} One reference value answers only one question. If the brain possessed the entire model, how low would the reverse score of a unit be? We simulate a recording of such a brain and match it to the real recording. The simulated $j$-th site is
	$
	\tilde y_j=\sqrt{q_j}\,\tilde r_j+\sqrt{1-q_j}\,\tilde\varepsilon_j .
	$
	The signal $\tilde r_j$ is a random combination of units in this layer, passed through the response function of the recording method, and then scaled to unit variance. The noise $\tilde\varepsilon_j$ is the part that the forward model cannot predict for the real site $j$, and the order of the stimuli is shuffled so that only its statistical properties remain. The share $q_j$ is chosen so that the model predicts the simulated site as well as it predicts the real site. The reference value $C(f)$ is the reverse score of $f$ given by the simulated recording, averaged over multiple samplings.
	
	There is shared tuning between sites in the real recording, which could be due to two reasons, and the data cannot distinguish between them (\cref{thm:identified}). Either the measurement mixes adjacent sites together, or the brain itself has very few dimensions. The independent site reference value $\Cind$ assigns each simulated site its own random direction, which is equivalent to attributing all sharing to the brain. The shared tuning reference value $\Csh$ allows simulated sites to have the tuning overlap exhibited by real sites, estimated using forward weights fitted to two halves of the stimuli, which is equivalent to attributing all sharing to the measurement. Therefore, a reverse score can be split into two parts,
	\begin{equation}
		\label{eq:split}
		1-R(f)=\big[1-C(f)\big]+\big[C(f)-R(f)\big].
	\end{equation}
	The first part represents the shortfall that a brain possessing the entire model would also exhibit. The second part is what remains for a difference between the brain and the model, which could be absence or weaker expression. The first part is computed for a brain that holds every unit of the layer. It does not test whether the real brain does. Units that the brain lacks are part of the simulated brain like all others, so they lower the reference value as they lower the score (\cref{sec:simulation}).
	
\begin{table*}[t]
	\caption{Observations of \citet{muzellec2026} and the same quantities in four simulated brains analysed with their protocol. The complete brain holds every unit of the model. The brain with other content holds every unit, and an eighth of the signal of its sites is unrelated to the model. The partial brain holds the first 512 of the 2,048 dimensions of the model, with the same share of unrelated signal. The incomplete brain holds 40 of the 2,048 dimensions of the model. Scores are medians in percent of explained variance, and the two rows on dimensionality are rank correlations over models that are compared with one brain. Common and unique units are the fifths of the units with the highest and the lowest reverse scores on further stimuli. The second instance of the model has the same dimensions and other units, and its four rows involve no brain. Published values are from \citet{muzellec2026}, who give reverse scores for two models. Where the study gives no value, the column says that a score is high, which of two scores is higher, or that the two are close. A dash marks a quantity that the study does not report.}
	\label{tab:published}
	\centering
	\small
	\setlength{\tabcolsep}{5pt}
	\begin{tabular}{@{}lccccc@{}}
		\toprule
		& & \multicolumn{4}{c}{Simulated brain} \\
		\cmidrule(l){3-6}
		Observation & Published & Complete & With other content & Partial & Incomplete \\
		\midrule
		Forward score, sites predicted from the model & about 50 & \val{t1.forward.complete} & \val{t1.forward.extra} & \val{t1.forward.partial} & \val{t1.forward.incomplete} \\
		Reverse score, units predicted from the sites & 5.7, 16.9 & \val{t1.reverse.complete} & \val{t1.reverse.extra} & \val{t1.reverse.partial} & \val{t1.reverse.incomplete} \\
		Agreement of unit scores between the two animals & 0.87 & \val{t1.agreement.complete} & \val{t1.agreement.extra} & \val{t1.agreement.partial} & \val{t1.agreement.incomplete} \\
		Least predictable 500 units, predicted from 300 sites & 2.9 & \val{t1.leastlast.complete} & \val{t1.leastlast.extra} & \val{t1.leastlast.partial} & \val{t1.leastlast.incomplete} \\
		The same units, predicted from the other units of the model & high & \val{t1.fromunits.complete} & \val{t1.fromunits.extra} & \val{t1.fromunits.partial} & \val{t1.fromunits.incomplete} \\
		Sites of animal 2, predicted from animal 1 & 35.5 & \val{t1.a12.complete} & \val{t1.a12.extra} & \val{t1.a12.partial} & \val{t1.a12.incomplete} \\
		Sites of animal 1, predicted from animal 2 & 43.2 & \val{t1.a21.complete} & \val{t1.a21.extra} & \val{t1.a21.partial} & \val{t1.a21.incomplete} \\
		Reverse score against effective dimensionality & $-$0.53 & \val{t1.fixrev.complete} & \val{t1.fixrev.extra} & \val{t1.fixrev.partial} & \val{t1.fixrev.incomplete} \\
		Forward score against effective dimensionality & 0.22 & \val{t1.fixfwd.complete} & \val{t1.fixfwd.extra} & \val{t1.fixfwd.partial} & \val{t1.fixfwd.incomplete} \\
		\addlinespace
		Forward score, model reduced to \val{t1.match.components} components & higher & \val{t1.matchfwd.complete} & \val{t1.matchfwd.extra} & \val{t1.matchfwd.partial} & \val{t1.matchfwd.incomplete} \\
		Reverse score, model reduced to \val{t1.match.components} components & lower & \val{t1.matchrev.complete} & \val{t1.matchrev.extra} & \val{t1.matchrev.partial} & \val{t1.matchrev.incomplete} \\
		Forward score without the common units & lower & \val{t1.cutfwdcommon.complete} & \val{t1.cutfwdcommon.extra} & \val{t1.cutfwdcommon.partial} & \val{t1.cutfwdcommon.incomplete} \\
		Forward score without the unique units & higher & \val{t1.cutfwdunique.complete} & \val{t1.cutfwdunique.extra} & \val{t1.cutfwdunique.partial} & \val{t1.cutfwdunique.incomplete} \\
		Reverse score without the common units & lower & \val{t1.cutrevcommon.complete} & \val{t1.cutrevcommon.extra} & \val{t1.cutrevcommon.partial} & \val{t1.cutrevcommon.incomplete} \\
		Reverse score without the unique units & higher & \val{t1.cutrevunique.complete} & \val{t1.cutrevunique.extra} & \val{t1.cutrevunique.partial} & \val{t1.cutrevunique.incomplete} \\
		Forward score, each site predicted from its best unit & higher & \val{t1.pairfwd.complete} & \val{t1.pairfwd.extra} & \val{t1.pairfwd.partial} & \val{t1.pairfwd.incomplete} \\
		Reverse score, each unit predicted from its best site & lower & \val{t1.pairrev.complete} & \val{t1.pairrev.extra} & \val{t1.pairrev.partial} & \val{t1.pairrev.incomplete} \\
		Second instance of the model, predicted from the first & close & \multicolumn{4}{c}{\val{t1.inst.allfwd}} \\
		First instance, predicted from the second & close & \multicolumn{4}{c}{\val{t1.inst.allrev}} \\
		\val{t1.inst.seen} noisy units of the second instance, predicted from the first & -- & \multicolumn{4}{c}{\val{t1.inst.seenfwd}} \\
		First instance, predicted from these \val{t1.inst.seen} noisy units & -- & \multicolumn{4}{c}{\val{t1.inst.seenrev}} \\
		\addlinespace
		Random 500 units, predicted from 300 sites & -- & \val{t1.randomlast.complete} & \val{t1.randomlast.extra} & \val{t1.randomlast.partial} & \val{t1.randomlast.incomplete} \\
		Expressible share of the fifth of the units with the lowest scores & -- & \val{t1.truebottom.complete} & \val{t1.truebottom.extra} & \val{t1.truebottom.partial} & \val{t1.truebottom.incomplete} \\
		Expressible share of the unique units that are removed & -- & \val{t1.cuttrueunique.complete} & \val{t1.cuttrueunique.extra} & \val{t1.cuttrueunique.partial} & \val{t1.cuttrueunique.incomplete} \\
		\bottomrule
	\end{tabular}
\end{table*}
	
	\section{Simulated Populations}
	\label{sec:simulation}
	
	\subsection{Design}
	
	\textbf{Formula.} A population has 3,000 dimensions, with variances falling as $i^{-\gamma}$, where $\gamma$ takes values of 0.6, 1, or 1.4. Sites follow a Gaussian distribution with a reliability between 0.5 and 1, and a single recording contains 10 to 1,280 sites. The prediction target either lies on a single component of the population or is a held-out site. In each setting, we calculate $R_n(f)$ precisely using \cref{eq:scores} for ten recordings and compare it with \cref{eq:resolved}.
	
	\textbf{Published protocol.} Next, we analyze four simulated brains following the approach \citet{muzellec2026} used to analyze the inferior temporal cortex. One model has 2,048 dimensions, variances proportional to $1/i$, and 4,096 units, each a random mixture of eight dimensions. Two animals were recorded on 1,320 stimuli, with 169 and 107 sites, respectively, and the two halves of each site's repetitions correlated at 0.75. Scores were calculated following the code published in that study, using ridge regression with a fixed penalty of 0.1 and 10-fold cross-validation, and normalized with split-half reliability. Sites in the complete brain read all dimensions of the model. Sites in the brain containing other content also read all dimensions, and one-eighth of the signal was unrelated to the model, a proportion that brought the forward scores close to the published values. Sites in the partial brain read the first quarter of the dimensions and carry the same share of unrelated signal. Sites in the incomplete brain read the first 40 dimensions. Each brain underwent various analyses as described in the study. These analyses included two scores, consistency between animals, curves showing changes with site count, predictions of units using other units of the model, predictions between animals, reducing the model to \val{t1.match.components} principal components, removing the top or bottom fifth of units by reverse score, and predicting the site using the single best unit of each site and predicting the unit using the single best site of each unit. A second instance of the model, with the same dimensions but different units, replaced the comparison between the two models. Another set of \val{t1.family} models, varying in spectrum and size, provided the relationship between scores and effective dimensionality. They were all compared to the same brain, as in the study, and each was also compared to a brain belonging to it. Each simulation was repeated \val{t1.repeats} times. See Appendix~\ref{app:details} for the complete setup.
	
	\textbf{One brain and multiple models.} This simulation keeps one brain and one recording constant, changing the models. Five groups of units were drawn, each containing 1,024 units, and each group had its own 512 dimensions. A site was a random combination of four groups of units. Groups 1 and 2 entered at full intensity, group 3 at half intensity, and group 4 at quarter intensity. Group 5 did not enter, so the brain had no units for it. The recording had \val{ob.sites} sites and \val{ob.stimuli} stimuli. Twelve models were compared to it (\cref{tab:one}). Four of these models contained 64, 128, 256, or 512 units from group 1, respectively, located on the dimensions with the largest variance. These were called the strongest units in that group. The remaining models contained group 1, group 1 plus another group, group 1 plus 256 or 512 units from group 5, and all five groups. Each model was scored using the published protocol and also using a procedure from \cref{sec:recordings}, which provides a reverse score $R$ and two reference values. Then, only one of the site count and the noise at the sites was changed each time.
	
	\begin{figure}[t]
		\centering
		\datarange{fig2a.csv}{rangeA}{xmin=10, xmax=1280, ymin=0, ymax=1}
		\datarange{fig2b.csv}{rangeB}{xmin=20, xmax=300, ymin=0, ymax=10}
		\begin{tikzpicture}
			\begin{axis}[paperaxis, rangeA, name=A, width=1.15in, xmode=log, ymin=0, ymax=1.08,
				xlabel={sites $n$}, ylabel={reverse score}, title={(a)}]
				\ifdata{fig2a.csv}{
					\addplot[black, only marks, mark=*, mark size=1.2pt] table[x=n, y=sim1, col sep=comma]{figdata/fig2a.csv};
					\addplot[black, only marks, mark=square*, mark size=1.1pt] table[x=n, y=sim16, col sep=comma]{figdata/fig2a.csv};
					\addplot[black, only marks, mark=triangle*, mark size=1.4pt] table[x=n, y=sim256, col sep=comma]{figdata/fig2a.csv};
					\addplot[black, line width=0.5pt] table[x=n, y=pred1, col sep=comma]{figdata/fig2a.csv};
					\addplot[black, line width=0.5pt, forget plot] table[x=n, y=pred16, col sep=comma]{figdata/fig2a.csv};
					\addplot[black, line width=0.5pt, forget plot] table[x=n, y=pred256, col sep=comma]{figdata/fig2a.csv};
				}{\nodata}
			\end{axis}
			\begin{axis}[paperaxis, rangeB, at={(A.south east)}, anchor=south west, xshift=0.47in, width=1.15in,
				xlabel={sites $n$}, ylabel={reverse score (\%)}, title={(b)}, ymin=0]
				\ifdata{fig2b.csv}{
					\addplot[black, line width=0.7pt, mark=square, mark size=1.1pt] table[x=n, y=completerandom, col sep=comma]{figdata/fig2b.csv};
					\addplot[black, line width=0.7pt, mark=*, mark size=1.1pt] table[x=n, y=completeleast, col sep=comma]{figdata/fig2b.csv};
					\addplot[black, densely dotted, line width=0.7pt, mark=square, mark size=1.1pt, mark options={solid}] table[x=n, y=extrarandom, col sep=comma]{figdata/fig2b.csv};
					\addplot[black, densely dotted, line width=0.7pt, mark=*, mark size=1.1pt, mark options={solid}] table[x=n, y=extraleast, col sep=comma]{figdata/fig2b.csv};
					\addplot[gray, line width=0.7pt, mark=square, mark size=1.1pt] table[x=n, y=partialrandom, col sep=comma]{figdata/fig2b.csv};
					\addplot[gray, line width=0.7pt, mark=*, mark size=1.1pt] table[x=n, y=partialleast, col sep=comma]{figdata/fig2b.csv};
					\addplot[gray, dashed, line width=0.7pt, mark=square, mark size=1.1pt, mark options={solid}] table[x=n, y=incompleterandom, col sep=comma]{figdata/fig2b.csv};
					\addplot[gray, dashed, line width=0.7pt, mark=*, mark size=1.1pt, mark options={solid}] table[x=n, y=incompleteleast, col sep=comma]{figdata/fig2b.csv};
				}{\nodata}
			\end{axis}
		\end{tikzpicture}
		\caption{Simulated populations. (a) Exact reverse scores (marks) and \cref{eq:resolved} (lines) for targets on component 1 (circles), 16 (squares) and 256 (triangles) of a population with variances $1/i$ and reliability 0.7. (b) Reverse score against the number of sites under the published protocol, for 500 random units (squares) and for the 500 least predictable units (circles), in the complete brain (black, solid), in the brain with other content (black, dotted), in the partial brain (gray, solid) and in the incomplete brain (gray, dashed).}
		\label{fig:simulation}
	\end{figure}
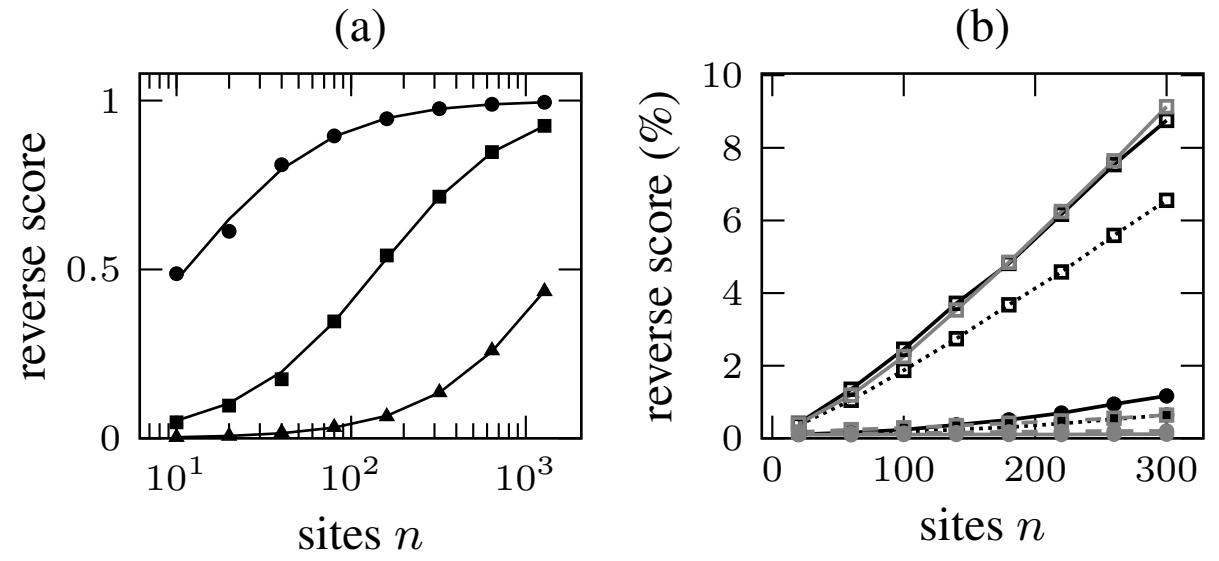

	\begin{table*}[t]
		\caption{One simulated brain and twelve models. The sites combine the units of the first and the second group at full strength, of the third group at half and of the fourth group at a quarter of that strength. The brain lacks the fifth group. The word in brackets is the strength at which the brain expresses the added group. $\bar\rho$ is the mean expressible share of the units of a model, and the fourth column is the effective dimensionality of the model. Scores of the published protocol are medians in percent of explained variance. The forward $R^2$, $\bar R$, $\bar\Cind$ and $\bar\Csh$ come from the workflow of \cref{sec:recordings}. The agreement is the rank correlation over units between reverse score and reference. The last column is the expressible share of the fifth of the units with the lowest reverse scores under the published protocol. Means over \val{ob.repeats} repetitions.}
		\label{tab:one}
		\centering
		\small
		\setlength{\tabcolsep}{3.3pt}
		\begin{tabular}{@{}lcccccccccccc@{}}
			\toprule
			& & & Effective & \multicolumn{2}{c}{Published protocol} & Forward & & \multicolumn{2}{c}{Reference} & \multicolumn{2}{c}{Agreement} & $\bar\rho$ of \\
			\cmidrule(lr){5-6}\cmidrule(lr){9-10}\cmidrule(lr){11-12}
			Model & Units & $\bar\rho$ & dim. & Forward & Reverse & $R^2$ & $\bar R$ & $\bar\Cind$ & $\bar\Csh$ & ind. & shared & bottom fifth \\
			\midrule
			First group, 64 strongest units & \val{ob.default.top64.units} & \val{ob.default.top64.true} & \val{ob.default.top64.ed} & \val{ob.default.top64.pubfwd} & \val{ob.default.top64.pubrev} & \val{ob.default.top64.forward} & \val{ob.default.top64.raw} & \val{ob.default.top64.indep} & \val{ob.default.top64.shared} & \val{ob.default.top64.ranki} & \val{ob.default.top64.ranks} & \val{ob.default.top64.pubbottom} \\
			First group, 128 strongest units & \val{ob.default.top128.units} & \val{ob.default.top128.true} & \val{ob.default.top128.ed} & \val{ob.default.top128.pubfwd} & \val{ob.default.top128.pubrev} & \val{ob.default.top128.forward} & \val{ob.default.top128.raw} & \val{ob.default.top128.indep} & \val{ob.default.top128.shared} & \val{ob.default.top128.ranki} & \val{ob.default.top128.ranks} & \val{ob.default.top128.pubbottom} \\
			First group, 256 strongest units & \val{ob.default.top256.units} & \val{ob.default.top256.true} & \val{ob.default.top256.ed} & \val{ob.default.top256.pubfwd} & \val{ob.default.top256.pubrev} & \val{ob.default.top256.forward} & \val{ob.default.top256.raw} & \val{ob.default.top256.indep} & \val{ob.default.top256.shared} & \val{ob.default.top256.ranki} & \val{ob.default.top256.ranks} & \val{ob.default.top256.pubbottom} \\
			First group, 512 strongest units & \val{ob.default.top512.units} & \val{ob.default.top512.true} & \val{ob.default.top512.ed} & \val{ob.default.top512.pubfwd} & \val{ob.default.top512.pubrev} & \val{ob.default.top512.forward} & \val{ob.default.top512.raw} & \val{ob.default.top512.indep} & \val{ob.default.top512.shared} & \val{ob.default.top512.ranki} & \val{ob.default.top512.ranks} & \val{ob.default.top512.pubbottom} \\
			\addlinespace
			First group & \val{ob.default.base.units} & \val{ob.default.base.true} & \val{ob.default.base.ed} & \val{ob.default.base.pubfwd} & \val{ob.default.base.pubrev} & \val{ob.default.base.forward} & \val{ob.default.base.raw} & \val{ob.default.base.indep} & \val{ob.default.base.shared} & \val{ob.default.base.ranki} & \val{ob.default.base.ranks} & \val{ob.default.base.pubbottom} \\
			First and second group (full) & \val{ob.default.gain1.units} & \val{ob.default.gain1.true} & \val{ob.default.gain1.ed} & \val{ob.default.gain1.pubfwd} & \val{ob.default.gain1.pubrev} & \val{ob.default.gain1.forward} & \val{ob.default.gain1.raw} & \val{ob.default.gain1.indep} & \val{ob.default.gain1.shared} & \val{ob.default.gain1.ranki} & \val{ob.default.gain1.ranks} & \val{ob.default.gain1.pubbottom} \\
			First and third group (half) & \val{ob.default.gain0.5.units} & \val{ob.default.gain0.5.true} & \val{ob.default.gain0.5.ed} & \val{ob.default.gain0.5.pubfwd} & \val{ob.default.gain0.5.pubrev} & \val{ob.default.gain0.5.forward} & \val{ob.default.gain0.5.raw} & \val{ob.default.gain0.5.indep} & \val{ob.default.gain0.5.shared} & \val{ob.default.gain0.5.ranki} & \val{ob.default.gain0.5.ranks} & \val{ob.default.gain0.5.pubbottom} \\
			First and fourth group (quarter) & \val{ob.default.gain0.25.units} & \val{ob.default.gain0.25.true} & \val{ob.default.gain0.25.ed} & \val{ob.default.gain0.25.pubfwd} & \val{ob.default.gain0.25.pubrev} & \val{ob.default.gain0.25.forward} & \val{ob.default.gain0.25.raw} & \val{ob.default.gain0.25.indep} & \val{ob.default.gain0.25.shared} & \val{ob.default.gain0.25.ranki} & \val{ob.default.gain0.25.ranks} & \val{ob.default.gain0.25.pubbottom} \\
			\addlinespace
			First group and 256 of the fifth & \val{ob.default.gain0-256.units} & \val{ob.default.gain0-256.true} & \val{ob.default.gain0-256.ed} & \val{ob.default.gain0-256.pubfwd} & \val{ob.default.gain0-256.pubrev} & \val{ob.default.gain0-256.forward} & \val{ob.default.gain0-256.raw} & \val{ob.default.gain0-256.indep} & \val{ob.default.gain0-256.shared} & \val{ob.default.gain0-256.ranki} & \val{ob.default.gain0-256.ranks} & \val{ob.default.gain0-256.pubbottom} \\
			First group and 512 of the fifth & \val{ob.default.gain0-512.units} & \val{ob.default.gain0-512.true} & \val{ob.default.gain0-512.ed} & \val{ob.default.gain0-512.pubfwd} & \val{ob.default.gain0-512.pubrev} & \val{ob.default.gain0-512.forward} & \val{ob.default.gain0-512.raw} & \val{ob.default.gain0-512.indep} & \val{ob.default.gain0-512.shared} & \val{ob.default.gain0-512.ranki} & \val{ob.default.gain0-512.ranks} & \val{ob.default.gain0-512.pubbottom} \\
			First and fifth group (absent) & \val{ob.default.gain0.units} & \val{ob.default.gain0.true} & \val{ob.default.gain0.ed} & \val{ob.default.gain0.pubfwd} & \val{ob.default.gain0.pubrev} & \val{ob.default.gain0.forward} & \val{ob.default.gain0.raw} & \val{ob.default.gain0.indep} & \val{ob.default.gain0.shared} & \val{ob.default.gain0.ranki} & \val{ob.default.gain0.ranks} & \val{ob.default.gain0.pubbottom} \\
			\addlinespace
			All five groups & \val{ob.default.all.units} & \val{ob.default.all.true} & \val{ob.default.all.ed} & \val{ob.default.all.pubfwd} & \val{ob.default.all.pubrev} & \val{ob.default.all.forward} & \val{ob.default.all.raw} & \val{ob.default.all.indep} & \val{ob.default.all.shared} & \val{ob.default.all.ranki} & \val{ob.default.all.ranks} & \val{ob.default.all.pubbottom} \\
			\bottomrule
		\end{tabular}
	\end{table*}
	
	\begin{figure}[t]
		\centering
		\datarange{figone_sites.csv}{rangeO}{xmin=42, xmax=676, ymin=0, ymax=0.4}
		\datarange{figone_noise.csv}{rangeP}{xmin=0.5, xmax=0.95, ymin=0, ymax=0.4}
		\newcommand{\groupcurves}[1]{%
			\addplot[black, line width=0.7pt, mark=*, mark size=1.1pt] table[x=x, y=baseraw, col sep=comma]{figdata/#1};
			\addplot[black, line width=0.7pt, mark=square, mark size=1.5pt] table[x=x, y=gain1raw, col sep=comma]{figdata/#1};
			\addplot[gray, line width=0.7pt, mark=triangle*, mark size=1.5pt] table[x=x, y=gain05raw, col sep=comma]{figdata/#1};
			\addplot[gray, line width=0.7pt, mark=diamond*, mark size=1.5pt] table[x=x, y=gain025raw, col sep=comma]{figdata/#1};
			\addplot[black, densely dotted, line width=0.7pt, mark=x, mark size=1.8pt, mark options={solid}] table[x=x, y=gain0raw, col sep=comma]{figdata/#1};
			\addplot[black, dashed, line width=0.6pt] table[x=x, y=baseindep, col sep=comma]{figdata/#1};}
		\begin{tikzpicture}
			\begin{axis}[paperaxis, rangeO, name=A, width=1.15in, xmode=log, log basis x=2,
				xtick={42,169,676}, xticklabels={42,169,676},
				xlabel={sites $n$}, ylabel={reverse score}, title={(a)}]
				\ifdata{figone_sites.csv}{\groupcurves{figone_sites.csv}}{\nodata}
			\end{axis}
			\begin{axis}[paperaxis, rangeP, at={(A.south east)}, anchor=south west, xshift=0.47in, width=1.15in,
				xtick={0.5,0.75,0.95},
				xlabel={correlation of the halves}, ylabel={reverse score}, title={(b)}]
				\ifdata{figone_noise.csv}{\groupcurves{figone_noise.csv}}{\nodata}
			\end{axis}
		\end{tikzpicture}
		\caption{One brain and several models. Mean reverse score of each group of units within the model that holds all five groups, against the number of sites (a) and against the correlation between the two halves of the repetitions of a site (b). The brain expresses the first and the second group at full strength (circles and squares), the third at half strength (triangles) and the fourth at a quarter of that strength (diamonds), and it lacks the fifth (crosses). The dashed line is the independent-site reference of the first group.}
		\label{fig:one}
	\end{figure}
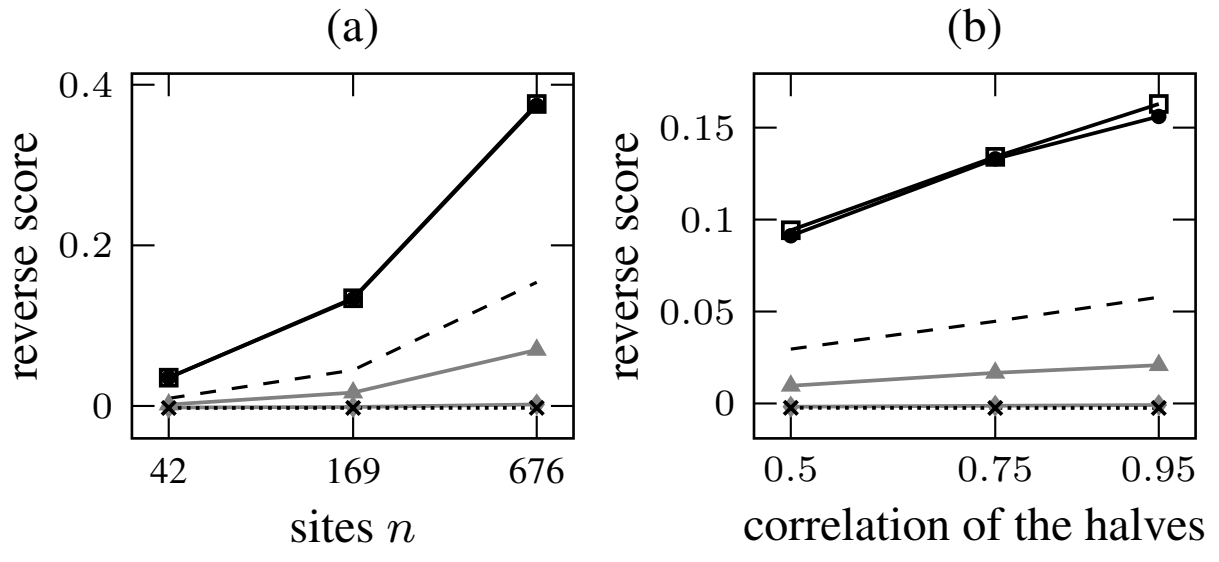
	
	\subsection{Results}

	Over \val{f.all.n} settings, the mean absolute difference between \cref{eq:resolved} and the exact score is \val{f.all.mean}, and the maximum difference is \val{f.all.max} (\cref{fig:simulation}a, \cref{tab:formula}). With 160 sites, targets falling on the 256th component have scores close to zero, while targets falling on the 1st component are almost completely recovered, although the population holds both.
	
	\cref{tab:published} compares the published observations with four simulated brains. The complete brain possesses every unit, and the expressible share of its lowest fifth of units is \val{t1.truebottom.complete}. Its forward score remains \val{t1.forward.complete} percent, and its reverse score is \val{t1.reverse.complete} percent. The agreement of its unit scores between the two animals is \val{t1.agreement.complete}. The units that were most difficult to predict remained at \val{t1.leastlast.complete} percent at 300 sites, while the other units of the model could predict them at \val{t1.fromunits.complete} percent. This value is limited by the stimuli, which are fewer than the dimensions of the simulated model (1,320 against 2,048). The brain containing other content also possessed every unit. The unrelated content caused its forward score to become \val{t1.forward.extra} percent and its reverse score to become \val{t1.reverse.extra} percent. Its agreement between the two animals was \val{t1.agreement.extra}, with the most difficult-to-predict units remaining at \val{t1.leastlast.extra} percent. Therefore, every observation that was taken to rule out limited sampling appeared in a brain that lacked nothing.
	
	Other analyses of the study yielded the same results (\cref{tab:published}). After reducing the model to \val{t1.match.components} components, the forward score for the complete brain remained \val{t1.matchfwd.complete}\%, and the reverse score remained \val{t1.matchrev.complete}\%. The two instances of the model predicted each other equally well, \val{t1.inst.allfwd}\% and \val{t1.inst.allrev}\%, because both were observed completely and without noise. When the second instance was observed as a recording, through \val{t1.inst.seen} of its units and with site noise, the two scores became \val{t1.inst.seenfwd}\% and \val{t1.inst.seenrev}\%. Therefore, the difference between the two directions stems from how one side is observed, and reducing the model to \val{t1.match.components} components does not eliminate this difference. Removing the units with the highest reverse score reduces the complete brain's forward score to \val{t1.cutfwdcommon.complete}\%, while removing the units with the lowest score keeps it at \val{t1.cutfwdunique.complete}\%, even though every removed unit is present in the brain. The reverse score of one unit does not involve other units, so removing the lowest fifth raises the median of the remaining units, here to \val{t1.cutrevunique.complete}\%. When each unit is paired with one site and each site is paired with one unit, the forward score is \val{t1.pairfwd.complete}\%, and the reverse score is \val{t1.pairrev.complete}\%.
	
	The model family provides the relationship between scores and dimensionality. When all \val{t1.fix.models} models are compared to the complete brain (which has every unit of each model), the reverse score decreases with increasing effective dimensionality (\val{t1.fixrev.complete}), while the forward score increases (\val{t1.fixfwd.complete}). These are the signs of the published values. In the seven brains we use as ``that one brain,'' the reverse score for each decreases with increasing dimensionality. The forward relation can be positive or negative. It is positive for the complete brain and a more spectrally flattened complete brain (\val{t1.fixfwd.decay0.7}), and negative for the other five brains (\cref{tab:dimensionality}). When each model has its own brain, both scores decrease with increasing dimensionality (\val{t1.edrev.complete} and \val{t1.edfwd.complete} for complete brains, \cref{fig:family}). Therefore, the reverse score decreasing with model dimensionality does not indicate missing units in the brain.
	
	The partial brain holds a quarter of the dimensions, and the expressible share of its lowest fifth of units is \val{t1.truebottom.partial}. Its forward score is \val{t1.forward.partial} percent and its reverse score is \val{t1.reverse.partial} percent. The agreement of its unit scores between the two animals is \val{t1.agreement.partial}, its most difficult-to-predict units remain at \val{t1.leastlast.partial} percent, and its two animals predict each other at \val{t1.a12.partial} and \val{t1.a21.partial} percent. The reduction of the model, the removal of units and the prediction from single units and sites give the same pattern as in the complete brain. The same observations therefore appear in a brain that lacks nothing and in a brain that lacks most of its lowest-scoring units, in line with \cref{thm:identified}.
	
	The incomplete brain is the one that deviates from the published values. Its unit scores show consistency between the two animals at \val{t1.agreement.incomplete}. The levels of mutual prediction between its two animals are \val{t1.a12.incomplete} and \val{t1.a21.incomplete} percent, respectively, because it is a low-dimensional population, making predictions easy with any part of the sample. The complete brain gives \val{t1.a12.complete} and \val{t1.a21.complete} percent, while the brain containing other content gives \val{t1.a12.extra} and \val{t1.a21.extra} percent, compared to published values of 35.5 and 43.2. The two simulated animals differ only in their number of sites, so the direction with more predicting sites scores higher, and the published values have the opposite order. Random units also set the incomplete brain apart. As sites were added, the score of random units rose to \val{t1.randomlast.complete} percent in the complete brain, \val{t1.randomlast.partial} percent in the partial brain and \val{t1.randomlast.incomplete} percent in the incomplete brain (\cref{fig:simulation}b). The most unpredictable units had a flat curve in all four brains, so the curve did not distinguish between them.
	
	\cref{tab:one} compares twelve models to a brain. The first group is owned by the brain at full strength, with an average reverse score of \val{ob.default.base.raw}, and reference values of \val{ob.default.base.indep} and \val{ob.default.base.shared}. Adding 256, 512, or 1,024 units not owned by the brain lowers the scores to \val{ob.default.gain0-256.raw}, \val{ob.default.gain0-512.raw}, and \val{ob.default.gain0.raw}, respectively, and the median score of the published protocol drops from \val{ob.default.base.pubrev}\% to \val{ob.default.gain0.pubrev}\%. Therefore, reverse predictivity responds to units not owned by the brain. The reference values also respond. The simulated brain has the added units just like any other unit, so the reference values drop to \val{ob.default.gain0.indep} and \val{ob.default.gain0.shared}, while the average score remains above them. Comparing the average does not reveal the added units. Unit-by-unit comparisons reveal them when there are many of these units. Agreement with independent site reference values is \val{ob.default.base.ranki} for the first group, \val{ob.default.gain0-256.ranki} with 256 added units, and \val{ob.default.gain0.ranki} with 1,024 added units.
	
	The units the brain possesses show the same pattern when the brain's expression is weak. A group of units with half the intensity lowers the average score to \val{ob.default.gain0.5.raw} and the published score to \val{ob.default.gain0.5.pubrev}\%. A group of units with quarter intensity reduces them to \val{ob.default.gain0.25.raw} and \val{ob.default.gain0.25.pubrev}\%, close to the values of the group the brain lacks. Agreement decreases to \val{ob.default.gain0.5.ranki} and \val{ob.default.gain0.25.ranki}. The model with the half-intensity group contains more of the brain and predicts the sites better than the first group alone (forward $R^2$ of \val{ob.default.gain0.5.forward} compared to \val{ob.default.base.forward}), but the first group would be chosen based on the reverse score. The forward score of the published protocol does not show this improvement because its fixed penalty term predicts poorly for models with a similar dimension to the number of training stimuli (Appendix~\ref{app:details}). In both models, the lowest-scoring fifth of units are all present in the brain (\cref{tab:one} last column). In the model containing all five groups, the brain possesses \val{ob.default.all.pubbottom} of that fifth. Therefore, comparing the mean is insensitive as a method for detecting missing units, and low scores, low agreement, and falling into the lowest 1/5 are not specific.
	
	\cref{fig:one} follows the five groups in the model containing all five groups. At 42, 169, and 676 sites, the scores for the half-intensity group are \val{ob.sites-42.all.gain0.5.all.raw}, \val{ob.default.all.gain0.5.all.raw}, and \val{ob.sites-676.all.gain0.5.all.raw}, respectively. The group the brain lacks stays at \val{ob.sites-676.all.gain0.all.raw}, while the group with quarter intensity remains at \val{ob.sites-676.all.gain0.25.all.raw} at 676 sites. More sites lower the resolution limit, separating weak expression and absence step by step according to intensity (\cref{prop:resolution}). Less noise improves the scores of the two full-intensity groups, with little change for other groups (\cref{tab:onerec}).
	
	The four models containing only the strongest units from the first group are fully represented by the brain. Their average score is lower than the independent site reference value, with the smallest being \val{ob.default.top64.raw} compared to \val{ob.default.top64.indep}. The reference value assumes that the sites randomly combine the units of the model. This brain's sites combine all the units from the first group, and these models are a subset of those units. In the six models that the brain possesses at full strength, the forward score increases with the effective dimensionality, while the reverse score decreases (rank correlations are \val{ob.default.full.edfwd} and \val{ob.default.full.edrev}), and the order in which the models are ranked by the average reverse score is consistent with the reference value (\val{ob.default.full.revindep}).
	
	\section{Brains Built from Language Models}
	\label{sec:known}
	
	\subsection{Design}
	
	Simulated populations cannot tell whether real models have units that fall below the resolution limit of real recordings. Therefore, we construct brains from language models. The stimuli are 20,000 words in context from WikiText-103 \citep{merity2017}, and the layer used is the middle layer of GPT-2 XL, Pythia 1.4B, Llama 3.2 3B, or Gemma 2 2B. The principal components of the layer on these words define seven brains (\cref{tab:known}). Each recording contains 256 sites, each a random combination of brain content observed on 2,000 words. Each word is presented twice, and noise is added to make the correlation between two presentations of a site 0.5. The recording is the mean of the two presentations. We vary the sampling method of the sites, the number of sites, the site noise, the number of words, and the layer depth, one at a time. In sparse sampling, a site reads eight units. In redundant sampling, eight sites share a common direction. In regional sampling, all sites read the same half of the units.
	
	For each unit, we compute $\rho(f)$, $\Rsig_n(f)$, and $R_n(f)$ exactly according to the construction method (Appendix~\ref{app:details}). The estimates are obtained using the same workflow we used for real-world recordings in \cref{sec:recordings}, including ridge regression with cross-validation, noise-corrected scores, and three simulated recordings for each reference value.
	
	\begin{table*}[t]
		\caption{Brains built from the middle layer of four language models (256 sites, 2,000 words). The second column says which principal components of the layer the brain holds. Means over units, models and recordings. $\bar\rho$ is the true expressible share, $\bar\Rsig$ the exact noise-free reverse score, Corrected the noise-corrected score, and the last column the true expressible share of the fifth of the units with the lowest reverse scores.}
		\label{tab:known}
		\centering
		\small
		\setlength{\tabcolsep}{5pt}
		\begin{tabular}{@{}llccccccc@{}}
			\toprule
			& & \multicolumn{2}{c}{Exact} & \multicolumn{2}{c}{Estimated} & \multicolumn{2}{c}{Reference} & $\bar\rho$ of \\
			\cmidrule(lr){3-4}\cmidrule(lr){5-6}\cmidrule(lr){7-8}
			Brain & Components held & $\bar\rho$ & $\bar\Rsig$ & Corrected & Reverse $\bar R$ & $\bar\Cind$ & $\bar\Csh$ & bottom fifth \\
			\midrule
			Complete & all & \val{t2.complete.true} & \val{t2.complete.signal} & \val{t2.complete.corrected} & \val{t2.complete.raw} & \val{t2.complete.indep} & \val{t2.complete.shared} & \val{t2.complete.bottom} \\
			With other content & all, plus unrelated content & \val{t2.extra.true} & \val{t2.extra.signal} & \val{t2.extra.corrected} & \val{t2.extra.raw} & \val{t2.extra.indep} & \val{t2.extra.shared} & \val{t2.extra.bottom} \\
			Top only & the first 64 & \val{t2.top.true} & \val{t2.top.signal} & \val{t2.top.corrected} & \val{t2.top.raw} & \val{t2.top.indep} & \val{t2.top.shared} & \val{t2.top.bottom} \\
			No top & all but the first 64 & \val{t2.notop.true} & \val{t2.notop.signal} & \val{t2.notop.corrected} & \val{t2.notop.raw} & \val{t2.notop.indep} & \val{t2.notop.shared} & \val{t2.notop.bottom} \\
			Half & a random half & \val{t2.half.true} & \val{t2.half.signal} & \val{t2.half.corrected} & \val{t2.half.raw} & \val{t2.half.indep} & \val{t2.half.shared} & \val{t2.half.bottom} \\
			Flatter & all, with variances $\mu_i^{0.5}$ & \val{t2.flatter.true} & \val{t2.flatter.signal} & \val{t2.flatter.corrected} & \val{t2.flatter.raw} & \val{t2.flatter.indep} & \val{t2.flatter.shared} & \val{t2.flatter.bottom} \\
			Steeper & all, with variances $\mu_i^{1.5}$ & \val{t2.steeper.true} & \val{t2.steeper.signal} & \val{t2.steeper.corrected} & \val{t2.steeper.raw} & \val{t2.steeper.indep} & \val{t2.steeper.shared} & \val{t2.steeper.bottom} \\
			\bottomrule
		\end{tabular}
	\end{table*}
	
	\begin{figure*}[t]
		\centering
		\datarange{fig3a.csv}{rangeC}{ymin=0, ymax=1}
		\datarange{fig3b_complete.csv}{rangeD}{xmin=0, xmax=1, ymin=0, ymax=1}
		\datarange{fig3c.csv}{rangeE}{ymin=0, ymax=1}
		\datarange{fig3d_complete.csv}{rangeF}{xmin=1, xmax=3000, ymin=0, ymax=1}
		\begin{tikzpicture}
			\begin{axis}[paperaxis, rangeC, name=A, ybar stacked, bar width=6pt, xmin=0.3, xmax=7.7, ymin=0, ymax=1,
				xtick={1,2,3,4,5,6,7}, xticklabels={complete, other, top, no top, half, flatter, steeper},
				x tick label style={rotate=40, anchor=north east, font=\scriptsize, inner sep=1pt},
				ylabel={share of unit variance}, title={(a)}]
				\ifdata{fig3a.csv}{
					\addplot[fill=black, draw=black, line width=0.3pt] table[x=x, y=reverse, col sep=comma]{figdata/fig3a.csv};
					\addplot[fill=gray!80, draw=black, line width=0.3pt] table[x=x, y=stimuli, col sep=comma]{figdata/fig3a.csv};
					\addplot[fill=gray!50, draw=black, line width=0.3pt] table[x=x, y=noise, col sep=comma]{figdata/fig3a.csv};
					\addplot[fill=gray!20, draw=black, line width=0.3pt] table[x=x, y=sites, col sep=comma]{figdata/fig3a.csv};
					\addplot[fill=white, draw=black, line width=0.3pt] table[x=x, y=absent, col sep=comma]{figdata/fig3a.csv};
				}{\nodata}
			\end{axis}
			\begin{axis}[paperaxis, rangeD, name=B, at={(A.south east)}, anchor=south west, xshift=0.455in,
				xmin=-0.05, xmax=1.02, ymin=-0.05, ymax=1.02,
				xlabel={reference $\Cind(f)$}, ylabel={reverse score $R(f)$}, title={(b)}]
				\addplot[black, line width=0.4pt, domain=0:1, samples=2] {x};
				\ifdata{fig3b_notop.csv}{
					\addplot[gray, only marks, mark=o, mark size=0.9pt] table[x=independent, y=raw, col sep=comma]{figdata/fig3b_notop.csv};
				}{}
				\ifdata{fig3b_complete.csv}{
					\addplot[black, only marks, mark=*, mark size=0.7pt] table[x=independent, y=raw, col sep=comma]{figdata/fig3b_complete.csv};
				}{\nodata}
			\end{axis}
			\begin{axis}[paperaxis, rangeE, name=C, at={(B.south east)}, anchor=south west, xshift=0.455in,
				ybar, bar width=3.6pt, xmin=0.4, xmax=4.6, ymin=0, ymax=1,
				xtick={1,2,3,4}, xticklabels={independent, sparse, redundant, regional},
				x tick label style={rotate=40, anchor=north east, font=\scriptsize, inner sep=1pt},
				ylabel={score}, title={(c)}]
				\ifdata{fig3c.csv}{
					\addplot[fill=black, draw=black, line width=0.3pt] table[x=x, y=raw, col sep=comma]{figdata/fig3c.csv};
					\addplot[fill=gray!55, draw=black, line width=0.3pt] table[x=x, y=independent, col sep=comma]{figdata/fig3c.csv};
					\addplot[fill=white, draw=black, line width=0.3pt] table[x=x, y=shared, col sep=comma]{figdata/fig3c.csv};
				}{\nodata}
			\end{axis}
			\begin{axis}[paperaxis, rangeF, at={(C.south east)}, anchor=south west, xshift=0.455in,
				xmode=log, ymin=-0.03, ymax=1.03,
				xlabel={component of the layer}, ylabel={score}, title={(d)}]
				\ifdata{fig3d_complete.csv}{
					\addplot[gray, line width=0.5pt] table[x=rank, y=signal, col sep=comma]{figdata/fig3d_complete.csv};
					\addplot[black, line width=0.9pt] table[x=rank, y=raw, col sep=comma]{figdata/fig3d_complete.csv};
					\addplot[black, dashed, line width=0.6pt] table[x=rank, y=independent, col sep=comma]{figdata/fig3d_complete.csv};
					\addplot[black, dotted, line width=0.8pt] table[x=rank, y=shared, col sep=comma]{figdata/fig3d_complete.csv};
				}{\nodata}
			\end{axis}
		\end{tikzpicture}
		\caption{Brains built from language models. (a) The variance of an average unit in each brain, from bottom to top the reverse score (black), the losses to stimuli, to noise and to sites that were not recorded (dark to light gray), and the part that the brain lacks (white). (b) Units of the complete brain (black) and of the brain without the top components (gray circles), reverse score against the independent-site reference. (c) Complete brain under four ways of sampling sites, reverse score (black), independent-site reference (gray) and shared-tuning reference (white). (d) Complete brain by principal component of the layer, exact noise-free score (gray), reverse score (black), independent-site reference (dashed) and shared-tuning reference (dotted).}
		\label{fig:known}
	\end{figure*}
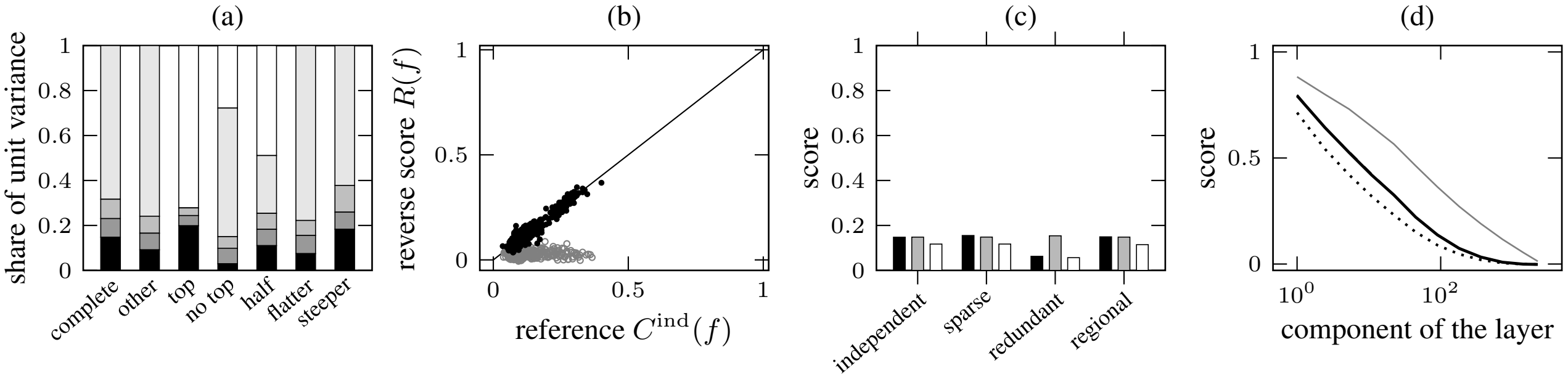
	
	\subsection{Results}
	
	\textbf{The complete brain scores very low.} Each unit of the complete brain has an expressible share of one, and its average reverse score is \val{t2.complete.raw} (\cref{tab:known}). \cref{fig:known}a shows where the rest went. At 256 sites, the noise-free score is \val{t2.complete.signal}, noise reduces it to \val{t2.complete.noisy}, and a finite number of words further reduces it to the measured value. The lowest-scoring fifth of units has an actual share of \val{t2.complete.bottom}. Scores decrease sequentially along the principal components of this layer, reaching zero well before the exact noise-free score reaches zero (\cref{fig:known}d). A large portion of the variance in real language models falls below the resolution limit of recordings of this size.
	
	\textbf{Noise-corrected scores.} The noise-corrected score for the complete brain is \val{t2.complete.corrected}, which lies between the reverse score and the exact noise-free score. It is lower than the noise-free score on all brains, which is exactly what \cref{prop:noise} requires. It recovers some of the loss caused by noise, but not any loss caused by sites.
	
	\textbf{Reference Values.} The independent site reference value for the complete brain is \val{t2.complete.indep}, and the shared tuning reference value is \val{t2.complete.shared}. The units lie along the diagonal of \cref{fig:known}b. Unrelated content in the brain similarly reduces the forward score (from \val{t2.complete.forward} to \val{t2.extra.forward}), the reverse score (\val{t2.extra.raw}), and the reference values (\val{t2.extra.indep} and \val{t2.extra.shared}). When the brain lacks the first 64 principal components of this layer, the reverse score drops to \val{t2.notop.raw}, while the reference values remain at \val{t2.notop.indep} and \val{t2.notop.shared}, and the units also move off the diagonal. At this point, the second part of \cref{eq:split} signals a difference between the brain and the model.
	
	\textbf{What the reference values do not show.} A brain with only the first 64 components holds a share of \val{t2.top.true} of an average unit. Its reverse score is \val{t2.top.raw}, while the reference values are \val{t2.top.indep} and \val{t2.top.shared}. What this brain is missing lies below the resolution limit and cannot be seen regardless of how the recording is interpreted. Two brains with altered variances possess each unit, with scores below and above the reference values (\val{t2.flatter.raw} and \val{t2.steeper.raw}, corresponding to reference values \val{t2.flatter.indep} and \val{t2.steeper.indep}, respectively). The distance from the reference value is evidence of differences in unit expression intensity. It only becomes evidence of absence if an assumption is made about this intensity (\cref{cor:ceiling}).
	
	\textbf{Shared tuning.} When each direction is shared by eight sites, the reverse score of the complete brain is \val{k.complete.sampling-redundant.raw}. The independent site reference value is \val{k.complete.sampling-redundant.indep}, and the shared tuning reference value is \val{k.complete.sampling-redundant.shared} (\cref{fig:known}c). When the sites are independent, the score instead matches the independent site reference value (\val{k.complete.default.raw}, compared to \val{k.complete.default.indep}). The shared tuning reference value then lies below the score (\val{k.complete.default.shared}), and further below with 1,024 sites (\val{k.complete.sites-1024.shared} against \val{k.complete.sites-1024.raw}). Therefore, the score of the complete brain is close to the independent site reference value when its sites are independent and close to the shared tuning reference value when its sites share tuning. See \cref{tab:settings} for all settings.
	
	\section{Recordings}
	\label{sec:recordings}
	
	\subsection{Design}
	
	\textbf{Inferior temporal cortex of monkeys.} We used the recordings from \citet{muzellec2026}. The publicly available version contains responses from one macaque monkey to 1,320 images, which were averaged over time by the authors. The sites were the \val{mk.m2.stored} sites selected by the authors. Similar to that study, we retained sites with a split-half reliability of at least 0.7, leaving \val{t3.monkey.sites} sites. The repetitions of an image were split alternately into two groups, giving two presentations. The models were the \val{mk.count} vision models whose activations the authors released.
	
	\textbf{Human electrocorticography (ECoG).} In the Podcast dataset, nine patients listened to a 30-minute story \citep{zada2025}. One site was represented by one electrode, described by its high gamma power over four 200-millisecond time windows following each word. Electrodes from all patients were analyzed together.
	
	\textbf{Human fMRI.} In the Narratives dataset, participants listened to \val{ds.narratives.stories} stories \citep{nastase2021}. A site is a vertex on the cortical surface after averaging across participants. Participants were split in half, resulting in two presentations.
	
	\textbf{Sites, Models, and Units.} In the human datasets, we retained the fifth of the response columns that other brains predict best, and this criterion does not involve any model. A column is a vertex, or the response of an electrode in one of its four time windows. We analyzed the middle layer of 21 language models ranging from Pythia 70M to Gemma 2 9B, three of which have random weights, and additionally analyzed two depths for four of these models (\cref{tab:models}). A unit is a neuron in a layer. We also scored the principal components of a layer and, for the two Gemma 2 models, the features of sparse autoencoders \citep{lieberum2024}.
	
	\textbf{Scores.} All scores were obtained using kernel ridge regression with nested cross-validation on the held-out stimuli, yielding $R^2$. Cross-validation folds were grouped stories in fMRI, chunks of words in ECoG, and randomly assigned pictures in the monkey data. The fMRI regression used a delay of one to five scan volumes. The ceiling for the forward direction is the score that other brains, or other repetitions, achieve for the same sites. Each reference uses three simulated recordings. See Appendix~\ref{app:details} for details.
	
	\begin{table*}[t]
		\caption{Recordings. Means over models. Samples are images, words and fMRI volumes. The ceiling is the $R^2$ with which other brains, or other repetitions in the monkey data, predict the sites, and the forward score is the $R^2$ with which a model predicts them. $\bar R$ is the mean reverse score over the units of a layer, $\bar\Cind$ and $\bar\Csh$ the two references. The share is $(1-\bar C)/(1-\bar R)$, the part of the shortfall from one that a brain holding the whole model would also show. A share above 100 means that the reference lies below the reverse score. The agreement is the rank correlation over units between reverse score and reference. The last three columns come from one analysis of the recordings with two presentations, and the last of them is the noise-corrected score (Appendix~\ref{app:details}).}
		\label{tab:recordings}
		\centering
		\small
		\setlength{\tabcolsep}{2.5pt}
		\begin{tabular}{@{}lcccccccccccccc@{}}
			\toprule
			& & & & & & \multicolumn{2}{c}{Reference} & \multicolumn{2}{c}{Share (\%)} & \multicolumn{2}{c}{Agreement} & \multicolumn{3}{c}{Two presentations} \\
			\cmidrule(lr){7-8}\cmidrule(lr){9-10}\cmidrule(lr){11-12}\cmidrule(l){13-15}
			Recording & Sites & Samples & Ceiling & Forward & $\bar R$ & $\bar\Cind$ & $\bar\Csh$ & ind. & shared & ind. & shared & Reverse & Normalised & Corrected \\
			\midrule
			Monkey IT & \val{t3.monkey.sites} & \val{t3.monkey.rows} & \val{t3.monkey.ceiling} & \val{t3.monkey.forwardraw} & \val{t3.monkey.reverse} & \val{t3.monkey.indep} & \val{t3.monkey.shared} & \val{t3.monkey.sharei} & \val{t3.monkey.shares} & \val{t3.monkey.ranki} & \val{t3.monkey.ranks} & \val{t3.monkey.raw} & \val{t3.monkey.normalised} & \val{t3.monkey.corrected} \\
			Human ECoG & \val{t3.podcast.sites} & \val{t3.podcast.rows} & \val{t3.podcast.ceiling} & \val{t3.podcast.forwardraw} & \val{t3.podcast.reverse} & \val{t3.podcast.indep} & \val{t3.podcast.shared} & \val{t3.podcast.sharei} & \val{t3.podcast.shares} & \val{t3.podcast.ranki} & \val{t3.podcast.ranks} & -- & -- & -- \\
			Human fMRI & \val{t3.narratives.sites} & \val{t3.narratives.rows} & \val{t3.narratives.ceiling} & \val{t3.narratives.forwardraw} & \val{t3.narratives.reverse} & \val{t3.narratives.indep} & \val{t3.narratives.shared} & \val{t3.narratives.sharei} & \val{t3.narratives.shares} & \val{t3.narratives.ranki} & \val{t3.narratives.ranks} & \val{t3.narratives.raw} & \val{t3.narratives.normalised} & \val{t3.narratives.corrected} \\
			\bottomrule
		\end{tabular}
	\end{table*}
	
	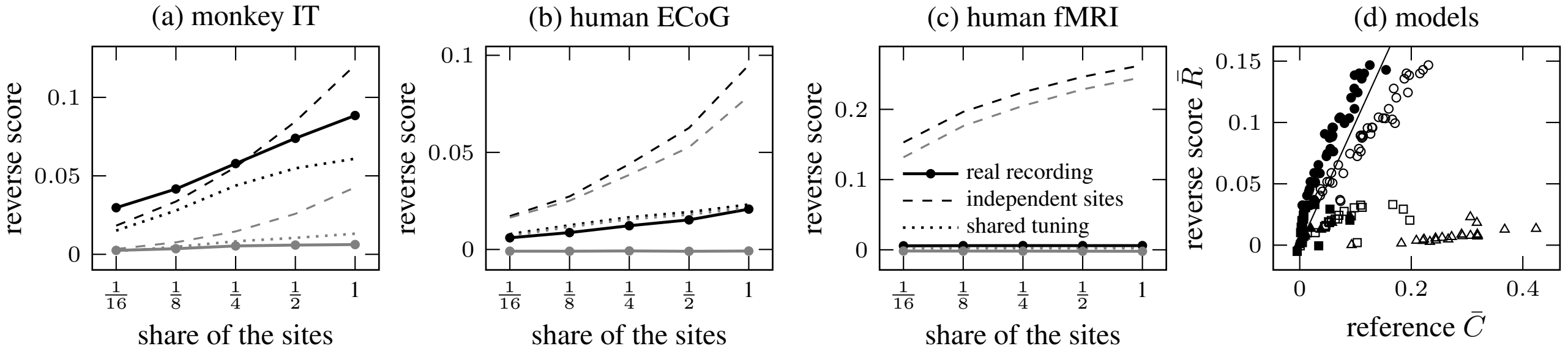
\begin{figure*}[t]
		\centering
		\datarange{fig4_monkey.csv}{rangeG}{xmin=0.0625, xmax=1, ymin=0, ymax=1}
		\datarange{fig4_podcast.csv}{rangeH}{xmin=0.0625, xmax=1, ymin=0, ymax=1}
		\datarange{fig4_narratives.csv}{rangeI}{xmin=0.0625, xmax=1, ymin=0, ymax=1}
		\ifdata{fig4d_monkey.csv}{\pgfplotsset{rangeJ/.style={}}}{\ifdata{fig4d_podcast.csv}{\pgfplotsset{rangeJ/.style={}}}{\datarange{fig4d_narratives.csv}{rangeJ}{xmin=0, xmax=1, ymin=0, ymax=1}}}
		\pgfplotsset{sitecurve/.style={paperaxis, xtick={0.0625,0.125,0.25,0.5,1},
				xticklabels={$\frac{1}{16}$,$\frac{1}{8}$,$\frac{1}{4}$,$\frac{1}{2}$,1}, xlabel={share of the sites}, ylabel={reverse score}}}
		\newcommand{\sitecurves}[1]{%
			\addplot[black, line width=0.9pt, mark=*, mark size=1.0pt] table[x=fraction, y=realall, col sep=comma]{figdata/#1};
			\addplot[black, dashed, line width=0.6pt] table[x=fraction, y=independentall, col sep=comma]{figdata/#1};
			\addplot[black, dotted, line width=0.8pt] table[x=fraction, y=sharedall, col sep=comma]{figdata/#1};
			\addplot[gray, line width=0.9pt, mark=*, mark size=1.0pt] table[x=fraction, y=realbottom, col sep=comma]{figdata/#1};
			\addplot[gray, dashed, line width=0.6pt] table[x=fraction, y=independentbottom, col sep=comma]{figdata/#1};
			\addplot[gray, dotted, line width=0.8pt] table[x=fraction, y=sharedbottom, col sep=comma]{figdata/#1};}
		\begin{tikzpicture}
			\begin{axis}[sitecurve, xmode=log, log basis x=2, rangeG, name=A, title={(a) monkey IT}]
				\ifdata{fig4_monkey.csv}{\sitecurves{fig4_monkey.csv}}{\nodata}
			\end{axis}
			\begin{axis}[sitecurve, xmode=log, log basis x=2, rangeH, name=B, at={(A.south east)}, anchor=south west, xshift=0.455in, title={(b) human ECoG}]
				\ifdata{fig4_podcast.csv}{\sitecurves{fig4_podcast.csv}}{\nodata}
			\end{axis}
			\begin{axis}[sitecurve, xmode=log, log basis x=2, rangeI, name=C, at={(B.south east)}, anchor=south west, xshift=0.455in, title={(c) human fMRI}, legend style={at={(0.04,0.31)}, anchor=west}]
				\ifdata{fig4_narratives.csv}{\sitecurves{fig4_narratives.csv}\legend{real recording, independent sites, shared tuning}}{\nodata}
			\end{axis}
			\begin{axis}[paperaxis, rangeJ, at={(C.south east)}, anchor=south west, xshift=0.455in,
				xlabel={reference $\bar C$}, ylabel={reverse score $\bar R$}, title={(d) models}]
				\draw[black, line width=0.4pt] (axis cs:0,0) -- (axis cs:1,1);
				\ifdata{fig4d_monkey.csv}{
					\addplot[black, only marks, mark=o, mark size=1.3pt] table[x=independent, y=reverse, col sep=comma]{figdata/fig4d_monkey.csv};
					\addplot[black, only marks, mark=*, mark size=1.3pt] table[x=shared, y=reverse, col sep=comma]{figdata/fig4d_monkey.csv};
				}{}
				\ifdata{fig4d_podcast.csv}{
					\addplot[black, only marks, mark=square, mark size=1.2pt] table[x=independent, y=reverse, col sep=comma]{figdata/fig4d_podcast.csv};
					\addplot[black, only marks, mark=square*, mark size=1.2pt] table[x=shared, y=reverse, col sep=comma]{figdata/fig4d_podcast.csv};
				}{}
				\ifdata{fig4d_narratives.csv}{
					\addplot[black, only marks, mark=triangle, mark size=1.6pt] table[x=independent, y=reverse, col sep=comma]{figdata/fig4d_narratives.csv};
					\addplot[black, only marks, mark=triangle*, mark size=1.6pt] table[x=shared, y=reverse, col sep=comma]{figdata/fig4d_narratives.csv};
				}{}
				\ifdata{fig4d_monkey.csv}{}{\ifdata{fig4d_podcast.csv}{}{\ifdata{fig4d_narratives.csv}{}{\nodata}}}
			\end{axis}
		\end{tikzpicture}
		\caption{Recordings. (a) to (c) Reverse score against the share of the sites that is used, averaged over all models for the monkey and over the four models of \cref{sec:known} for the human recordings, for all units (black) and for the fifth of the units with the lowest scores in the full recording (gray). Solid lines with marks are the real recording, dashed lines the independent-site reference and dotted lines the shared-tuning reference. (d) Every model as a point, mean reverse score against the independent-site reference (open marks) and the shared-tuning reference (filled marks), for the monkey recordings (circles), ECoG (squares) and fMRI (triangles).}
		\label{fig:recordings}
	\end{figure*}
	
	\subsection{Results}
	
	\textbf{Forward and Reverse.} \cref{tab:recordings} lists the scores. In monkey, ECoG, and fMRI recordings, the mean $R^2$ for model-predicted sites are \val{t3.monkey.forwardraw}, \val{t3.podcast.forwardraw}, and \val{t3.narratives.forwardraw}, respectively, with corresponding ceilings \val{t3.monkey.ceiling}, \val{t3.podcast.ceiling}, and \val{t3.narratives.ceiling}. The mean $R^2$ for site-predicted units are \val{t3.monkey.reverse}, \val{t3.podcast.reverse}, and \val{t3.narratives.reverse}. Without reference values, these reverse scores would suggest that the brain lacks most of each model.
	
	\textbf{Reference Values.} If the brain held the entire model and the model predicted its sites as well as it predicts the real ones, it would score \val{t3.monkey.indep} (independent site) and \val{t3.monkey.shared} (shared tuning) in monkey recordings, \val{t3.podcast.indep} and \val{t3.podcast.shared} in ECoG, and \val{t3.narratives.indep} and \val{t3.narratives.shared} in fMRI. At least \val{abs.sharelow} percent of the difference between the reverse scores and 1 can be explained by the reference values. This share does not test whether a brain holds the model. In the brains of \cref{sec:known} it lies between \val{t2.notop.sharei} and \val{t2.top.shares} percent, including the brains that lack a part of the model. In \cref{tab:one} the mean score of a model stays above its reference values when the brain lacks half of its units. The two reference values are closest in monkey recordings, where the site is a single electrode in the same brain region, and furthest apart in fMRI, where adjacent vertices share most of the signal. In the three recordings, the average reverse score minus the independent site reference value was \val{t3.monkey.diffi}, \val{t3.podcast.diffi}, and \val{t3.narratives.diffi}, and the average reverse score minus the shared tuning reference value was \val{t3.monkey.diffs}, \val{t3.podcast.diffs}, and \val{t3.narratives.diffs} (\cref{fig:recordings}d, \cref{tab:intervals}). In each recording, the reverse score was lower than the independent site reference value. In the two human recordings, it did not differ significantly from the shared tuning reference value, while in the monkey recording, it fell between the two reference values. Across models, the mean reverse score follows the reference values, with rank correlations of \val{rel.monkey.reverse.indep}, \val{rel.podcast.reverse.indep} and \val{rel.narratives.reverse.indep} with the independent site reference value and \val{rel.monkey.reverse.shared}, \val{rel.podcast.reverse.shared} and \val{rel.narratives.reverse.shared} with the shared tuning reference value (\cref{fig:recordings}d). The reference values depend on how well each model predicts the sites. Among the four models of \cref{tab:one} that add another group to the first, the forward score, the reverse score and the reference values are all lower when the brain expresses the added group weakly or lacks it.
	
	\textbf{Which units scored low.} The reference values also indicate which units scored low. Over all units in one layer, the rank correlations of the reverse scores with independent site reference values are \val{t3.monkey.ranki}, \val{t3.podcast.ranki}, and \val{t3.narratives.ranki}, respectively, and with shared tuning reference values are \val{t3.monkey.ranks}, \val{t3.podcast.ranks}, and \val{t3.narratives.ranks}, respectively. The reference values are calculated without any knowledge of what these sites are tuned to. For comparison, the brains of \cref{sec:known} give \val{t2.complete.ranki} and \val{t2.complete.ranks} when they are complete and \val{t2.notop.ranki} and \val{t2.notop.ranks} when they lack the top components. In \cref{tab:one} the agreement is \val{ob.default.base.ranki} for a model that the brain holds at full strength, \val{ob.default.gain0-256.ranki} when a fifth of its units are absent, \val{ob.default.gain0.5.ranki} when half of its units are expressed at half strength, and \val{ob.default.gain0.ranki} when half are absent. To this extent, the order in which units are ranked by reverse scores is a property of the model and of the size of the recording.
	
	\textbf{More sites.} In monkey and ECoG recordings, the reverse scores increase with the number of sites, and their reference values also increase (\cref{fig:recordings}a and b). The monkey reverse score lies above the shared tuning reference value at every share of the sites, and above the independent site reference value as well when an eighth of the sites or fewer are used. The brain with steeper variances in \cref{sec:known} also scores above both reference values although it holds every unit. In fMRI, the reverse scores and shared tuning reference values remain flat, with only independent site reference values increasing (\cref{fig:recordings}c). In each recording, the lowest fifth of units remained near zero, as expected from \cref{sec:simulation}.
	
	\textbf{Noise.} Both the monkey and fMRI recordings have two presentations. In the monkey recordings, the reverse score was \val{t3.monkey.raw}, the normalized score from previous work was \val{t3.monkey.normalised}, and the noise-corrected score was \val{t3.monkey.corrected}. In fMRI, these three values were \val{t3.narratives.raw}, \val{t3.narratives.normalised}, and \val{t3.narratives.corrected}, respectively. The correction left the difference from 1 largely intact.
	
	\textbf{Other Units and Layers.} Principal components and autoencoder features showed the same results as neurons, and the same for other depths (\cref{tab:families}). Along the principal components of a layer, the reverse score drops to zero along with the shared tuning reference, whereas in human recordings, the reverse score drops to zero much earlier than the independent site reference (\cref{fig:profile}).
	
	\section{Discussion}
	\label{sec:discussion}
	
	\textbf{What should be reported.} A reverse score should be reported with four things. First, the scale of the recording, i.e., the number of sites and their reliability, because the score is a property of the recording, and also a property of the brain. Second, a reference value for a brain that holds the entire model, given for both independent-site and shared-tuning cases. Third, a curve showing the score of all units as a function of the number of sites. The curve is flat for the most unpredictable units, regardless of whether the brain has them. Fourth, the assumption used to interpret low scores as absence, such as the minimum variance $v$ in \cref{cor:ceiling}, because \cref{thm:identified} shows that some assumption is essential.
	
	\textbf{Scope of application.} Our results are for linear readouts, which are the readouts currently in use. The formula for the resolution limit assumes that the sites are randomly selected from a brain region and that their noise is independent across sites. The reference values keep the noise of the real sites with its correlations. If all sites are located in one part of the brain region, the rest of the brain region is not recorded. In a constructed brain, this did not change the score (\cref{fig:known}c). Reference values use a forward model to set the noise and signal at the simulated sites, so their quality depends on the quality of the forward model. They also assume that a brain holding the entire model expresses its units with the variances that they have in the model. Two limits follow (\cref{tab:one}). Units that the brain lacks lower the mean reference value as they lower the mean score, so the two means can agree for a model of which the brain lacks a part. A model that is a selected part of what the sites combine receives reference values above its score although the brain holds all of it. None of our results shows that the brain holds the units with low scores. They show that a recording leaves this open.
	
	\textbf{One-to-one mapping and behavior.} The original study also maps each unit to its single best site and finds the same gap between the two directions \citep{muzellec2026}. A single site is a linear readout, so its score is a floor on the expressible share as well (Proposition~\ref{prop:losses}), and the complete brain of \cref{tab:published} shows the same gap. \cref{thm:identified} is a statement about the data of a recording, so it applies to every analysis of them. \citet{muzellec2026} further report that units with high reverse scores agree better with the behavior of primates than units with low reverse scores. We did not analyze behavior. In a brain that holds the entire model, the units with high reverse scores are those that lie on the components of the layer with the largest variance (Proposition~\ref{prop:resolution}), so a comparison of the two groups of units is also a comparison of components with large and small variance.
	
	\textbf{Language Models.} The question that drives research on reverse predictivity is a good one. For language models, it asks what the model has learned that humans have not. With the current scale of recordings, this question can be answered for the top-ranking components in the model, and the rest remains unanswered. To answer the rest, recordings with more sites are needed, or a causal test—that is, teaching a human what a unit encodes.
	
	\section{Conclusion}
	
	Model units that brain activity cannot predict may not exist in the brain, or they may not exist in the recordings, and the recorded data cannot make a judgment between the two. The reverse score measures the presence that can be distinguished at the resolution limit of this recording. We provide a formula for this resolution limit, a noise-corrected score, and two reference values that illustrate what a brain with the entire model would present. Read against these reference values, the reverse scores show where the model and the brain differ in what they express strongly. They do not list what the brain lacks.
	
	\section*{Software and Data}
	
	The code of all experiments, the summary tables and the numbers of this paper will be released with the final version. The recordings are public. The monkey data are those released with \citet{muzellec2026}, the Podcast dataset is OpenNeuro ds005574 and the Narratives dataset is OpenNeuro ds002345.
	
	\section*{AI Use Statement}
	
	Generative AI tools were used to aid and polish the writing of this paper, to support the retrieval and discovery of related work, to assist in checking and revising portions of the mathematical arguments and proofs, and to support the verification of portions of the experimental code. All AI-assisted mathematical content was independently verified by the authors against the stated assumptions and derivations, and all references identified with AI assistance were checked against the original publications. The authors take full responsibility for the final content of this work.
	
	\section*{Impact Statement}
	
	This paper presents work whose goal is to advance the field of machine learning and its links with neuroscience. It concerns the reading of a measurement and uses public recordings from earlier studies. We do not see societal consequences that must be highlighted here.
	
	\section*{Acknowledgements}
	We thank Sabine Muzellec and Kohitij Kar for making their recordings and code public, and Kohitij Kar for comments on an earlier version that led to the simulations with one brain and several models.
	
	\bibliography{example_paper}
	\bibliographystyle{icml2027}
	
	\newpage
	\appendix
	\onecolumn
	\section{Proofs}
	\label{app:proofs}
	
	\subsection{Setting}
	
	Functions of the stimulus live in $L^2_0$, the space of functions with mean zero and finite variance under the stimulus distribution. We assume that this space is separable. A site consists of a signal $r\in L^2_0$ and a noise distribution with mean zero and variance $\sigma^2$, and a population $Q$ is a distribution over sites. The covariance operator of a population is $T=\E_Q[r\otimes r]$, where $(r\otimes r)g=\langle r,g\rangle\,r$. It is positive and has finite trace because $\E_Q\|r\|^2<\infty$.
	
	\begin{lemma}
		\label{lem:span}
		The closure of the range of $T$ is the smallest closed subspace that contains $r$ with probability one.
	\end{lemma}
	
	\begin{proof}
		A function $g$ lies in the kernel of $T$ exactly when $\langle g,Tg\rangle=\E_Q\langle g,r\rangle^2=0$, that is, when $g\perp r$ with probability one. Let $V$ be a closed subspace with $Q(r\in V)=1$. Every $g\perp V$ is then in the kernel of $T$, so $(\ker T)^\perp\subseteq V$. It remains to show that $(\ker T)^\perp$ contains $r$ with probability one. Take a countable dense subset $g_1,g_2,\dots$ of $\ker T$. Each event $\langle g_k,r\rangle=0$ has probability one, so all of them hold at once with probability one, and then $r\perp\ker T$. The claim follows because $(\ker T)^\perp$ is the closure of the range of $T$.
	\end{proof}
	
	The subspace of \cref{lem:span} is the span $\Hsp$ of the population. The eigenfunctions $\varphi_i$ of $T$ with positive eigenvalues $\lambda_i$ form an orthonormal basis of $\Hsp$, so the weights $\beta_i$ of a unit sum to one.
	
	\subsection{Proof of \cref{prop:losses}}
	
	Fix the recorded sites and write $r(s)=(r_1(s),\dots,r_n(s))^\top$. A readout $w\in\mathbb{R}^n$ predicts $f$ by $w^\top y$ with $y=r(s)+\varepsilon$. The noise has mean zero and is independent of the stimulus, so
	\begin{equation}
		\label{eq:mse}
		\E\big(f-w^\top y\big)^2=\|f-w^\top r\|^2+w^\top\Sigma w .
	\end{equation}
	The right side equals $\|f\|^2-2w^\top a+w^\top(A+\Sigma)w$. Its minimum over $w$ is $\|f\|^2-a^\top(A+\Sigma)^{-1}a$, which gives the first expression in \cref{eq:scores}. With $\Sigma=0$ the minimum is $\|f\|^2-a^\top A^{+}a=\|f\|^2-\|\Pi_S f\|^2$, where $S$ is the span of $r_1,\dots,r_n$. This gives the second expression.
	
	A readout $\hat w$ that is fitted on a finite sample explains the share $1-\E(f-\hat w^\top y)^2/\|f\|^2$ on new stimuli, which is at most $R_n(f)$ for every $\hat w$ and therefore in expectation. \cref{eq:mse} shows that $\E(f-w^\top y)^2\ge\|f-w^\top r\|^2\ge\|f\|^2(1-\Rsig_n(f))$ for every $w$, so $R_n(f)\le\Rsig_n(f)$. Every $r_j$ lies in $\Hsp$ with probability one, so $S\subseteq\Hsp$ and $\|\Pi_Sf\|\le\|\Pi_{\Hsp}f\|$, which is $\Rsig_n(f)\le\rho(f)$. The last inequality holds because a projection does not increase the norm.
	
	Each inequality is strict in simple cases. A readout fitted on few stimuli makes the first strict. Noise with positive variance at every site makes the second inequality strict whenever $\Rsig_n(f)>0$. A Gaussian population with more than $n$ dimensions, all of which carry a part of $f$, makes the third strict with probability one. A unit with a part outside $\Hsp$ makes the fourth strict.
	
	In the forward direction let $\mathcal{U}$ be the span of the units. For $g\in\mathcal{U}$ we have $\E(y_j-g)^2=\|r_j-g\|^2+\sigma_j^2$, which is smallest at $g=\Pi_{\mathcal{U}}r_j$. The explained share of the variance of $y_j$ is then $\|\Pi_{\mathcal{U}}r_j\|^2/(\|r_j\|^2+\sigma_j^2)$, the product of $F(r_j)$ and the reliability of the site. The sites are drawn independently from $Q$, so the mean of $F(r_j)$ over the recorded sites has expectation $\E_QF(r)$.
	
	\subsection{Proof of \cref{thm:identified} and \cref{cor:test}}
	
	\textbf{Part (i).} Let $\tilde Q$ be a population whose support contains $r_1,\dots,r_n$, and let $\tilde\Hsp$ be its span. By \cref{lem:span} the subspace $\tilde\Hsp$ is closed and has probability one, so it contains the support of $\tilde Q$ and with it $S$. Hence $\|\Pi_{\tilde\Hsp}f\|^2\ge\|\Pi_Sf\|^2=\|f\|^2\,\Rsig_n(f)$. For the second claim, the noise of two presentations is independent, has mean zero and is independent of the stimulus. Therefore $a_j=\E[f(s)\,y_j^{(1)}]$ and $A_{jl}=\E[y_j^{(1)}y_l^{(2)}]$ for all $j$ and $l$, including $j=l$. These moments belong to the joint distribution of $f$ and two presentations, and $\Rsig_n(f)=a^\top A^{+}a/\|f\|^2$ is a function of them.
	
	\textbf{Part (ii).} Let $e=f-\Pi_{\Hsp}f$. If $e=0$ the population $Q$ holds $f$ and we take $Q'=Q$. Otherwise let $\eta=\delta/n$ and
	\[
	Q'=(1-\eta)\,Q+\eta\,Q_e ,
	\]
	where $Q_e$ puts all its mass on one site with signal $e$ and any noise distribution. The span of $Q'$ contains $\Hsp$, $e$ and hence $f=\Pi_{\Hsp}f+e$, so $Q'$ holds $f$. To compare the data, draw $n$ sites from $Q$ and $n$ independent indicators that equal one with probability $\eta$. A recording from $Q'$ uses the site from $Q$ where the indicator is zero and the site with signal $e$ where it is one. Give both recordings the same stimuli, and the same noise at the sites that they share. With probability $(1-\eta)^n\ge1-n\eta=1-\delta$ all indicators are zero, and the two recordings then produce the same data, whatever the number of stimuli and presentations. The total variation distance between the two distributions of the data is therefore at most $\delta$.
	
	\textbf{\cref{cor:test}.} Let $Q$ be any population and $\delta>0$, and let $Q'$ be the population of part (ii). The hypothesis is true under $Q'$, so the test rejects with probability at most $\alpha$ under $Q'$. The probabilities of any event under $Q$ and $Q'$ differ by at most $\delta$, so the test rejects with probability at most $\alpha+\delta$ under $Q$. This holds for every $\delta>0$.
	
	\begin{proposition}[The interval is sharp]
		\label{prop:sharp}
		Let $r_1,\dots,r_n$ be the recorded signals and let $L^2_0$ have dimension at least $n+2$. For every value $v$ with $\Rsig_n(f)\le v\le1$ there is a population whose support contains $r_1,\dots,r_n$ and for which $\rho(f)=v$.
	\end{proposition}
	
	\begin{proof}
		The uniform distribution on $r_1,\dots,r_n$ has span $S$ and gives $\rho(f)=\Rsig_n(f)$. Let $e_S=f-\Pi_Sf$. If $e_S=0$ then $\Rsig_n(f)=1$ and nothing is left to show. Otherwise choose a function $u$ of unit norm that is orthogonal to $S$ and to $f$, which exists because $S$ and $f$ span at most $n+1$ dimensions. For $t\ge0$ let $h_t=e_S+t\,u$ and let $Q_t$ be the uniform distribution on $r_1,\dots,r_n,h_t$. The function $h_t$ is orthogonal to $S$ and $\langle f,h_t\rangle=\|e_S\|^2$, so the projection of $f$ onto the span of $Q_t$ has squared norm $\|\Pi_Sf\|^2+\|e_S\|^4/(\|e_S\|^2+t^2)$. With $\|e_S\|^2=\|f\|^2(1-\Rsig_n(f))$ this gives
		\[
		\rho_t(f)=\Rsig_n(f)+\big(1-\Rsig_n(f)\big)\frac{\|e_S\|^2}{\|e_S\|^2+t^2},
		\]
		which equals one at $t=0$, decreases continuously in $t$ and tends to $\Rsig_n(f)$ as $t$ grows.
	\end{proof}
	
	Given the recorded sites, the distribution of the data does not depend on the population. \cref{prop:sharp} and part (i) of \cref{thm:identified} therefore show that the values of $\rho(f)$ compatible with a recording are exactly those of the interval from $\Rsig_n(f)$ to one, the identified set in the sense of \citet{manski2003}.
	
	\subsection{Proof of \cref{prop:resolution}}
	
	Let $Z$ map a vector $z\in\mathbb{R}^n$ to $\sum_jz_jr_j$. Then $A=Z^*Z$ and $a=Z^*f$, and the operator $\hat T=ZZ^*/n=\frac1n\sum_jr_j\otimes r_j$ has expectation $T$. With $\tau=\sigma^2/n$,
	\begin{equation}
		\label{eq:pushthrough}
		\|f\|^2R_n(f)=\big\langle f,Z(Z^*Z+\sigma^2I)^{-1}Z^*f\big\rangle=\big\langle f,\hat T(\hat T+\tau)^{-1}f\big\rangle=\|f\|^2-\tau\,\big\langle f,(\hat T+\tau)^{-1}f\big\rangle .
	\end{equation}
	
	\textbf{Part (a).} For a positive definite operator $M$ and every function $h$ we have $\langle f,M^{-1}f\rangle\ge2\langle f,h\rangle-\langle h,Mh\rangle$, with equality at $h=M^{-1}f$. Apply this to $M=\hat T+\tau$ with the fixed choice $h=(T+\tau)^{-1}f$ and take expectations. Since $\E\hat T=T$,
	\[
	\E\big\langle f,(\hat T+\tau)^{-1}f\big\rangle\ge2\langle f,h\rangle-\langle h,(T+\tau)h\rangle=\big\langle f,(T+\tau)^{-1}f\big\rangle .
	\]
	With \cref{eq:pushthrough} this gives $\|f\|^2\,\E R_n(f)\le\langle f,T(T+\tau)^{-1}f\rangle=\sum_i\langle f,\varphi_i\rangle^2\lambda_i/(\lambda_i+\tau)$, and $\langle f,\varphi_i\rangle^2=\rho(f)\,\|f\|^2\beta_i$. The argument uses only that every site is drawn from $Q$. It does not need the sites to be independent of each other.
	
	\textbf{Part (b).} The statement is an instance of a known result on sample covariance matrices, which we restate for Gaussian data and one test matrix.
	
	\begin{theorem}[\citealp{bach2024}, Proposition 1]
		\label{thm:bach}
		Let $\Lambda$ be a diagonal covariance matrix of size $d$ with entries $\lambda_1,\dots,\lambda_d$, let $r_1,\dots,r_n$ be independent Gaussian vectors with mean zero and covariance $\Lambda$, and let $\hat\Lambda=\frac1n\sum_jr_jr_j^\top$. Let $n$ and $d$ grow such that $d/n$ converges to a positive number. Assume that the $\lambda_i$ stay in a fixed interval of positive numbers, that the measures $\frac1d\sum_i\delta_{\lambda_i}$ converge weakly, and that $B$ is a symmetric matrix of bounded operator norm for which the measures $\sum_iB_{ii}\,\delta_{\lambda_i}$ converge weakly to a measure of bounded total variation. Then for every $\tau>0$, almost surely,
		\[
		\frac{\operatorname{tr}\big[B\hat\Lambda(\hat\Lambda+\tau I)^{-1}\big]}{\operatorname{tr}\big[B\Lambda(\Lambda+\kappa I)^{-1}\big]}\longrightarrow1,
		\qquad\text{where }\kappa>0\text{ solves }\ \kappa-\tau=\frac{\kappa}{n}\operatorname{tr}\big[\Lambda(\Lambda+\kappa I)^{-1}\big].
		\]
	\end{theorem}
	
	\citet{bach2024} states the result for a general covariance matrix, with $B$ in the basis of its eigenvectors, and defines $\kappa$ through the limits of the two measures. The equation above is the same equation with the measures at finite $n$. Divided by $\kappa$, its two sides differ by a function of $\kappa$ that is continuous and strictly monotone and that converges uniformly on compact sets, so its solution is unique and has the same limit. The result rests on the analysis of \citet{ledoit2011}.
	
	The conditions of part (b) are those of \cref{thm:bach}. The population has $d$ dimensions and $d/n$ converges to a positive number, the $\lambda_i$ stay in a fixed interval of positive numbers, the ratio $\tau=\sigma^2/n$ is fixed, and the measures $\frac1d\sum_i\delta_{\lambda_i}$ and $\sum_i\beta_i\delta_{\lambda_i}$ converge weakly. The signal variance of a site is then of order $n$, so a fixed $\tau$ corresponds to a fixed reliability of the sites. In the coordinates of the $\varphi_i$ the site signals are Gaussian vectors with covariance $\Lambda$ and $\hat T$ is $\hat\Lambda$. Let $B$ be the matrix of rank one built from the coordinates of $\Pi_{\Hsp}f/\|\Pi_{\Hsp}f\|$, so that $B_{ii}=\beta_i$. By \cref{eq:pushthrough} the numerator of \cref{thm:bach} is $R_n(f)/\rho(f)$, the denominator is $\sum_i\beta_i\lambda_i/(\lambda_i+\kappa)$, and the equation for $\kappa$ is \cref{eq:kappa} divided by $n$ and multiplied by $\kappa$. The ratio of $R_n(f)$ to the right side of \cref{eq:resolved} therefore tends to one almost surely.
	
	Results of this kind also hold without the proportional limit, with explicit error bounds under conditions on the spectrum \citep{cheng2024,misiakiewicz2024,defilippis2024}. \cref{tab:formula} checks \cref{eq:resolved} for spectra that decay as a power law and for as few as ten sites.
	
	\textbf{Number of components above the resolution limit.} Every term of the sum in \cref{eq:kappa} with $\lambda_i\ge\kappa_n$ is at least one half. The sum is at most $n$, so at most $2n$ components have a variance of $\kappa_n$ or more.
	
	\textbf{Proof of \cref{cor:ceiling}.} The function $\lambda\mapsto\lambda/(\lambda+\kappa_n)$ increases in $\lambda$ and the weights $\beta_i$ sum to one, so $\sum_i\beta_i\lambda_i/(\lambda_i+\kappa_n)\ge v/(v+\kappa_n)$ under the assumption. With \cref{eq:resolved} this gives $R_n(f)\ge(1+o(1))\,\rho(f)\,v/(v+\kappa_n)$, which is the first claim. For the second, every term of the sum in \cref{eq:kappa} is at most $\lambda_i/\kappa_n$, so $n\le(\sum_i\lambda_i+\sigma^2)/\kappa_n$.
	
	\subsection{A Brain that Holds the Whole Model}
	
	Let a layer have $p$ units $x_1,\dots,x_p$ of unit variance. Let $\mu_1\ge\mu_2\ge\dots$ be the eigenvalues of their covariance matrix, $V$ the matrix of its eigenvectors, and $\psi_i=\mu_i^{-1/2}\sum_mV_{mi}x_m$ the principal components, which have unit norm. A site of the complete brain is $r=\sum_m\theta_mx_m$ with independent standard normal weights $\theta_m$. Its covariance operator is $T=\sum_mx_m\otimes x_m$, with eigenfunctions $\psi_i$ and eigenvalues $\mu_i$, and $r=\sum_i\sqrt{\mu_i}\,\xi_i\psi_i$ with $\xi_i=\sum_mV_{mi}\theta_m$ independent and standard normal. The span of the population is the span of the units, so $\rho(x_m)=1$ for every unit and $F(r)=1$ for every site. Part (b) of \cref{prop:resolution} applies with $\lambda_i=\mu_i$, and unit $x_m$ has $\beta_i=\mu_iV_{mi}^2$. Write $G(f,\kappa)=\sum_i\beta_i\mu_i/(\mu_i+\kappa)$ for the sum in \cref{eq:resolved}.
	
	\textbf{Low scores.} A unit that lies on components $k$ and beyond has $G(f,\kappa_n)\le\mu_k/(\mu_k+\kappa_n)$, and at most $2n$ components reach $\kappa_n$.
	
	\textbf{Agreement between recordings.} A second recording of the area differs from the first in its resolution limit only. For a unit on a single component the score $\mu_i/(\mu_i+\kappa)$ increases with $\mu_i$ at every $\kappa$, so two recordings order such units in the same way. A general unit averages these increasing functions with the same weights $\beta_i$ in both recordings.
	
	\textbf{More sites.} A unit on component $k$ reaches a score of one half only when $\mu_k\ge\kappa_n$, which requires $k\le2n$. A unit on components of rank $k$ and beyond therefore stays below one half as long as the recording has fewer than $k/2$ sites.
	
	\textbf{Other units.} The other units of the layer are $p-1$ predictors without noise, and their span lacks at most one dimension. They predict a unit except for the part that no other unit carries.
	
	\textbf{Two recordings.} A site of a second recording has the signal $r_0=\sum_i\sqrt{\mu_i}\,\xi_i\psi_i$, with weights $\beta_i=\mu_i\xi_i^2/\|r_0\|^2$. The first recording explains the share $G(r_0,\kappa_n)$ of this signal, and the share of the noisy response that it explains is $G(r_0,\kappa_n)$ times the reliability of the predicted site. Averaged over sites with their variances as weights, $G(r_0,\kappa_n)$ equals $\sum_i\mu_i^2/(\mu_i+\kappa_n)\big/\sum_i\mu_i$. This number is below one. It is the same in both directions when the two recordings have the same number of sites and the same noise.
	
	\textbf{Layers.} The mean of $G(x_m,\kappa_n)$ over the units of a layer is $\frac1p\sum_i\mu_i^2/(\mu_i+\kappa_n)$, because the weights $\beta_i$ of the $p$ units add up to $\mu_i$ and $\sum_i\mu_i=p$. It is the mean of $\mu_i/(\mu_i+\kappa_n)$ with the variances as weights. A flat spectrum places most of the variance on components below the resolution limit. For a fixed $\kappa$ the function $\mu\mapsto\mu/(\mu+\kappa)$ is concave and $\mu\mapsto\mu^2/(\mu+\kappa)$ is convex. Let a second spectrum with the same sum be flatter, in the sense that the first majorizes it. The sum in \cref{eq:kappa} is then at least as large at every $\kappa$, so the resolution limit of the flatter spectrum is at least as large. The sum $\sum_i\mu_i^2/(\mu_i+\kappa)$ is at most as large at every $\kappa$ and decreases in $\kappa$, so the mean score of the flatter spectrum is at most as large.
	
	\subsection{Proof of \cref{prop:noise}}
	
	Let $w$ be fixed. The prediction from presentation $k$ is $\hat f^{(k)}=w^\top r(s)+w^\top\varepsilon^{(k)}$. The noise has mean zero and is independent of the stimulus and of the other presentation, so $\Cov(\hat f,f)=\langle w^\top r,f\rangle$ and $\Cov(\hat f^{(1)},\hat f^{(2)})=\|w^\top r\|^2$. Sample covariances over stimuli that were not used to fit $w$ are unbiased for these two quantities, and
	\[
	2\langle w^\top r,f\rangle-\|w^\top r\|^2=\|f\|^2-\|f-w^\top r\|^2 .
	\]
	The right side is at most $\|f\|^2-\min_w\|f-w^\top r\|^2=\|f\|^2\,\Rsig_n(f)$, and $w=A^{+}a$ attains the minimum.
	
	\textbf{Relation to the normalised score.} The normalised score of prior work is the squared correlation between $\hat f$ and $f$ divided by the reliability of $\hat f$. The variance of $\hat f$ times its reliability is $\|w^\top r\|^2$. The normalised score therefore estimates $\langle w^\top r,f\rangle^2/(\|w^\top r\|^2\|f\|^2)$, the share of $f$ that the best multiple of $w^\top r$ explains. It removes the noise from the evaluation of a readout, as the noise-corrected score does, and it keeps a readout that was fitted to noisy sites.
	
	\section{Details of the Experiments}
	\label{app:details}
	
	\subsection{Simulated Populations}
	
	\textbf{Formula.} The variances of the population are $\lambda_i\propto i^{-\gamma}$ for $i\le3{,}000$ and sum to one. A site with reliability $q$ has noise variance $(1-q)/q$. For each of the three exponents, four reliabilities (1, 0.9, 0.7 and 0.5) and eight numbers of sites (10 to 1,280 in factors of two) we draw ten recordings and compute $R_n(f)$ from \cref{eq:scores} for targets on components 1, 4, 16, 64, 256 and 1,024 and for a held-out site, for which $\beta_i\propto\lambda_i$. \cref{tab:formula} reports the difference between the mean over the ten recordings and \cref{eq:resolved}. The largest amount by which a mean over ten recordings exceeds the bound of part (a) of \cref{prop:resolution} is \val{f.all.above}.
	
	\begin{table}[h]
		\caption{Absolute difference between exact reverse scores and \cref{eq:resolved}, over eight numbers of sites and seven targets per cell.}
		\label{tab:formula}
		\centering
		\small
		\begin{tabular}{@{}lcccccccc@{}}
			\toprule
			& \multicolumn{4}{c}{Mean, by reliability} & \multicolumn{4}{c}{Largest, by reliability} \\
			\cmidrule(lr){2-5}\cmidrule(l){6-9}
			Exponent $\gamma$ & 1 & 0.9 & 0.7 & 0.5 & 1 & 0.9 & 0.7 & 0.5 \\
			\midrule
			0.6 & \val{f.0.6.1.mean} & \val{f.0.6.0.9.mean} & \val{f.0.6.0.7.mean} & \val{f.0.6.0.5.mean} & \val{f.0.6.1.max} & \val{f.0.6.0.9.max} & \val{f.0.6.0.7.max} & \val{f.0.6.0.5.max} \\
			1 & \val{f.1.1.mean} & \val{f.1.0.9.mean} & \val{f.1.0.7.mean} & \val{f.1.0.5.mean} & \val{f.1.1.max} & \val{f.1.0.9.max} & \val{f.1.0.7.max} & \val{f.1.0.5.max} \\
			1.4 & \val{f.1.4.1.mean} & \val{f.1.4.0.9.mean} & \val{f.1.4.0.7.mean} & \val{f.1.4.0.5.mean} & \val{f.1.4.1.max} & \val{f.1.4.0.9.max} & \val{f.1.4.0.7.max} & \val{f.1.4.0.5.max} \\
			\bottomrule
		\end{tabular}
	\end{table}
	
	\textbf{Estimates.} A population with 400 dimensions and variances proportional to $1/i$ is recorded with 40 or 160 sites and two presentations of every stimulus. The noise is set so that the two presentations of a site correlate at 0.9, 0.5 or 0.2. Targets lie on components 1, 4, 16, 64 and 256. \cref{tab:estimates} compares the three estimates of \cref{sec:reading} with the exact values, for 500 to 20,000 stimuli and ten recordings per row. The estimates use the procedure of Appendix~\ref{app:recdetails} with five folds.
	
	\begin{table}[h]
		\caption{Estimates from two presentations against exact values in a simulated population. Means over five targets and ten recordings. $\Rsig_n$ and $R_n$ are exact, and $R_n$ refers to the mean of the two presentations.}
		\label{tab:estimates}
		\centering
		\small
		\begin{tabular}{@{}cccccccc@{}}
			\toprule
			& Correlation between & & \multicolumn{2}{c}{Exact} & \multicolumn{3}{c}{Estimated} \\
			\cmidrule(lr){4-5}\cmidrule(l){6-8}
			Sites & presentations & Stimuli & $\Rsig_n$ & $R_n$ & Reverse & Normalised & Noise-corrected \\
			\midrule
			40 & 0.9 & 500 & \val{e.40.0.9.500.signal} & \val{e.40.0.9.500.noisy} & \val{e.40.0.9.500.raw} & \val{e.40.0.9.500.normalised} & \val{e.40.0.9.500.corrected} \\
			40 & 0.9 & 1{,}320 & \val{e.40.0.9.1320.signal} & \val{e.40.0.9.1320.noisy} & \val{e.40.0.9.1320.raw} & \val{e.40.0.9.1320.normalised} & \val{e.40.0.9.1320.corrected} \\
			40 & 0.9 & 5{,}000 & \val{e.40.0.9.5000.signal} & \val{e.40.0.9.5000.noisy} & \val{e.40.0.9.5000.raw} & \val{e.40.0.9.5000.normalised} & \val{e.40.0.9.5000.corrected} \\
			40 & 0.9 & 20{,}000 & \val{e.40.0.9.20000.signal} & \val{e.40.0.9.20000.noisy} & \val{e.40.0.9.20000.raw} & \val{e.40.0.9.20000.normalised} & \val{e.40.0.9.20000.corrected} \\
			\addlinespace
			40 & 0.5 & 500 & \val{e.40.0.5.500.signal} & \val{e.40.0.5.500.noisy} & \val{e.40.0.5.500.raw} & \val{e.40.0.5.500.normalised} & \val{e.40.0.5.500.corrected} \\
			40 & 0.5 & 1{,}320 & \val{e.40.0.5.1320.signal} & \val{e.40.0.5.1320.noisy} & \val{e.40.0.5.1320.raw} & \val{e.40.0.5.1320.normalised} & \val{e.40.0.5.1320.corrected} \\
			40 & 0.5 & 5{,}000 & \val{e.40.0.5.5000.signal} & \val{e.40.0.5.5000.noisy} & \val{e.40.0.5.5000.raw} & \val{e.40.0.5.5000.normalised} & \val{e.40.0.5.5000.corrected} \\
			40 & 0.5 & 20{,}000 & \val{e.40.0.5.20000.signal} & \val{e.40.0.5.20000.noisy} & \val{e.40.0.5.20000.raw} & \val{e.40.0.5.20000.normalised} & \val{e.40.0.5.20000.corrected} \\
			\addlinespace
			40 & 0.2 & 500 & \val{e.40.0.2.500.signal} & \val{e.40.0.2.500.noisy} & \val{e.40.0.2.500.raw} & \val{e.40.0.2.500.normalised} & \val{e.40.0.2.500.corrected} \\
			40 & 0.2 & 1{,}320 & \val{e.40.0.2.1320.signal} & \val{e.40.0.2.1320.noisy} & \val{e.40.0.2.1320.raw} & \val{e.40.0.2.1320.normalised} & \val{e.40.0.2.1320.corrected} \\
			40 & 0.2 & 5{,}000 & \val{e.40.0.2.5000.signal} & \val{e.40.0.2.5000.noisy} & \val{e.40.0.2.5000.raw} & \val{e.40.0.2.5000.normalised} & \val{e.40.0.2.5000.corrected} \\
			40 & 0.2 & 20{,}000 & \val{e.40.0.2.20000.signal} & \val{e.40.0.2.20000.noisy} & \val{e.40.0.2.20000.raw} & \val{e.40.0.2.20000.normalised} & \val{e.40.0.2.20000.corrected} \\
			\addlinespace
			160 & 0.9 & 500 & \val{e.160.0.9.500.signal} & \val{e.160.0.9.500.noisy} & \val{e.160.0.9.500.raw} & \val{e.160.0.9.500.normalised} & \val{e.160.0.9.500.corrected} \\
			160 & 0.9 & 1{,}320 & \val{e.160.0.9.1320.signal} & \val{e.160.0.9.1320.noisy} & \val{e.160.0.9.1320.raw} & \val{e.160.0.9.1320.normalised} & \val{e.160.0.9.1320.corrected} \\
			160 & 0.9 & 5{,}000 & \val{e.160.0.9.5000.signal} & \val{e.160.0.9.5000.noisy} & \val{e.160.0.9.5000.raw} & \val{e.160.0.9.5000.normalised} & \val{e.160.0.9.5000.corrected} \\
			160 & 0.9 & 20{,}000 & \val{e.160.0.9.20000.signal} & \val{e.160.0.9.20000.noisy} & \val{e.160.0.9.20000.raw} & \val{e.160.0.9.20000.normalised} & \val{e.160.0.9.20000.corrected} \\
			\addlinespace
			160 & 0.5 & 500 & \val{e.160.0.5.500.signal} & \val{e.160.0.5.500.noisy} & \val{e.160.0.5.500.raw} & \val{e.160.0.5.500.normalised} & \val{e.160.0.5.500.corrected} \\
			160 & 0.5 & 1{,}320 & \val{e.160.0.5.1320.signal} & \val{e.160.0.5.1320.noisy} & \val{e.160.0.5.1320.raw} & \val{e.160.0.5.1320.normalised} & \val{e.160.0.5.1320.corrected} \\
			160 & 0.5 & 5{,}000 & \val{e.160.0.5.5000.signal} & \val{e.160.0.5.5000.noisy} & \val{e.160.0.5.5000.raw} & \val{e.160.0.5.5000.normalised} & \val{e.160.0.5.5000.corrected} \\
			160 & 0.5 & 20{,}000 & \val{e.160.0.5.20000.signal} & \val{e.160.0.5.20000.noisy} & \val{e.160.0.5.20000.raw} & \val{e.160.0.5.20000.normalised} & \val{e.160.0.5.20000.corrected} \\
			\addlinespace
			160 & 0.2 & 500 & \val{e.160.0.2.500.signal} & \val{e.160.0.2.500.noisy} & \val{e.160.0.2.500.raw} & \val{e.160.0.2.500.normalised} & \val{e.160.0.2.500.corrected} \\
			160 & 0.2 & 1{,}320 & \val{e.160.0.2.1320.signal} & \val{e.160.0.2.1320.noisy} & \val{e.160.0.2.1320.raw} & \val{e.160.0.2.1320.normalised} & \val{e.160.0.2.1320.corrected} \\
			160 & 0.2 & 5{,}000 & \val{e.160.0.2.5000.signal} & \val{e.160.0.2.5000.noisy} & \val{e.160.0.2.5000.raw} & \val{e.160.0.2.5000.normalised} & \val{e.160.0.2.5000.corrected} \\
			160 & 0.2 & 20{,}000 & \val{e.160.0.2.20000.signal} & \val{e.160.0.2.20000.noisy} & \val{e.160.0.2.20000.raw} & \val{e.160.0.2.20000.normalised} & \val{e.160.0.2.20000.corrected} \\
			\bottomrule
		\end{tabular}
	\end{table}
	
	\textbf{Published protocol.} A model has 2,048 dimensions with variances proportional to $1/i$. Each of its 4,096 units mixes eight randomly chosen dimensions with Gaussian weights and is scaled to unit variance. Each animal has 300 simulated sites. A site is a Gaussian combination of all dimensions in the complete brain and of the first 40 in the incomplete brain. In the brain with other content a site adds a Gaussian combination of 2,048 further dimensions that are independent of the model and have the same variances, and both animals record these same dimensions. This content makes up an eighth of the signal variance of every site, a share that brings the forward score of the simulated brain close to the published value. The forward score falls by more than an eighth because the regression is fitted on fewer stimuli than units, and the unrelated content then adds to its error as noise does. A site of the partial brain is a Gaussian combination of the first quarter of the dimensions, 512 in this model, and adds the same unrelated content with the same share. Every site is scaled to unit variance, and the noise of its two halves of repetitions is set so that the halves correlate at 0.75. The first 169 sites of animal 1 and the first 107 sites of animal 2 enter all analyses except the curves of \cref{fig:simulation}b, which draw 20 to 300 sites of animal 1 at random, three times per size.
	
	Scores follow the code released with \citet{muzellec2026}. Ridge regression with a penalty of 0.1 is fitted in ten folds on the 1,320 stimuli, with the fold assignment of that code. The score is 100 times the squared correlation between the prediction and the target, divided by the split-half reliability of the prediction and by the split-half reliability of the target. The first is obtained by refitting on each half of the repetitions, and the second equals one for a unit of the model. Both reliabilities carry the Spearman-Brown correction. The least predictable 500 units are those with the lowest reverse scores from the 169 sites of animal 1. The same units are also predicted from all other units of the model without any correction. The agreement between animals is the rank correlation of the unit scores of the two animals over all units. The bottom fifth of the units are the fifth with the lowest reverse scores from animal 1, and their expressible share follows from the dimensions that each unit mixes.
	
	The model was reduced to its first \val{t1.match.components} principal components, which is the number of recorded sites given in the study. These components are the principal components of the units' covariance on the stimulus distribution, replacing the units in both directions. When removing units, both animals were shown \val{t1.cut.stimuli} additional stimuli, and the units were ranked by their reverse scores using their respective first \val{t1.cut.sitesfirst} and first \val{t1.cut.sitessecond} sites, which are the numbers of sites for the study's additional pictures. The fifth of units with the highest (common) or lowest (unique) scores were removed, with \val{t1.cut.units} units removed for each animal. The forward scores were recalculated on 1,320 stimuli using the remaining units. The reverse scores of the remaining units remained unchanged, and the reverse score after removal was the median of the remaining units. For predictions made using a single unit, each site was paired with the unit that was most correlated with it on the training stimuli in each fold. The score is the square of the correlation coefficient between the two on the test stimuli, divided by the split-half reliability of that site, and then averaged over the folds. In the reverse direction, each unit is paired with a site using the same method. A second instance of the model combines the same 2,048 dimensions into another 4,096 units. The two instances predict each other using all units, without reliability correction, as was done with models in the study. The second instance was also observed like Animal 1. For this purpose, \val{t1.inst.seen} of its units were randomly selected, site noise was added, and scores were calculated according to the recording procedure.
	
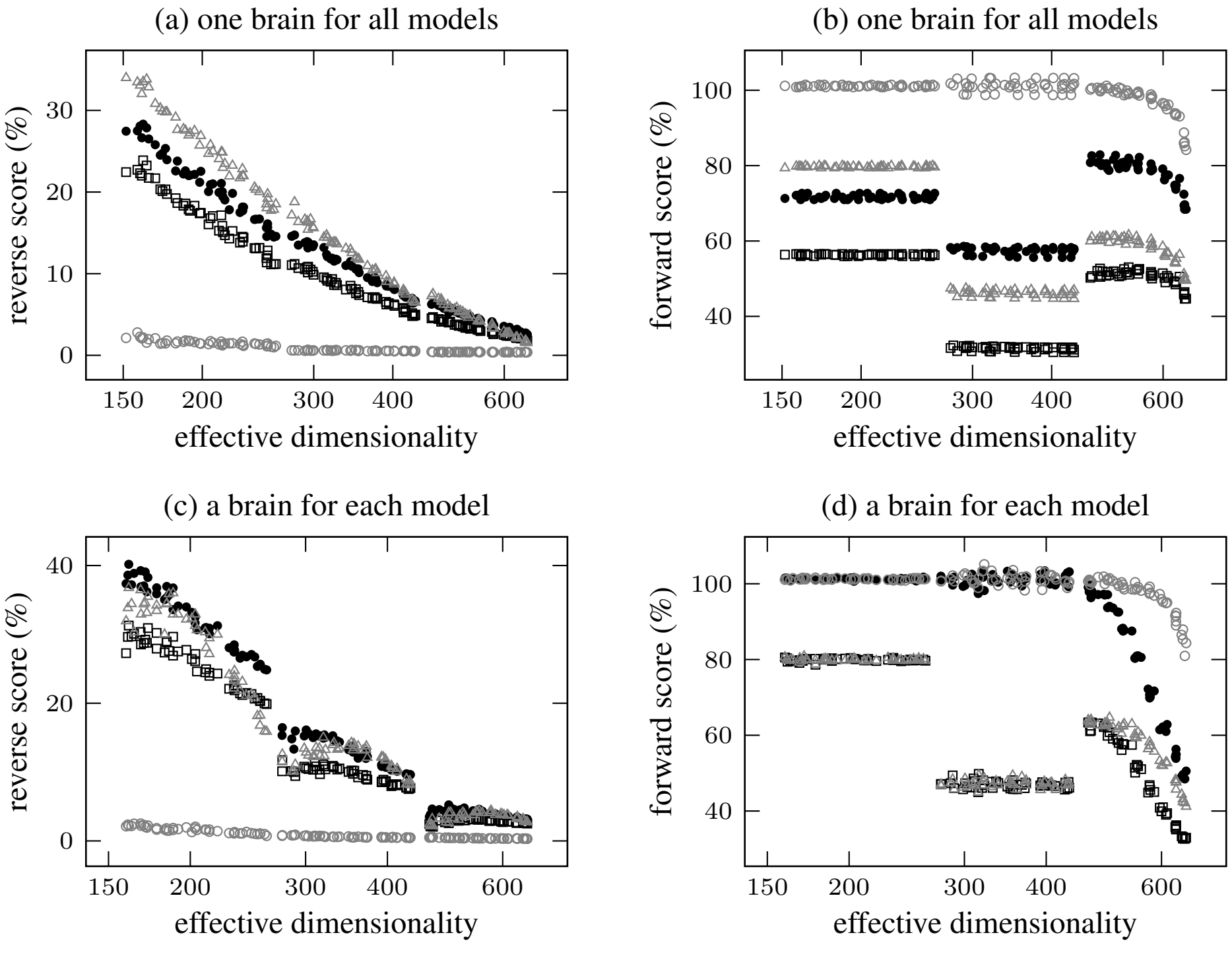
\begin{figure}[h]
		\centering
		\datarange{fig2d_complete.csv}{rangeQ}{xmin=1, xmax=1000, ymin=0, ymax=100}
		\datarange{fig2c_complete.csv}{rangeK}{xmin=1, xmax=1000, ymin=0, ymax=100}
		\pgfplotsset{familyticks/.style={xtick={150,200,300,400,600}, log ticks with fixed point}}
		\newcommand{\familypoints}[2]{%
			\ifdata{#1_complete.csv}{
				\addplot[black, only marks, mark=*, mark size=1.1pt] table[x=dimension, y=#2, col sep=comma]{figdata/#1_complete.csv};
			}{\nodata}
			\ifdata{#1_extra.csv}{
				\addplot[black, only marks, mark=square, mark size=1.2pt] table[x=dimension, y=#2, col sep=comma]{figdata/#1_extra.csv};
			}{}
			\ifdata{#1_incomplete.csv}{
				\addplot[gray, only marks, mark=o, mark size=1.3pt] table[x=dimension, y=#2, col sep=comma]{figdata/#1_incomplete.csv};
			}{}
			\ifdata{#1_partial.csv}{
				\addplot[gray, only marks, mark=triangle, mark size=1.5pt] table[x=dimension, y=#2, col sep=comma]{figdata/#1_partial.csv};
			}{}}
		\begin{tikzpicture}
			\begin{axis}[paperaxis, rangeQ, name=A, width=1.9in, height=1.3in, xmode=log, familyticks,
				xlabel={effective dimensionality}, ylabel={reverse score (\%)}, title={(a) one brain for all models}]
				\familypoints{fig2d}{reverse}
			\end{axis}
			\begin{axis}[paperaxis, rangeQ, name=B, at={(A.south east)}, anchor=south west, xshift=0.7in, width=1.9in, height=1.3in, xmode=log, familyticks,
				xlabel={effective dimensionality}, ylabel={forward score (\%)}, title={(b) one brain for all models}]
				\familypoints{fig2d}{forward}
			\end{axis}
			\begin{axis}[paperaxis, rangeK, name=C, at={(A.south west)}, anchor=north west, yshift=-0.62in, width=1.9in, height=1.3in, xmode=log, familyticks,
				xlabel={effective dimensionality}, ylabel={reverse score (\%)}, title={(c) a brain for each model}]
				\familypoints{fig2c}{reverse}
			\end{axis}
			\begin{axis}[paperaxis, rangeK, at={(C.south east)}, anchor=south west, xshift=0.7in, width=1.9in, height=1.3in, xmode=log, familyticks,
				xlabel={effective dimensionality}, ylabel={forward score (\%)}, title={(d) a brain for each model}]
				\familypoints{fig2c}{forward}
			\end{axis}
		\end{tikzpicture}
		\caption{The family of simulated models under the published protocol. Median reverse score and median forward score against the effective dimensionality of the model, when all models are compared with one brain (a, b) and when each model has a brain of its own (c, d), for complete brains (black dots), brains with other content (squares), partial brains (gray triangles) and incomplete brains (gray circles). Every point is one model in one repetition.}
		\label{fig:family}
	\end{figure}

	The model family has nine exponents from 0.4 to 1.6 and three sizes (512, 1,024, and 2,048 dimensions, with twice the number of units). The effective dimensionality of the model is the participation ratio of its unit covariance eigenvalues. The model family is compared to the brain in two ways. When using a single brain, these models share a common set of dimensions. A model with $d$ dimensions has the first $d$ of them, with variance following its own exponent. All models are scored against the same 169-site recording. This brain is one of four brains in \cref{tab:published}, or a complete brain with variance decreasing as $i^{-0.7}$ or $i^{-1.3}$, or a complete brain with 8,192 dimensions. When each model is paired with a brain, the model uses 169-site recordings from four brains, each constructed from the model's dimensions and variances. \cref{tab:dimensionality} gives the rank correlation between effective dimensionality and median scores on \val{t1.family} models in two ways, and \cref{fig:family} shows these models. The published protocol uses a fixed penalty of 0.1. When the penalty is so small, if the dimensionality of the predictor variable is similar to the number of training stimuli, the regression will predict poorly \citep{hastie2022}. Here, each fold has 1,188 training stimuli. Therefore, for the complete brain, the 1,024-dimensional models have lower forward scores than the smaller and larger models (\cref{fig:family}b). All values in \cref{tab:published}, \cref{tab:dimensionality}, and \cref{fig:simulation}b are the average of the entire simulation repeated \val{t1.repeats} times with new models and recordings.
	
	\begin{table}[h]
		\caption{Rank correlation between effective dimensionality and the median score over the \val{t1.family} models of the family. With one brain, all models are scored against the same recording. With a brain for each model, the brain takes its dimensions and variances from the model, so the last three brains exist with one brain only. The second column is the mean expressible share of the units of all models in the one brain.}
		\label{tab:dimensionality}
		\centering
		\small
		\begin{tabular}{@{}lccccc@{}}
			\toprule
			& Expressible & \multicolumn{2}{c}{One brain for all models} & \multicolumn{2}{c}{A brain for each model} \\
			\cmidrule(lr){3-4}\cmidrule(l){5-6}
			Brain & share & Forward & Reverse & Forward & Reverse \\
			\midrule
			Complete & \val{t1.fixtrue.complete} & \val{t1.fixfwd.complete} & \val{t1.fixrev.complete} & \val{t1.edfwd.complete} & \val{t1.edrev.complete} \\
			With other content & \val{t1.fixtrue.extra} & \val{t1.fixfwd.extra} & \val{t1.fixrev.extra} & \val{t1.edfwd.extra} & \val{t1.edrev.extra} \\
			Partial & \val{t1.fixtrue.partial} & \val{t1.fixfwd.partial} & \val{t1.fixrev.partial} & \val{t1.edfwd.partial} & \val{t1.edrev.partial} \\
			Incomplete & \val{t1.fixtrue.incomplete} & \val{t1.fixfwd.incomplete} & \val{t1.fixrev.incomplete} & \val{t1.edfwd.incomplete} & \val{t1.edrev.incomplete} \\
			Complete, variances $i^{-0.7}$ & \val{t1.fixtrue.decay0.7} & \val{t1.fixfwd.decay0.7} & \val{t1.fixrev.decay0.7} & -- & -- \\
			Complete, variances $i^{-1.3}$ & \val{t1.fixtrue.decay1.3} & \val{t1.fixfwd.decay1.3} & \val{t1.fixrev.decay1.3} & -- & -- \\
			Complete, 8{,}192 dimensions & \val{t1.fixtrue.dims8192} & \val{t1.fixfwd.dims8192} & \val{t1.fixrev.dims8192} & -- & -- \\
			\bottomrule
		\end{tabular}
	\end{table}
	
	\textbf{One brain and multiple models.} Five groups of units are drawn, each with \val{ob.group.units} units. Each group has its own \val{ob.group.dims} dimensions, with variance proportional to $1/i$. Each unit mixes eight of these dimensions using Gaussian weights and is scaled to unit variance, as above. The signal for a site is a Gaussian combination of the first four groups of units. The first and second groups enter with a coefficient of 1, the third with a coefficient of 0.5, and the fourth with a coefficient of 0.25, so these four groups account for approximately 43\%, 43\%, 11\%, and 3\% of the signal variance, respectively. The fifth group does not enter. The site is scaled to unit variance, and the noise from its split repetitions is set to ensure a correlation of 0.75 between the two halves. The recording has \val{ob.sites} sites and \val{ob.stimuli} stimuli. The strongest units in a group are those whose maximum component lies in the dimensions with the largest variance. The scores for the published protocol are calculated using the method described above. A two-group model has 1,024 dimensions, close to the 1,188 training stimuli in a fold, so its forward score is lower for the reasons mentioned above under the fixed penalty of this protocol. The forward $R^2$, reverse score, and two reference values are calculated according to the procedure in Appendix~\ref{app:recdetails}, using five folds, with the average of the two halves as the recording, and each reference value using three simulated recordings. There are four additional recordings, each different from this one in one factor. They have 42 or 676 sites, or a correlation of 0.5 or 0.95 between the two halves, and the analysis uses the first group and all five groups as models (\cref{tab:onerec}). All values are the average of \val{ob.repeats} repetitions with new groups and recordings.
	
	\begin{table}[h]
		\caption{One brain under five recordings. The default recording has \val{ob.sites} sites whose halves correlate at 0.75, and each of the others differs from it in one factor. With all five groups as the model, the mean reverse score $\bar R$ and the median score of the published protocol in percent are given for the units of each group. The brain expresses the first and the second group at full strength, the third at half and the fourth at a quarter of that strength, and it lacks the fifth. The agreement is that with the independent-site reference.}
		\label{tab:onerec}
		\centering
		\small
		\setlength{\tabcolsep}{3.2pt}
		\begin{tabular}{@{}lcccccccccccccc@{}}
			\toprule
			& \multicolumn{4}{c}{First group as the model} & \multicolumn{10}{c}{All five groups as the model} \\
			\cmidrule(lr){2-5}\cmidrule(l){6-15}
			& & & & & \multicolumn{5}{c}{$\bar R$ of group} & \multicolumn{5}{c}{Published score of group} \\
			\cmidrule(lr){6-10}\cmidrule(l){11-15}
			Recording & $\bar R$ & $\bar\Cind$ & $\bar\Csh$ & Agreement & 1 & 2 & 3 & 4 & 5 & 1 & 2 & 3 & 4 & 5 \\
			\midrule
			Default & \val{ob.default.base.raw} & \val{ob.default.base.indep} & \val{ob.default.base.shared} & \val{ob.default.base.ranki} & \val{ob.default.all.base.all.raw} & \val{ob.default.all.gain1.all.raw} & \val{ob.default.all.gain0.5.all.raw} & \val{ob.default.all.gain0.25.all.raw} & \val{ob.default.all.gain0.all.raw} & \val{ob.default.all.base.all.pubrev} & \val{ob.default.all.gain1.all.pubrev} & \val{ob.default.all.gain0.5.all.pubrev} & \val{ob.default.all.gain0.25.all.pubrev} & \val{ob.default.all.gain0.all.pubrev} \\
			42 sites & \val{ob.sites-42.base.raw} & \val{ob.sites-42.base.indep} & \val{ob.sites-42.base.shared} & \val{ob.sites-42.base.ranki} & \val{ob.sites-42.all.base.all.raw} & \val{ob.sites-42.all.gain1.all.raw} & \val{ob.sites-42.all.gain0.5.all.raw} & \val{ob.sites-42.all.gain0.25.all.raw} & \val{ob.sites-42.all.gain0.all.raw} & \val{ob.sites-42.all.base.all.pubrev} & \val{ob.sites-42.all.gain1.all.pubrev} & \val{ob.sites-42.all.gain0.5.all.pubrev} & \val{ob.sites-42.all.gain0.25.all.pubrev} & \val{ob.sites-42.all.gain0.all.pubrev} \\
			676 sites & \val{ob.sites-676.base.raw} & \val{ob.sites-676.base.indep} & \val{ob.sites-676.base.shared} & \val{ob.sites-676.base.ranki} & \val{ob.sites-676.all.base.all.raw} & \val{ob.sites-676.all.gain1.all.raw} & \val{ob.sites-676.all.gain0.5.all.raw} & \val{ob.sites-676.all.gain0.25.all.raw} & \val{ob.sites-676.all.gain0.all.raw} & \val{ob.sites-676.all.base.all.pubrev} & \val{ob.sites-676.all.gain1.all.pubrev} & \val{ob.sites-676.all.gain0.5.all.pubrev} & \val{ob.sites-676.all.gain0.25.all.pubrev} & \val{ob.sites-676.all.gain0.all.pubrev} \\
			Halves correlate at 0.5 & \val{ob.noise-0.5.base.raw} & \val{ob.noise-0.5.base.indep} & \val{ob.noise-0.5.base.shared} & \val{ob.noise-0.5.base.ranki} & \val{ob.noise-0.5.all.base.all.raw} & \val{ob.noise-0.5.all.gain1.all.raw} & \val{ob.noise-0.5.all.gain0.5.all.raw} & \val{ob.noise-0.5.all.gain0.25.all.raw} & \val{ob.noise-0.5.all.gain0.all.raw} & \val{ob.noise-0.5.all.base.all.pubrev} & \val{ob.noise-0.5.all.gain1.all.pubrev} & \val{ob.noise-0.5.all.gain0.5.all.pubrev} & \val{ob.noise-0.5.all.gain0.25.all.pubrev} & \val{ob.noise-0.5.all.gain0.all.pubrev} \\
			Halves correlate at 0.95 & \val{ob.noise-0.95.base.raw} & \val{ob.noise-0.95.base.indep} & \val{ob.noise-0.95.base.shared} & \val{ob.noise-0.95.base.ranki} & \val{ob.noise-0.95.all.base.all.raw} & \val{ob.noise-0.95.all.gain1.all.raw} & \val{ob.noise-0.95.all.gain0.5.all.raw} & \val{ob.noise-0.95.all.gain0.25.all.raw} & \val{ob.noise-0.95.all.gain0.all.raw} & \val{ob.noise-0.95.all.base.all.pubrev} & \val{ob.noise-0.95.all.gain1.all.pubrev} & \val{ob.noise-0.95.all.gain0.5.all.pubrev} & \val{ob.noise-0.95.all.gain0.25.all.pubrev} & \val{ob.noise-0.95.all.gain0.all.pubrev} \\
			\bottomrule
		\end{tabular}
	\end{table}
	
	\subsection{Brains Built from Language Models}
	
	\textbf{Stimuli and layers.} We draw passages of 120 to 300 words from the training part of WikiText-103 until they hold 200,000 words and choose 20,000 of these words at random, leaving out the first ten words of every passage. A model reads each passage from its start, and the state of a word is the hidden state at its last token. These 20,000 words play the part of the stimulus distribution. The units of a layer are scaled to unit variance over them, and all exact values are computed over them.
	
	\textbf{Brains.} Let $\mu_i$ and $\psi_i$ be the variances and the principal components of the layer over the 20,000 words. A brain is given by a factor $b_i$ for every component. The factor is one for a component that the brain holds as the model does and zero for a component that it lacks, and it equals $\mu_i^{-1/4}$ and $\mu_i^{1/4}$ in the brains with flatter and steeper variances. A site with weights $\theta_m$ on the units has the signal $\sum_ib_i\langle r,\psi_i\rangle\psi_i$ with $r=\sum_m\theta_mx_m$. The weights are standard normal for all units under independent sampling and for eight randomly chosen units under sparse sampling. Under redundant sampling the sites form groups of eight that share their weights. Under regional sampling all sites give weight zero to the same random half of the units. The brain with other content adds to every site a Gaussian combination of 512 further dimensions that are orthogonal to all units, with variances proportional to $1/k$ that sum to the total variance of the layer.
	
	\textbf{Recordings.} Every word is presented twice. The noise of a presentation has the variance that makes the two presentations of a site correlate at the stated value, 0.5 by default. The recording is the mean of the two presentations, and $R_n$ in \cref{tab:known} and \cref{fig:known} refers to this mean. A recording uses 2,000 words drawn from the 20,000, and five folds that keep the words of a passage in one fold.
	
	\textbf{Exact values.} The expressible share of a unit is the share of its variance on the components with $b_i>0$. The exact scores $\Rsig_n$ and $R_n$ follow from \cref{eq:scores}, with $A$ and $a$ computed over the 20,000 words in the coordinates of the principal components. The program checks this algebra against a direct least-squares computation before it runs.
	
	\textbf{Settings.} \cref{tab:settings} lists the default recording and the eleven recordings that differ from it in one factor. Each entry of \cref{tab:known,tab:settings} is a mean over four models and three recordings per model.
	
	\begin{table}[h]
		\caption{Brains built from language models under all settings, for the complete brain and for the brain without the first 64 components. Means over units, four models and three recordings.}
		\label{tab:settings}
		\centering
		\small
		\begin{tabular}{@{}lcccccccc@{}}
			\toprule
			& \multicolumn{4}{c}{Complete} & \multicolumn{4}{c}{No top} \\
			\cmidrule(lr){2-5}\cmidrule(l){6-9}
			Setting & $\bar\rho$ & $\bar R$ & $\bar\Cind$ & $\bar\Csh$ & $\bar\rho$ & $\bar R$ & $\bar\Cind$ & $\bar\Csh$ \\
			\midrule
			Default & \val{k.complete.default.true} & \val{k.complete.default.raw} & \val{k.complete.default.indep} & \val{k.complete.default.shared} & \val{k.notop.default.true} & \val{k.notop.default.raw} & \val{k.notop.default.indep} & \val{k.notop.default.shared} \\
			Sparse sampling & \val{k.complete.sampling-sparse.true} & \val{k.complete.sampling-sparse.raw} & \val{k.complete.sampling-sparse.indep} & \val{k.complete.sampling-sparse.shared} & \val{k.notop.sampling-sparse.true} & \val{k.notop.sampling-sparse.raw} & \val{k.notop.sampling-sparse.indep} & \val{k.notop.sampling-sparse.shared} \\
			Redundant sampling & \val{k.complete.sampling-redundant.true} & \val{k.complete.sampling-redundant.raw} & \val{k.complete.sampling-redundant.indep} & \val{k.complete.sampling-redundant.shared} & \val{k.notop.sampling-redundant.true} & \val{k.notop.sampling-redundant.raw} & \val{k.notop.sampling-redundant.indep} & \val{k.notop.sampling-redundant.shared} \\
			Regional sampling & \val{k.complete.sampling-regional.true} & \val{k.complete.sampling-regional.raw} & \val{k.complete.sampling-regional.indep} & \val{k.complete.sampling-regional.shared} & \val{k.notop.sampling-regional.true} & \val{k.notop.sampling-regional.raw} & \val{k.notop.sampling-regional.indep} & \val{k.notop.sampling-regional.shared} \\
			64 sites & \val{k.complete.sites-64.true} & \val{k.complete.sites-64.raw} & \val{k.complete.sites-64.indep} & \val{k.complete.sites-64.shared} & \val{k.notop.sites-64.true} & \val{k.notop.sites-64.raw} & \val{k.notop.sites-64.indep} & \val{k.notop.sites-64.shared} \\
			1{,}024 sites & \val{k.complete.sites-1024.true} & \val{k.complete.sites-1024.raw} & \val{k.complete.sites-1024.indep} & \val{k.complete.sites-1024.shared} & \val{k.notop.sites-1024.true} & \val{k.notop.sites-1024.raw} & \val{k.notop.sites-1024.indep} & \val{k.notop.sites-1024.shared} \\
			Presentations correlate at 0.2 & \val{k.complete.noise-0.2.true} & \val{k.complete.noise-0.2.raw} & \val{k.complete.noise-0.2.indep} & \val{k.complete.noise-0.2.shared} & \val{k.notop.noise-0.2.true} & \val{k.notop.noise-0.2.raw} & \val{k.notop.noise-0.2.indep} & \val{k.notop.noise-0.2.shared} \\
			Presentations correlate at 0.8 & \val{k.complete.noise-0.8.true} & \val{k.complete.noise-0.8.raw} & \val{k.complete.noise-0.8.indep} & \val{k.complete.noise-0.8.shared} & \val{k.notop.noise-0.8.true} & \val{k.notop.noise-0.8.raw} & \val{k.notop.noise-0.8.indep} & \val{k.notop.noise-0.8.shared} \\
			1{,}000 words & \val{k.complete.items-1000.true} & \val{k.complete.items-1000.raw} & \val{k.complete.items-1000.indep} & \val{k.complete.items-1000.shared} & \val{k.notop.items-1000.true} & \val{k.notop.items-1000.raw} & \val{k.notop.items-1000.indep} & \val{k.notop.items-1000.shared} \\
			4{,}000 words & \val{k.complete.items-4000.true} & \val{k.complete.items-4000.raw} & \val{k.complete.items-4000.indep} & \val{k.complete.items-4000.shared} & \val{k.notop.items-4000.true} & \val{k.notop.items-4000.raw} & \val{k.notop.items-4000.indep} & \val{k.notop.items-4000.shared} \\
			Layer at depth 0.25 & \val{k.complete.depth-0.25.true} & \val{k.complete.depth-0.25.raw} & \val{k.complete.depth-0.25.indep} & \val{k.complete.depth-0.25.shared} & \val{k.notop.depth-0.25.true} & \val{k.notop.depth-0.25.raw} & \val{k.notop.depth-0.25.indep} & \val{k.notop.depth-0.25.shared} \\
			Layer at depth 0.75 & \val{k.complete.depth-0.75.true} & \val{k.complete.depth-0.75.raw} & \val{k.complete.depth-0.75.indep} & \val{k.complete.depth-0.75.shared} & \val{k.notop.depth-0.75.true} & \val{k.notop.depth-0.75.raw} & \val{k.notop.depth-0.75.indep} & \val{k.notop.depth-0.75.shared} \\
			\bottomrule
		\end{tabular}
	\end{table}
	
	\subsection{Recordings}
	\label{app:recdetails}
	
	\textbf{Monkey recordings.} The data are the responses released with \citet{muzellec2026}. The release holds one macaque, 1,320 images and the \val{mk.m2.stored} sites that the authors selected, with every response averaged over time and up to \val{mk.m2.repeats} repetitions of every image. For every image and site the repetitions are assigned in turn to two presentations. A site is kept when the rank correlation between its two presentations, after the Spearman-Brown correction, is at least 0.7, the criterion of the original study. This leaves \val{mk.m2.kept} sites. The recording is the mean of the two presentations. The models are those whose activations were released with the data (\val{mk.models}). A layer with more than 20,000 units enters with a random sample of 20,000 units. The ten folds are random sets of images.
	
	\textbf{ECoG.} The Podcast dataset provides the high-gamma power of the electrodes of nine patients during one story \citep{zada2025}. We scale each electrode to unit variance and average its power in four windows of 200 ms after the onset of each word, which gives four response columns per electrode. The five folds are consecutive blocks of words, and words within two seconds of a test block are not used for fitting.
	
	\textbf{fMRI.} The Narratives dataset provides responses on the cortical surface at a repetition time of 1.5 s \citep{nastase2021}. We use \val{ds.narratives.stories} stories, the surface data without spatial smoothing in the fsaverage6 space, and the exclusions that the dataset recommends, which leaves \val{ds.narratives.scans} scans of single participants. For each story the participants are split at random into two halves. The responses of each half are averaged, which gives two presentations of every story, and the recording is their mean. The five folds are sets of stories, and versions of one story stay in one fold.
	
	\textbf{Sites.} All monkey sites that pass the reliability criterion enter the analysis. In the human datasets we score every response column by how well other brains predict it, with the regression described below. For an electrode the other brains are the other eight patients. For a vertex they are the other half of the participants, with their whole cortex at lags of $-1$, 0 and 1 volumes, and the score is the mean of the two directions. We keep the fifth of the columns with the highest scores, separately in each patient in ECoG. In ECoG this leaves \val{t3.podcast.columns} columns of \val{t3.podcast.sites} electrodes. The same scores are the ceilings of \cref{tab:recordings}. For the monkey the ceiling is the score with which the sites of one presentation predict a site in the other, averaged over the two directions.
	
	\textbf{Language models.} \cref{tab:models} lists the models \citep{radford2019,biderman2023,olmo2,llama3,gemma2}. The models with random weights keep the architecture and the tokenizer of the trained model. A model reads a story in windows of 200 words with the 200 preceding words as context. The state of a word is the hidden state at its last token. In ECoG the rows of the data are words. For fMRI the word states are resampled to the times of the volumes with a Lanczos filter \citep{huth2016}. Units are scaled to unit variance within each story. The features of the sparse autoencoders are those of the 16k residual-stream autoencoders of Gemma Scope at the middle layer \citep{lieberum2024}, restricted to features that are active on at least 0.5 percent of the words.
	
	\begin{table}[h]
		\caption{Language models. A layer is given by the number of the block whose output is read. Parameter counts exclude the output layer. The four models with three layers build the brains of \cref{sec:known}.}
		\label{tab:models}
		\centering
		\small
		\begin{tabular}{@{}lcccccc@{}}
			\toprule
			& Parameters & & & \multicolumn{3}{c}{Layer at depth} \\
			\cmidrule(l){5-7}
			Model & (millions) & Blocks & Width & 0.25 & 0.5 & 0.75 \\
			\midrule
			GPT-2 small & \val{m.gpt2-small.params} & \val{m.gpt2-small.blocks} & \val{m.gpt2-small.width} & -- & \val{m.gpt2-small.layer.0.5} & -- \\
			GPT-2 medium & \val{m.gpt2-medium.params} & \val{m.gpt2-medium.blocks} & \val{m.gpt2-medium.width} & -- & \val{m.gpt2-medium.layer.0.5} & -- \\
			GPT-2 large & \val{m.gpt2-large.params} & \val{m.gpt2-large.blocks} & \val{m.gpt2-large.width} & -- & \val{m.gpt2-large.layer.0.5} & -- \\
			GPT-2 XL & \val{m.gpt2-xl.params} & \val{m.gpt2-xl.blocks} & \val{m.gpt2-xl.width} & \val{m.gpt2-xl.layer.0.25} & \val{m.gpt2-xl.layer.0.5} & \val{m.gpt2-xl.layer.0.75} \\
			Pythia 70M & \val{m.pythia-70m.params} & \val{m.pythia-70m.blocks} & \val{m.pythia-70m.width} & -- & \val{m.pythia-70m.layer.0.5} & -- \\
			Pythia 160M & \val{m.pythia-160m.params} & \val{m.pythia-160m.blocks} & \val{m.pythia-160m.width} & -- & \val{m.pythia-160m.layer.0.5} & -- \\
			Pythia 410M & \val{m.pythia-410m.params} & \val{m.pythia-410m.blocks} & \val{m.pythia-410m.width} & -- & \val{m.pythia-410m.layer.0.5} & -- \\
			Pythia 1B & \val{m.pythia-1b.params} & \val{m.pythia-1b.blocks} & \val{m.pythia-1b.width} & -- & \val{m.pythia-1b.layer.0.5} & -- \\
			Pythia 1.4B & \val{m.pythia-1.4b.params} & \val{m.pythia-1.4b.blocks} & \val{m.pythia-1.4b.width} & \val{m.pythia-1.4b.layer.0.25} & \val{m.pythia-1.4b.layer.0.5} & \val{m.pythia-1.4b.layer.0.75} \\
			Pythia 2.8B & \val{m.pythia-2.8b.params} & \val{m.pythia-2.8b.blocks} & \val{m.pythia-2.8b.width} & -- & \val{m.pythia-2.8b.layer.0.5} & -- \\
			Pythia 6.9B & \val{m.pythia-6.9b.params} & \val{m.pythia-6.9b.blocks} & \val{m.pythia-6.9b.width} & -- & \val{m.pythia-6.9b.layer.0.5} & -- \\
			OLMo 2 1B & \val{m.olmo2-1b.params} & \val{m.olmo2-1b.blocks} & \val{m.olmo2-1b.width} & -- & \val{m.olmo2-1b.layer.0.5} & -- \\
			OLMo 2 7B & \val{m.olmo2-7b.params} & \val{m.olmo2-7b.blocks} & \val{m.olmo2-7b.width} & -- & \val{m.olmo2-7b.layer.0.5} & -- \\
			Llama 3.2 1B & \val{m.llama3.2-1b.params} & \val{m.llama3.2-1b.blocks} & \val{m.llama3.2-1b.width} & -- & \val{m.llama3.2-1b.layer.0.5} & -- \\
			Llama 3.2 3B & \val{m.llama3.2-3b.params} & \val{m.llama3.2-3b.blocks} & \val{m.llama3.2-3b.width} & \val{m.llama3.2-3b.layer.0.25} & \val{m.llama3.2-3b.layer.0.5} & \val{m.llama3.2-3b.layer.0.75} \\
			Llama 3.1 8B & \val{m.llama3.1-8b.params} & \val{m.llama3.1-8b.blocks} & \val{m.llama3.1-8b.width} & -- & \val{m.llama3.1-8b.layer.0.5} & -- \\
			Gemma 2 2B & \val{m.gemma2-2b.params} & \val{m.gemma2-2b.blocks} & \val{m.gemma2-2b.width} & \val{m.gemma2-2b.layer.0.25} & \val{m.gemma2-2b.layer.0.5} & \val{m.gemma2-2b.layer.0.75} \\
			Gemma 2 9B & \val{m.gemma2-9b.params} & \val{m.gemma2-9b.blocks} & \val{m.gemma2-9b.width} & -- & \val{m.gemma2-9b.layer.0.5} & -- \\
			GPT-2 small, random weights & \val{m.gpt2-small-random.params} & \val{m.gpt2-small-random.blocks} & \val{m.gpt2-small-random.width} & -- & \val{m.gpt2-small-random.layer.0.5} & -- \\
			Pythia 410M, random weights & \val{m.pythia-410m-random.params} & \val{m.pythia-410m-random.blocks} & \val{m.pythia-410m-random.width} & -- & \val{m.pythia-410m-random.layer.0.5} & -- \\
			Llama 3.2 1B, random weights & \val{m.llama3.2-1b-random.params} & \val{m.llama3.2-1b-random.blocks} & \val{m.llama3.2-1b-random.width} & -- & \val{m.llama3.2-1b-random.layer.0.5} & -- \\
			\bottomrule
		\end{tabular}
	\end{table}
	
	\textbf{Regression.} Both directions use kernel ridge regression with linear kernels that are scaled to a mean diagonal of one. Each fold is held out once. The penalty of every target is chosen from 15 values between $10^{-4}$ and $10^{3}$ on one further fold that is set aside from the training data, and the regression is then refitted on all training data. A score is the $R^2$ of the held-out predictions over all folds. In fMRI the forward regression uses the units at delays of one to five volumes, and the reverse regression uses the sites one to five volumes after the unit. The other two datasets use no delays.
	
	\textbf{References.} The references of \cref{sec:reading} are computed for every model and recording in five steps.
	\begin{enumerate}
		\item The forward regression gives a score and a held-out prediction for every site. The residual of a site is its response minus this prediction, scaled to unit variance within each story.
		\item The noise $\tilde\varepsilon_j$ of simulated site $j$ is the residual of real site $j$ with the order of the stimuli shuffled. Images are permuted. A story is rotated in time by a random shift of at least a fifth of its length, which keeps the autocorrelation of the residual. All sites share one shuffle, which keeps the correlations between their residuals.
		\item For the independent-site reference, the signal $\tilde r_j$ is a Gaussian combination of the units, drawn anew for every site. The columns of one electrode share the combination. In fMRI the signal is convolved with the canonical haemodynamic response at the delays of the regression.
		\item For the shared-tuning reference, we fit the forward weights of all sites on two disjoint halves of the stimuli, with the penalties of step 1. The products of the weights of two sites across the two halves estimate the overlap in tuning of every pair of sites without the errors of fitting, which are independent between the halves. We keep the eigenvectors of this matrix whose eigenvalues exceed the size of its most negative eigenvalue. The simulated signals are Gaussian combinations of the units that are mixed by these eigenvectors times the square roots of their eigenvalues, so that pairs of simulated sites overlap as the real pairs do. On average over models this keeps \val{t3.monkey.directions}, \val{t3.podcast.directions} and \val{t3.narratives.directions} eigenvectors in the monkey, ECoG and fMRI recordings.
		\item The share $q_j$ is read from a curve that gives the mean forward score of simulated sites as a function of their share, computed for twelve shares and 200 simulated sites each. A common factor then adjusts all shares until the mean forward score of the simulated recording is within two percent of the real one, in at most three rounds.
	\end{enumerate}
	Reverse scores of a simulated recording are computed as for the real recording, and a reference is the mean over three simulated recordings. The curves of \cref{fig:recordings} use random subsets of a sixteenth, an eighth, a quarter and a half of the sites, three subsets per size, and the same subsets for the real and the simulated recordings. They cover all models in the monkey recordings and the four models of \cref{sec:known} in the human recordings.
	
	\textbf{Scores from two presentations.} Within each training fold the sites are reduced to their first 1,024 principal components. Let $\bar y$ be the mean of the two presentations. The standard readout solves $(\hat K+\zeta I)w=\hat a$, where $\hat K$ is the sample covariance of $\bar y$, $\hat a$ the sample covariance of $\bar y$ with the unit, and $\zeta$ the penalty. Its $R^2$ on held-out stimuli is the reverse score, and its squared correlation with the unit, divided by the reliability of its prediction, is the normalised score. The noise-corrected readout replaces $\hat K$ by the sample covariance between the two presentations, made symmetric. This matrix estimates $A$ without the noise. Its eigenvalues scatter around zero in directions without signal, and the readout uses the eigenvectors whose eigenvalues exceed the size of the most negative one. The penalty of this readout is chosen by the left side of \cref{eq:corrected} on the fold that is set aside, among readouts whose two predictions have a positive covariance. \cref{tab:estimates} checks the three scores against exact values.
	
	\textbf{Uncertainty.} \cref{tab:intervals} gives intervals for the means over models from 2,000 bootstrap samples of the models. The test of a difference between reverse scores and a reference flips the sign of the difference of every model at random.
	
	\begin{table}[h]
		\caption{Means over models with 95 percent bootstrap intervals, and the mean difference between reverse score and reference with the $P$ value of a sign-flip test over models.}
		\label{tab:intervals}
		\centering
		\small
		\begin{tabular}{@{}lccccccc@{}}
			\toprule
			Recording & $\bar R$ & $\bar\Cind$ & $\bar\Csh$ & $\bar R-\bar\Cind$ & $P$ & $\bar R-\bar\Csh$ & $P$ \\
			\midrule
			Monkey IT & \val{t3.monkey.reverse} [\val{t3.monkey.reverse.lo}, \val{t3.monkey.reverse.hi}] & \val{t3.monkey.indep} [\val{t3.monkey.indep.lo}, \val{t3.monkey.indep.hi}] & \val{t3.monkey.shared} [\val{t3.monkey.shared.lo}, \val{t3.monkey.shared.hi}] & \val{t3.monkey.diffi} & \val{t3.monkey.pi} & \val{t3.monkey.diffs} & \val{t3.monkey.ps} \\
			Human ECoG & \val{t3.podcast.reverse} [\val{t3.podcast.reverse.lo}, \val{t3.podcast.reverse.hi}] & \val{t3.podcast.indep} [\val{t3.podcast.indep.lo}, \val{t3.podcast.indep.hi}] & \val{t3.podcast.shared} [\val{t3.podcast.shared.lo}, \val{t3.podcast.shared.hi}] & \val{t3.podcast.diffi} & \val{t3.podcast.pi} & \val{t3.podcast.diffs} & \val{t3.podcast.ps} \\
			Human fMRI & \val{t3.narratives.reverse} [\val{t3.narratives.reverse.lo}, \val{t3.narratives.reverse.hi}] & \val{t3.narratives.indep} [\val{t3.narratives.indep.lo}, \val{t3.narratives.indep.hi}] & \val{t3.narratives.shared} [\val{t3.narratives.shared.lo}, \val{t3.narratives.shared.hi}] & \val{t3.narratives.diffi} & \val{t3.narratives.pi} & \val{t3.narratives.diffs} & \val{t3.narratives.ps} \\
			\bottomrule
		\end{tabular}
	\end{table}
	
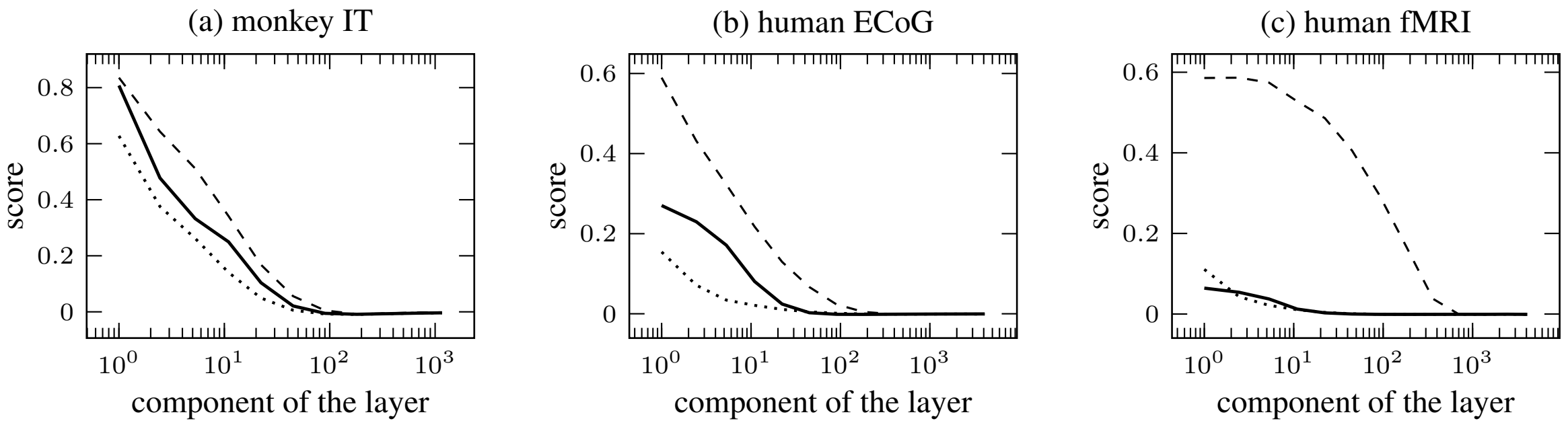
\begin{figure}[h]
		\centering
		\datarange{fig5_monkey.csv}{rangeL}{xmin=1, xmax=4000, ymin=0, ymax=1}
		\datarange{fig5_podcast.csv}{rangeM}{xmin=1, xmax=4000, ymin=0, ymax=1}
		\datarange{fig5_narratives.csv}{rangeN}{xmin=1, xmax=4000, ymin=0, ymax=1}
		\newcommand{\profilecurves}[1]{%
			\addplot[black, line width=0.9pt] table[x=rank, y=reverse, col sep=comma]{figdata/#1};
			\addplot[black, dashed, line width=0.6pt] table[x=rank, y=independent, col sep=comma]{figdata/#1};
			\addplot[black, dotted, line width=0.8pt] table[x=rank, y=shared, col sep=comma]{figdata/#1};}
		\begin{tikzpicture}
			\begin{axis}[paperaxis, rangeL, name=A, width=1.5in, height=1.1in, xmode=log,
				xlabel={component of the layer}, ylabel={score}, title={(a) monkey IT}]
				\ifdata{fig5_monkey.csv}{\profilecurves{fig5_monkey.csv}}{\nodata}
			\end{axis}
			\begin{axis}[paperaxis, rangeM, name=B, at={(A.south east)}, anchor=south west, xshift=0.6in, width=1.5in, height=1.1in, xmode=log,
				xlabel={component of the layer}, ylabel={score}, title={(b) human ECoG}]
				\ifdata{fig5_podcast.csv}{\profilecurves{fig5_podcast.csv}}{\nodata}
			\end{axis}
			\begin{axis}[paperaxis, rangeN, at={(B.south east)}, anchor=south west, xshift=0.6in, width=1.5in, height=1.1in, xmode=log,
				xlabel={component of the layer}, ylabel={score}, title={(c) human fMRI}]
				\ifdata{fig5_narratives.csv}{\profilecurves{fig5_narratives.csv}}{\nodata}
			\end{axis}
		\end{tikzpicture}
		\caption{Scores along the principal components of a layer, ordered by variance and averaged over models. Reverse score of the real recording (solid), independent-site reference (dashed) and shared-tuning reference (dotted).}
		\label{fig:profile}
	\end{figure}

	\textbf{Other units and layers.} \cref{tab:families} repeats the comparison of \cref{tab:recordings} for the principal components of a layer, for autoencoder features and for other depths. \cref{fig:profile} shows reverse scores and references along the principal components of a layer.
	
	\begin{table}[h]
		\caption{Reverse scores and references for other kinds of units and other depths. The first three rows are means over all models with such units at the middle layer. The last three rows are means over the four models that are read at three depths. A dash marks a combination that does not exist.}
		\label{tab:families}
		\centering
		\small
		\begin{tabular}{@{}lccccccccc@{}}
			\toprule
			& \multicolumn{3}{c}{Monkey IT} & \multicolumn{3}{c}{Human ECoG} & \multicolumn{3}{c}{Human fMRI} \\
			\cmidrule(lr){2-4}\cmidrule(lr){5-7}\cmidrule(l){8-10}
			Units & $\bar R$ & $\bar\Cind$ & $\bar\Csh$ & $\bar R$ & $\bar\Cind$ & $\bar\Csh$ & $\bar R$ & $\bar\Cind$ & $\bar\Csh$ \\
			\midrule
			Neurons & \val{t3.monkey.reverse} & \val{t3.monkey.indep} & \val{t3.monkey.shared} & \val{t3.podcast.reverse} & \val{t3.podcast.indep} & \val{t3.podcast.shared} & \val{t3.narratives.reverse} & \val{t3.narratives.indep} & \val{t3.narratives.shared} \\
			Principal components & \val{fam.monkey.pc.reverse} & \val{fam.monkey.pc.indep} & \val{fam.monkey.pc.shared} & \val{fam.podcast.pc.reverse} & \val{fam.podcast.pc.indep} & \val{fam.podcast.pc.shared} & \val{fam.narratives.pc.reverse} & \val{fam.narratives.pc.indep} & \val{fam.narratives.pc.shared} \\
			Autoencoder features & -- & -- & -- & \val{fam.podcast.sae.reverse} & \val{fam.podcast.sae.indep} & \val{fam.podcast.sae.shared} & \val{fam.narratives.sae.reverse} & \val{fam.narratives.sae.indep} & \val{fam.narratives.sae.shared} \\
			\addlinespace
			Neurons, depth 0.25 & -- & -- & -- & \val{dep.podcast.0.25.reverse} & \val{dep.podcast.0.25.indep} & \val{dep.podcast.0.25.shared} & \val{dep.narratives.0.25.reverse} & \val{dep.narratives.0.25.indep} & \val{dep.narratives.0.25.shared} \\
			Neurons, depth 0.5 & -- & -- & -- & \val{dep.podcast.0.5.reverse} & \val{dep.podcast.0.5.indep} & \val{dep.podcast.0.5.shared} & \val{dep.narratives.0.5.reverse} & \val{dep.narratives.0.5.indep} & \val{dep.narratives.0.5.shared} \\
			Neurons, depth 0.75 & -- & -- & -- & \val{dep.podcast.0.75.reverse} & \val{dep.podcast.0.75.indep} & \val{dep.podcast.0.75.shared} & \val{dep.narratives.0.75.reverse} & \val{dep.narratives.0.75.indep} & \val{dep.narratives.0.75.shared} \\
			\bottomrule
		\end{tabular}
	\end{table}
	
\end{document}